\documentclass[iop]{emulateapj}
\usepackage{amsmath}
\usepackage{natbib}
\usepackage{threeparttable}
\usepackage{threeparttablex}

\def\soned{{$\sigma_{\rm 1D}$}}
\newcommand{\be}{\begin{equation}}
\newcommand{\ee}{\end{equation}}
\newcommand{\ba}{\begin{eqnarray}}
\newcommand{\ea}{\end{eqnarray}}

\newcommand{\ra}{$\rightarrow$}

\def\eg{{e.g.,}}
\def\ie{{\it i.e. }}
\def\hp{{H$^{+}$}}
\def\hep{{He$^{+}$}}
\def\hetwo{{He$^{2+}$}}

\def\ctwo{{C$^{2+}$}}
\def\cthree{{C$^{3+}$}}
\def\cfour{{C$^{4+}$}}
\def\cfive{{C$^{5+}$}}
\def\oone{{O$^{+}$}}

\def\othree{{O$^{3+}$}}

\def\ofive{{O$^{5+}$}}

\def\oseven{{O$^{7+}$}}

\def\mgthree{{Mg$^{3+}$}}

\def\natwo{{Na$^{2+}$}}
\def\np{{N$^{+}$}}
\def\ntwop{{N$^{2+}$}}
\def\nthreep{{N$^{3+}$}}

\def\nsixp{{N$^{6+}$}}

\def\sitwop{{Si$^{2+}$}}
\def\sithreep{{Si$^{3+}$}}
\def\sifourp{{Si$^{4+}$}}
\def\sifivep{{Si$^{5+}$}}

\def\fefourp{{Fe$^{4+}$}}
\def\neone{{Ne$^{+}$}}
\def\netwo{{Ne$^{2+}$}}

\def\nefive{{Ne$^{5+}$}}

\def\nenine{{Ne$^{9+}$}}

\def\sfour{{S$^{4+}$}}

\def\cafour{{Ca$^{4+}$}}
\def\elec{{e$^{-}$}}

\begin{document}
\title{Atomic Chemistry in Turbulent Astrophysical Media II: Effect of the Redshift Zero Metagalactic Background}
\author{William J. Gray\altaffilmark{1} \& Evan Scannapieco\altaffilmark{2}}
\altaffiltext{1}{Lawrence Livermore National Laboratory, P.O. Box 808, L-038, Livermore, CA 94550, USA}
\altaffiltext{2}{School of Earth and Space Exploration, Arizona State University, P.O. Box 871404, Tempe, AZ 85287-1494, USA}
%\altaffiltext{3}{Lawrence Berkeley National Laboratory, Berkeley, CA 94720, USA}
%\altaffiltext{4}{Department of Physics, University of California, Berkeley, CA 94720, USA}
\keywords{ISM: abundances, ISM: atoms, astrochemistry, turbulence}

\begin{abstract}

We carry out direct numerical simulations of turbulent astrophysical media exposed to the redshift zero metagalactic background. The simulations assume solar composition and explicitly track ionizations, recombinations, and ion-by-ion radiative cooling for hydrogen, helium, carbon, nitrogen, oxygen, neon, sodium, magnesium, silicon, sulfur, calcium, and iron.   Each run reaches a global steady state that not only depends on the ionization parameter, $U,$ and mass-weighted average temperature, $T_{\rm MW},$ but also on the the one-dimensional turbulent velocity dispersion, \soned. We carry out runs that span a grid of models with $U$ ranging from 0 to 10$^{-1}$ and \soned\ ranging from 3.5 to 58 km s$^{-1}$, and we vary the product of the mean density and the driving scale of the turbulence, $nL,$  which determines the average temperature of the medium, from $nL =10^{16}$ to $nL =10^{20}$ cm$^{-2}$. The turbulent Mach numbers of our simulations vary from $M \approx 0.5$ for the lowest velocity dispersions cases to $M \approx 20$ for the largest velocity dispersion cases. When $M \lesssim1,$ turbulent effects are minimal, and the species abundances are reasonably described as those of a uniform photoionized medium at a fixed temperature.  On the other hand, when $M \gtrsim 1,$ dynamical simulations such as the ones carried out here are required to accurately predict the species abundances. We gather our results into a set of tables, to allow future redshift zero studies of the intergalactic medium to account for turbulent effects.

\end{abstract}

\section{Introduction}

Turbulence is omnipresent in astrophysics, where the Reynolds number, the ratio of the inertial forces to the viscous forces, is often orders of magnitudes higher than found on the Earth.  In many astrophysical regimes, efficient cooling causes turbulent velocities to exceed the local sound speed, which develop into supersonic turbulence. These motions compress a fraction of the medium to very high densities \citep[\eg][]{Padoan1997,MacLow2004,Federrath2008,Vazquez2012}, producing a complex, multi-phase medium. 

In addition, for many species, the recombination time and collisional ionization times are long compared to the ``eddy turn-over time" on which existing turbulent motions decay and new turbulent motions are added \citep{deAvillez2012}. Therefore, conditions experienced by a parcel of gas may change before any sort of equilibrium is reached \citep[\eg][]{Gray2015}. The ionization structure of the parcel, then, depends not only the  temperature, density, and chemical makeup but on the velocity distribution as well. For these reasons, the turbulent structure of the gas can significantly impact line emission and absorption diagnostics. Most interpretations of observed spectra, however, do not take into account these multi-phase and nonequilibrium effects.

At $z \approx 0,$ for example, \cite{Werk2014} used the Cosmic Origins Spectrograph \citep{Green2012} to measure low and intermediate ionization state ions in the circumgalactic medium (CGM) within 100 kpc of galaxies of various types \citep[see also][]{Tumlinson2013,Werk2013,Peeples2014}.  To match the observations with models that did not include turbulence, they had to adopt large ionization parameters, defined as the ratio of ionizing photon density to the hydrogen density.  Given the observed range of metagalactic and host galaxy fluxes, these ratios corresponded to densities and pressures over two orders of magnitude lower than expected from hydrostatic balance. However, given the long recombination and cooling times in the diffuse CGM, turbulent heating may be sufficient to substantially change this picture.

It is with these issues in mind that we have carried out the first exact numerical calculations of turbulent astrophysical media exposed to the redshift zero metagalactic background.  The simulations assume solar composition and explicitly track ionizations, recombinations, and ion-by-ion radiative cooling for hydrogen, helium, carbon, nitrogen, oxygen, neon, sodium, magnesium, silicon, sulfur, calcium, and iron.  The work presented here continues the work presented in \cite{Gray2015}.  Taking advantage of the scaling properties of cooling, turbulence, and the UV background, we are able to fully span the relevant range of conditions experienced in atomically cooled astrophysical plasmas: cataloging their hydrodynamical and chemical properties for comparison with more idealized simulations and tabulating their species mass fractions and Doppler parameters for use in the interpretation of future theoretical and observational studies. 

The paper is organized as follows. In \S2, we outline our numerical techniques, concentrating on the improvements to the chemical network and the addition of the UV background. We also present validation tests of the network. In \S3 we present the simulation setup and initial conditions, and in \S4 we describe our results, taking particular note of the probability density functions and the effect of the photoionizing background. Concluding remarks are given in \S5. 

\section{Numerical Method}

All simulations were performed with FLASH version 4.3 \citep{Fryxell2000}, a publicly-available hydrodynamics code. We solved the hydrodynamics equations using an unsplit solver with third-order reconstructions, which is based on the method presented in \cite{Lee2013}. To ensure the stability of the code as turbulence develops, we employed a hybrid Riemann solver which uses both an extremely accurate but somewhat fragile Harten-Lax-van Leer-Contact (HLLC) solver \citep[\eg][]{Toro1994, Toro1999} and a more robust, but more diffusive Harten Lax and van Leer (HLL) solver \citep{Einfeldt1991}. The HLLC solver is a modification to the HLL solver that includes the missing shear and contact waves, and it produces solutions that most accurately capture contact discontinuities. However, in regions with strong shocks or rarefactions, the HLLC solver can fail, and in such situations, we switch to the positivity-preserving HLL solver.  Magnetohydrodynamic effects were not included in this study, and we also ignored electron conduction, which was shown in Paper I to be unimportant in determining the species mass fractions.

Accurately determining the atomic properties of turbulent media requires the use of two code modules not available in the current public version of FLASH: a non-equilibrium ionization package that tracks the ionization state of each of the atomic species of interest, and a cooling routine that takes into account the cooling from each of these ionized states individually.  Here we describe our numerical implementation of each of these capabilities.

\subsection{Atomic Chemistry}

Our ionization network is an updated and expanded version of the network presented in Paper 1. The network now follows 240 reactions between 65 species of 12 elements: hydrogen (H and \hp), helium (He-\hetwo), carbon (C-\cfive), nitrogen (N-\nsixp), oxygen (O-\oseven), neon (Ne-\nenine), sodium (Na and \natwo), magnesium (Mg-\mgthree), silicon (Si-\sifivep), sulfur (S-\sfour), calcium (Ca-\cafour), iron (Fe-\fefourp), and electrons (\elec). As in our previous work, for each species we consider collisional ionizations by electrons, radiative and dielectronic recombinations, but unlike in Paper 1, our expanded network also includes charge transfer reactions, as well as photoionizations due to a UV background. Appendix~\ref{chemtable} lists the reactions we include, along with the reference from which each reaction rate was obtained. 

Recent observations of the circumgalactic medium around low redshift galaxies have found that dust makes up a non-zero portion of the gas \citep[\eg][]{Peek2015,Lehner2015}. Therefore, absorption of UV photons by dust can be an important issue. However, for the purposes of this study and the models presented here, we have not included this process.   Recent work has also considered the impact of non-Maxwellian electron energy distributions in single-zone models for nebular spectra \citep{Nicholls2012,Nicholls2013}.  Here we fully include the range of temperatures that occur in astrophysical turbulence, but assume that the electron energy distribution in each cell of the simulation can be adequately described as Maxwellian.

In this paper, the photoionizing reaction rates 107-158 are computed by assuming the gas is illuminated by the $z=0$ metagalactic background radiation field with a spectral shape taken from \cite{Haardt2012}. For each ion $i$, the photoionization rate is computed as
\be
\Gamma_i \equiv \int d\nu \frac{4\pi\mathcal{J}_{\nu}}{h\nu}\sigma_i(\nu),
\ee
where $\mathcal{J}_{\nu}$ is the background intensity in units of ergs s$^{-1}$ cm$^{-2}$ Hz$^{-1}$ sr$^{-1}$ and $\sigma_i(\nu)$ is the photoionization cross section, taken from \cite{Verner1996B}. 

In order to account for material under the full range of conditions seen in the intergalactic medium, we have chosen to parameterize the background in term of the flux at the Lyman limit, $\mathcal{J}_{\nu}=$10$^{-21}J_{21}$ ergs s$^{-1}$ cm$^{-2}$ Hz$^{-1}$ sr$^{-1}$. This allows us to vary the background during runtime, such that a single run is able to generate results applicable to a wide range of background levels. The ionization parameter $U,$ defined as is the ratio of the number of ionizing photons to the number density of hydrogen, is then given by
\be
U \equiv \frac{\Phi({\rm H})}{ {\rm n(H)} c} = \frac{4\pi}{ {\rm n(H) } c} \int \frac{\mathcal{J}_{\nu}}{h\nu} d\nu = \frac{n_{\gamma}}{{\rm n(H)}},
\ee
where $\Phi({\rm H})$ is the surface density of ionizing photons, ${\rm n(H)}$ is the total hydrogen number density, and $n_{\gamma}$ is equivalent to $\Phi({\rm H})/c$.
For a given background shape, $n_{\gamma}$ is precomputed and stored, such that for a target ionization parameter, $J_{21}$ can be computed and used in both the chemical and photoheating rates. 
To ensure that the computed photoionization rates are correct, we compared the difference between the calculated photoionization rates and those obtained by \cite{Haardt2012} for H, He, and \hep, and found a nearly identical match, with differences less than 2\%.  The method of solving this network is the same as described in Paper 1, to which we refer the interested reader for details. 

\begin{figure*}
\begin{center}
\includegraphics[trim=0.0mm 0.0mm 0.0mm 0.0mm, clip, scale=0.75]{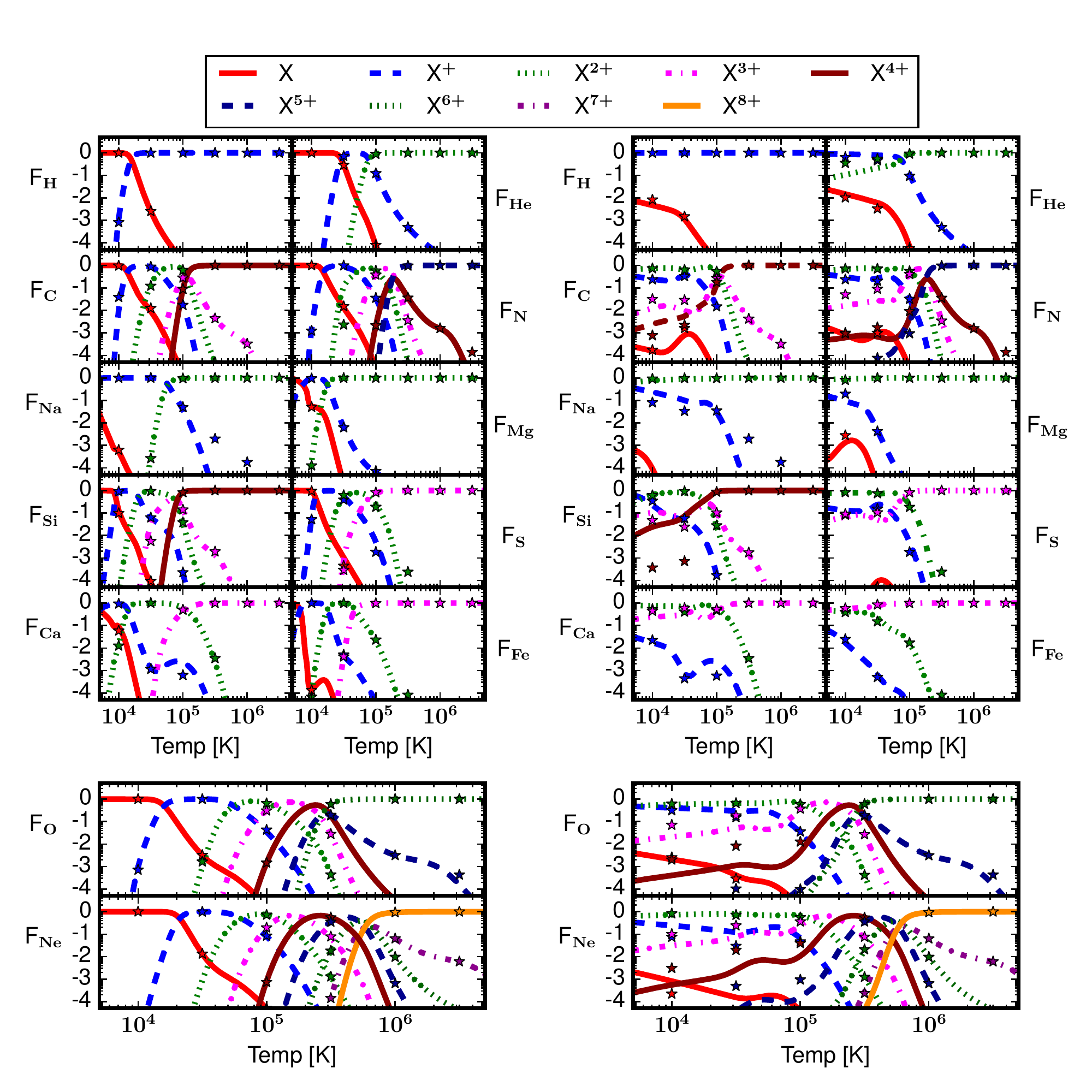}
\caption{Comparison of the species abundances between Cloudy and FLASH. {\it Left:} Comparison with a zero ionization parameter and {\it Right:} with an ionization parameter of $U = 1\times10^{-3}$. Each panel shows the results for a different element, as labeled. The lines corresponds to the Cloudy results while the points are from FLASH. We plot the equilibrium temperature along the $x$-axis and the fractional abundance of each species, \ie $F_i = n_i/n_s$ where $n_s$ is the elemental abundance, for each species as the $y$-axis. A universal legend is given at the top of the figure and the same ionization state is given by the same color and line style in each panel.}
\label{fig:chemtest}
\end{center}
\end{figure*}

To test our new network, we calculated a series of simple models with a fixed density and temperature, running each case until the species reached equilibrium. We ran a series of seven models, each with a fixed density of 2.0$\times$10$^{-20}$ g cm$^{-3},$ but with temperatures that spanned the range between 10$^4$ and 10$^7$ K. All species were assumed to be slightly ionized initially, with 90\% of each species neutral and 10\% singly ionized. 

 Tests of the network were then conducted both with and without the inclusion of an ionizing background, and the final abundances were compared to equilibrium models from Cloudy  \citep[version 10.01][]{Ferland1998}. As a test of our code in a case without an ionizing background, we used the ``coronal equilibrium'' command in Cloudy, which enforces only collisional ionizations. As a test of our code including an ionizing background, we compared our results to those from Cloudy assuming an ionization parameter of U=10$^{-3}$. To closely match the conditions implemented in Cloudy, we assumed optically-thin conditions and used the \cite{Badnell2006RR} and \cite{Badnell2006H} rates for  H$^{+}$    +  e$^{-}$    \ra  H,  He$^{+}$   +  e$^{-}$    \ra  He, and  He$^{2+}$  +  e$^{-}$    \ra  He$^{+},$ rather than the \cite{Glover2008} and \cite{Glover2009} rates taken for the rest of this study.

The left panel of Fig.~\ref{fig:chemtest} shows the results from collisional ionization test. We find that our new network gives results that are nearly identical to those obtained with Cloudy. The right panel of Fig.~\ref{fig:chemtest} shows the results from the UV background test. Again, we find very good agreement with Cloudy. Although there are a slight differences for some ions, such as He$^{2+}$ at low temperatures, it is likely this is due to the slight differences in the $z=0$ background implemented in FLASH, which uses the most recent  Haardt \& Madau spectrum, whereas Cloudy uses the spectrum from Haardt \& Madau (2005, private communication). 

\subsection{Cooling \& Heating}

We have also expanded the number of cooling terms from those in Paper I, to follow the cooling coming from our expanded species network. The cooling rates are computed in the same way as in Paper 1 and \cite{Gnat2012}, \citep[see also][]{Oppenheimer2013}, which takes advantage of several physical processes implemented in Cloudy \citep[][]{Ferland1998,Ferland2013}. 

With the inclusion of a UV background, photoheating of the gas also becomes important. The photoheating rate for each ion is computed as
\be
\mathcal{H}_i \equiv \int d\nu \frac{4\pi\mathcal{J}_{\nu}}{h\nu}\sigma_i(\nu)h(\nu-\nu_i),
\ee
where $h\nu_i$ is the ionization potential of ion $i$. In addition, we include the heating due to Compton scattering in which high-energy photons lose energy to low energy electrons. As we did for the photoionization rates, we have compared the calculated photoheating rates for H, He, \hep\, and Compton heating from those presented in \cite{Haardt2012}. In all cases, we obtain results within 1\% of those in \cite{Haardt2012}. 

\begin{figure}[t]
\begin{center}
\includegraphics[trim=0.0mm 0.0mm 0.0mm 0.0mm, clip, scale=0.40]{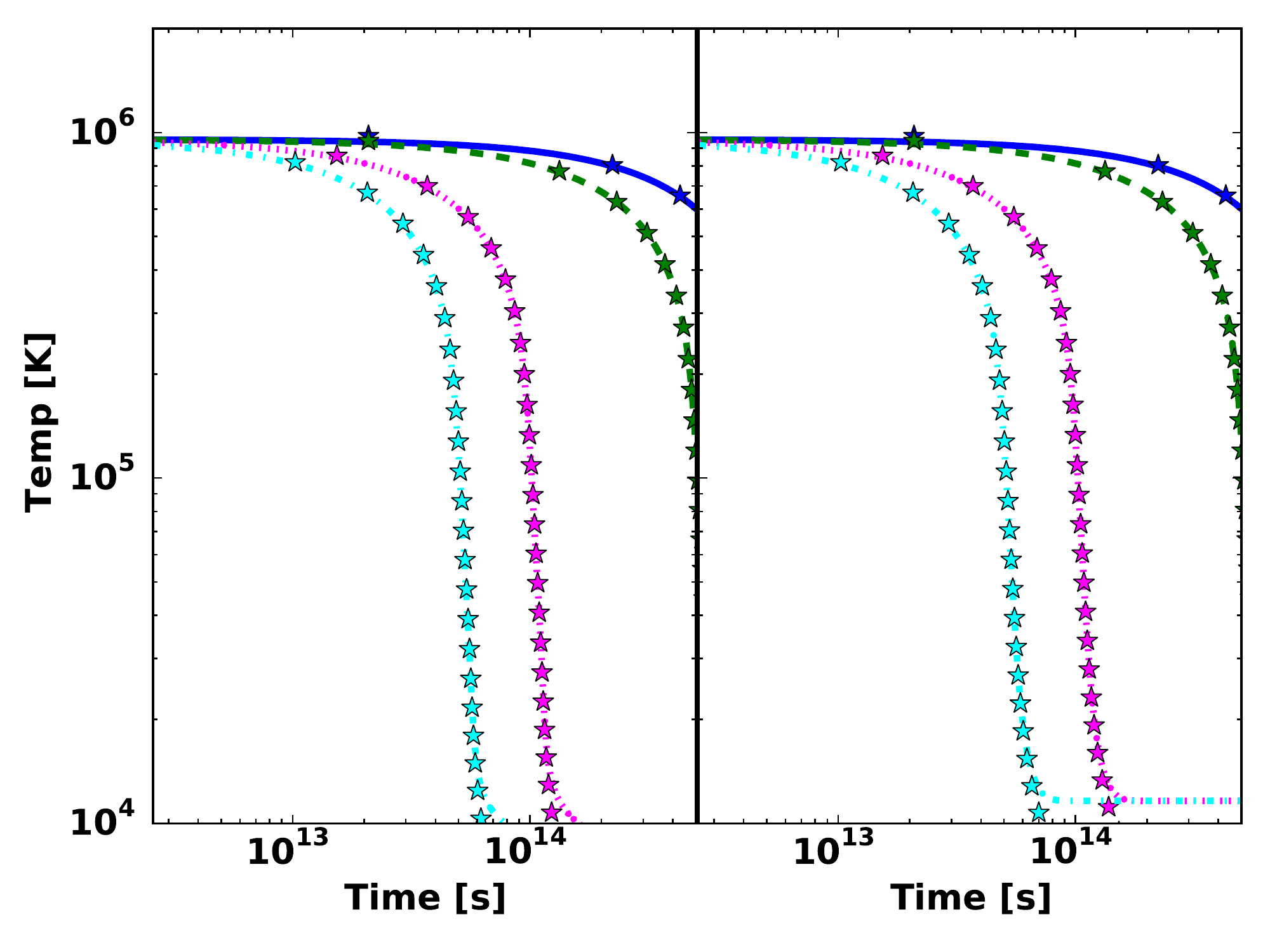}
\caption{Cooling curve comparison between Cloudy and FLASH. {\it Left Panel:} $U$=0 and {\it Right Panel:} $U$=10$^{-3}$. The $x$-axis is the logarithm of time and the $y$-axis is the gas temperature. The solid lines are the FLASH results while the star symbols are from Cloudy. The blue (solid) line shows $\rho$=5$\times$10$^{-27}$ g cm$^{-3}$, green (dashed) line shows $\rho$=1$\times$10$^{-26}$ g cm$^{-3}$, magenta (dotted) line shows $\rho$=5$\times$10$^{-26}$g cm$^{-3}$, and cyan (dash-dotted) line shows $\rho$=1$\times$10$^{-25}$g cm$^{-3}$.}
\label{fig:cooltest}
\end{center}
\end{figure}

To test our updated cooling routines, we ran a set simulations in which we initialized the gas at a temperature of $T = 10^{6}$ K, set the species  to their local ionization equilibrium values, and set the initial density  between 10$^{-27}$ and 10$^{-25}$ g cm$^{-3}$. Two cases are considered, one with no UV background and one with an $U=10^{-3}$ background. Fig.~\ref{fig:cooltest} shows the results of these runs, as compared to similar models run with Cloudy.  Again, we find that the temperature curves obtained from FLASH and Cloudy match closely. Furthermore, since all the cooling rates under consideration are two-body reactions, the cooling time should scale linearly with density, such that increasing the gas number density by ten, for example, will lead to cooling that occurs ten times faster.  For all runs, the temperature evolution captures this behavior.

\section{Model Framework \& Initial Conditions}

With the updated routines in place, we conducted a suite of simulations of turbulent media over a wide range of conditions. Each of these simulations was carried out in a $128^3$ periodic box, a resolution which is low compared to most current turbulence simulations \citep[e.g.][]{Ostriker2001,Haugen2004,Kritsuk2007,Federrath2008,Pan2010,Federrath2010,Downes2012,Gazol2013,Folini2014,Sur2014,Federrath2015} but nevertheless sufficient to obtain accurate results for the species abundances we are interested in here, as shown in Paper 1.  This allows us to significantly reduce computational costs and span a large parameter space. As in Paper 1, turbulence was continuously driven throughout the runtime, as a stochastic Ornstein-Uhlenbeck (OU) process \citep{Eswaran1988,Benzi2008}. This driving was carried out solely though solenoidal modes (\ie $\nabla \cdot {\bf F} = 0$) in the range of wavenumbers 1 $\le L_{\rm box} |{\bf k}|/2 \pi \le$ 3, such that the average forcing wavenumber was $k_f^{-1} \simeq 2 L_{\rm box}/2 \pi,$ where $L_{\rm box}$ is the size of our turbulent box, which is fixed at 100 parsecs on a side. As shown in \cite{Pan2010}, most of the turbulent kinetic energy is found in the solenoidal modes at smaller scales over a wide range of mach numbers. This justifies our choice of solenoidal only driving.  

Altogether, our simulations solve the following equations:
%\begin{widetext}
%\begin{minipage}{0.95\textwidth}
\begin{eqnarray}
\partial_t \rho + \nabla\cdot\left(\rho {\bf v}\right)&=&0 , \label{eqn:rhoeq}\\
\rho \left[
\partial_t {\bf v} + ({\bf v} \cdot \nabla) {\bf v} \right] 
&=& 
- \nabla p  +  \rho{\bf F}, \\
\partial_t E + \nabla\cdot\left[E {\bf v} \right]  & = & - 
 \nabla\cdot \left( p {\bf v} \right) + {\bf v} \rho{\bf F} \\ \nonumber
 & & - \frac{\rho^2 X_e}{m_{\rm H} m_e} \sum_i \Lambda_i(T) \frac{X_i}{\mu_i} \\ \nonumber
 & & + \chi U  \frac{\rho^2 X_{\rm H}}{m_{\rm H}^2} \sum_i \mathcal{H}_i  \frac{X_i}{\mu_i}, \\
\partial_t \rho X_i + \nabla\cdot\left(X_i \rho {\bf v}\right)&=& \rho A_i \dot R_i(U) ,
\label{eqn:xieq}
\end{eqnarray}
%\end{minipage}
%\end{widetext}
where $\rho$, ${\bf v}$, $p$, $E$, and $X_i$ denote the density, velocity, pressure, total (internal and kinetic) energy density, and mass fraction of species $i,$
m$_{\rm H}$ is the mass of hydrogen, $\mu_i$ is the atomic mass of the species $i$, $\chi$ is a constant determined by the UV background, and 
$\rho A_i \dot R_i(U)$  is the change in the mass density of species $i$ due to ionizations and recombinations for a given ionization parameter. $\dot R_i(U)$ has a form of
\ba
\dot R_{i} &=& \frac{\rho}{m_{\rm H}} \sum_{j,k} \left( \frac{X_l X_k}{\mu_l \mu_k}k_{l,k} - \frac{X_i X_j}{\mu_i \mu_j}k_{i,j} \right) \nonumber \\
& & -  \chi U \frac{\rho X_{\rm H}}{m_{\rm H}} \sum_i  \Gamma_i \frac{ X_i }{\mu_i} ,
\label{eq:Ri}
\ea
where the first term on the right hand side accounts for the formation of species $i$ through the collision of species $l$ and $k$ with rate $k_{l,k}$, the second term accounts for the destruction of $i$ by collisions with species $j$ with rate $k_{l,k}$. The third term corresponds to the photoionization of species $i$ by a background with ionization parameter $U.$

To ensure a consistent $\sigma$ driving throughout each simulation, we have implemented an adaptive method driving scheme to update the driving conditions. In essence, at each time step we compute the global average $\sigma$ and compare that to the target $\sigma,$ adjusting the forcing accordingly. At each timestep the Ornstein-Uhlenbeck process evolves the six separate phases (real and imaginary in each spatial direction) for each mode by first decaying the previous value by an exponential $y=e^{\Delta t/\tau}$, then adding a Gaussian random variable with a given variance, weighted by a driving factor of the from $\sqrt{1-y^2}$. The variance is defined at the square root of the specific energy input rate divided by the decay time. Therefore, by updating the specific energy input rate we can ensure that $\sigma$ remains nearly constant throughout each simulation. In the simulation presented below we have updated the specific energy input rate as $\propto ({ \sigma_i}/{\sigma_t})^{-5}$, where $\sigma_i$ is the current global velocity dispersion and $\sigma_t$ is the target value.

For all runs, the material was assumed to have solar metallicity, and each run was defined by an initial uniform density and the strength of turbulent forcing. The ionization state of each element was initially set to be consistent with a 10$^5$ K gas in collisional ionization equilibrium. For most cases, the eddy turnover timescale was much shorter than the timescale for the chemistry to come to equilibrium. Each model was then run normally until it reached a global steady state in terms of both the hydrodynamic variables and the chemical abundances. 

In reactions that involve free electrons recombining with ions, the optical depth of the environment becomes important. If the environment was optically thin \citep[Case A;][]{Osterbrock1989}, the ionizing photons were allowed to escape, while in the optically thick case (Case B), the photons were reabsorbed by a nearby neutral atom, which has the effect of lowering overall recombination rate. We included these effects for hydrogen, helium, and singly ionized helium which have the highest number densities and provide most of the free electrons. To best estimate the appropriate case for each run, at each time step we calculated the optical depth as,
\be
\tau_i = \bar{X}_{i} \bar{\rho} \sigma_{\nu,i} L_{\rm box} / \mu_i m_{\rm H},
\ee
where $\bar{X}_i$ is the global species mass fraction, $\bar{\rho}$ is the mean ambient density, $\sigma_{\nu,i}$ are the photoionization cross sections, and m$_{\rm H}$ is the mass of hydrogen, and $\mu$ is the atomic mass of species $i$. The recombination rate for these species is then
\be
k_{\rm rec} = e^{-\tau_i} k_{\rm A} + (1.0 - e^{-\tau_i}) k_{\rm B},
\ee
where k$_{\rm rec}$ is the new recombination rate and k$_{\rm A}$ and k$_{\rm B}$ are the Case A and B recombination rates respectively, which differ by a factor $\lesssim 2$. When the optical depth is low, e$^{-\tau_i} \approx 1.0$ and we defaulted to the Case A rate. Conversely when the optical depth is large, e$^{-\tau_{\rm H}} \approx 0.0$ and we used the Case B rate. 

Particular care must be taken for \hep recombination since some of the photons produced from recombination are used to photoionze hydrogen rather than He. As mentioned in \cite{Glover2009} and \cite{Osterbrock1989}, the net recombination rate is larger than the Case B rate. For primordial gas with low molecular abundances \cite{Osterbrock1989} found that around 68\% of the photons are absorbed by H and 32\% is absorbed by He. Therefore, we define an effective Case B \hep recombination rate as,
\be
k_{2} = 0.68 k_{A} + 0.32 k_{B} + k_{dia},
\ee
where $k_{A}$ is the Case A \hep recombination rate, $k_{B}$ is the Case B \hep recombination rate, and $k_{dia}$ is the dielectronic recombination rate. To include the effect of these photons on the photoioniation rate of H, we include an additional rate with the form of
\begin{align*}
k_{53,Case A} & = k_{A} + k_{dia} \\
k_{53,Case B} & = 0.68 (k_{A}-k_{B}) + 0.96 k_{B} + 2 k_{dia},\\ 
k_{53,Case B} & = 0.68 k_{A} + 0.28 k_{B} + 2 k_{dia},
\end{align*}
where $k_{A}$, $k_{B}$, and $k_{dia}$ are the \hep\ Case A, Case B, and dielectronic recombination rates respectively. The first term in the $k_{53,Case B}$ comes from the direct recombination to to the ground state, the second comes the nominal Case B rates, and the third comes from dielectric recombination. The factor of 0.96 for the Case B rates comes from the fact that in the low density limit 96\% of all recombinations to excited states will produce photons capable of ionizing hydrogen \cite{Glover2009,Osterbrock1989}. In addition, every dielectric recombination will produce two photons that can ionize hydrogen, one as the captured electron falls to the ground state, and one due to the radiative stabilization of the process \citep{Glover2009}.

As noted above, we have defined the cooling rates for each species between 5000 and 10$^{8}$ K. In regions where the temperature is below this range, we turned off cooling while allowing the chemistry to evolve, and enforced an absolute temperature floor at 100 K. 

\subsection{Adaptive Background}

For each combination of initial density and \soned\ we aim to study the steady state ionization fractions over a range in ionization parameter of 0 (\ie no background) to 1$\times$10$^{-1}$. For gas temperatures of T=10$^{5}$ and 10$^{6}$ K \cite{Wiersma2009} showed that the photoionizing background reduces the metal cooling efficiencies when the ionization parameter is $\gtrsim$10$^{-3}$ and $\gtrsim$10$^{-1}$ respectively. As we will show below, most of our gas is $\approx$10$^{4}$K, this justifies our upper limit on $U$. However, to avoid running a series of models in a three dimensional parameter space ($\rho$-\soned\ -U), we have devised a method of adaptively changing the background during runtime.

Each model is initialized with zero background. As the turbulence develops, the gas will reach some steady state where the abundances of each species may vary slightly over time. To compute the change in the species abundances, we compute the average species abundance over a number of time steps. This is done twice with each average done after a large number of time steps. The difference is then computed as
\be
\frac{\Delta X_i}{X_i} = \frac{\overline{X^a_i}-\overline{X^b_i}}{X^a_i},
\ee
where X$_i$ is the abundance of species $i$, $\overline{X^a_i}$ and $\overline{X^b_i}$ are the averaged species abundances. We then check and see if any $\Delta X_i$/$X_i$ is above a cutoff value. If so, we continue with the same value for the background, if not then we raise background to the next ionization parameter. To ensure that a species with a negligible abundance does not control this process, we employ a species cutoff in which only those species with abundances greater than the cutoff are used in the above equation. This saves on computation cost since as soon as a new equilibrium is reached, the simulation will raise the background without having to start the simulation from scratch. We nominally use 0.03 for both the $\Delta X_i$/$X_i$ and species cutoffs.

%As shown in Paper 1, the effect of electron conduction and increased resolution have little to no effect on either the PDFs or the final abundances. Therefore, we have ignored those processes in this work. 

\section{Results}

\subsection{Model Parameters}

Our goal is to  study the steady state ionization structure of a gas that is being stirred with 1D velocity dispersions between 12 and 58 km $\rm s^{-1}$ (corresponding to 3D velocity dispersions between 20 and 100 km $\rm s^{-1}$), over a wide range of densities. In particular, we are interested in  cases in which the heating from the turbulent stirring is balanced by atomic cooling.

As in Paper 1, the parameter space spanned by our simulations is greatly simplified by the dependencies of turbulent decay and cooling on density and length scale. In particular eqns.\ (\ref{eqn:rhoeq}-\ref{eq:Ri}) are invariant under transformations in which  $x \rightarrow \lambda x,$ $t \rightarrow \lambda t,$  $\rho \rightarrow \rho/\lambda.$  This means that the species fractions and thermal state of the gas will only depend on $U$, \soned, and the product of the mean density and the turbulent driving scale, $nL$. For the velocity dispersions we are interested in, for which cooling balances heating, this column density ranges between 10$^{16}$ and 10$^{20}$ cm$^{-2}$. 

As mentioned in Paper 1, the ratio between the recombination time and the eddy turnover time, defined as $\tau_{\rm eddy}$ = $L_{\rm box}$/(2$\sigma$) can be very large. This means that many, if not all, of the species of interest can experience a range of conditions during a single recombination time, which leads to abundance ratios that cannot be predicted from simpler local equilibrium calculations. 

A summary of the simulation runs is given in Table~\ref{tab:simruns}, with the name of each run referring to its column density and 1D velocity dispersion. For the detailed analysis that follows, we have chosen three representative models N2E17\_S12, N4E18\_S35, and N2E19\_S58, which span a range of density and velocity dispersion. We will refer to these models as Low, Medium, and High, respectively. In order to quantitatively discuss the changes due to the background, we choose four values that span the range of ionization parameters: no background (U=0), Low (U$=$10$^{-6}$), Medium (U$=$10$^{-3}$), and High (U$=$10$^{-1}$). 

\begin{figure*}
\begin{center}
\includegraphics[trim=0.0mm 0.0mm 0.0mm 0.0mm, clip, scale=0.8]{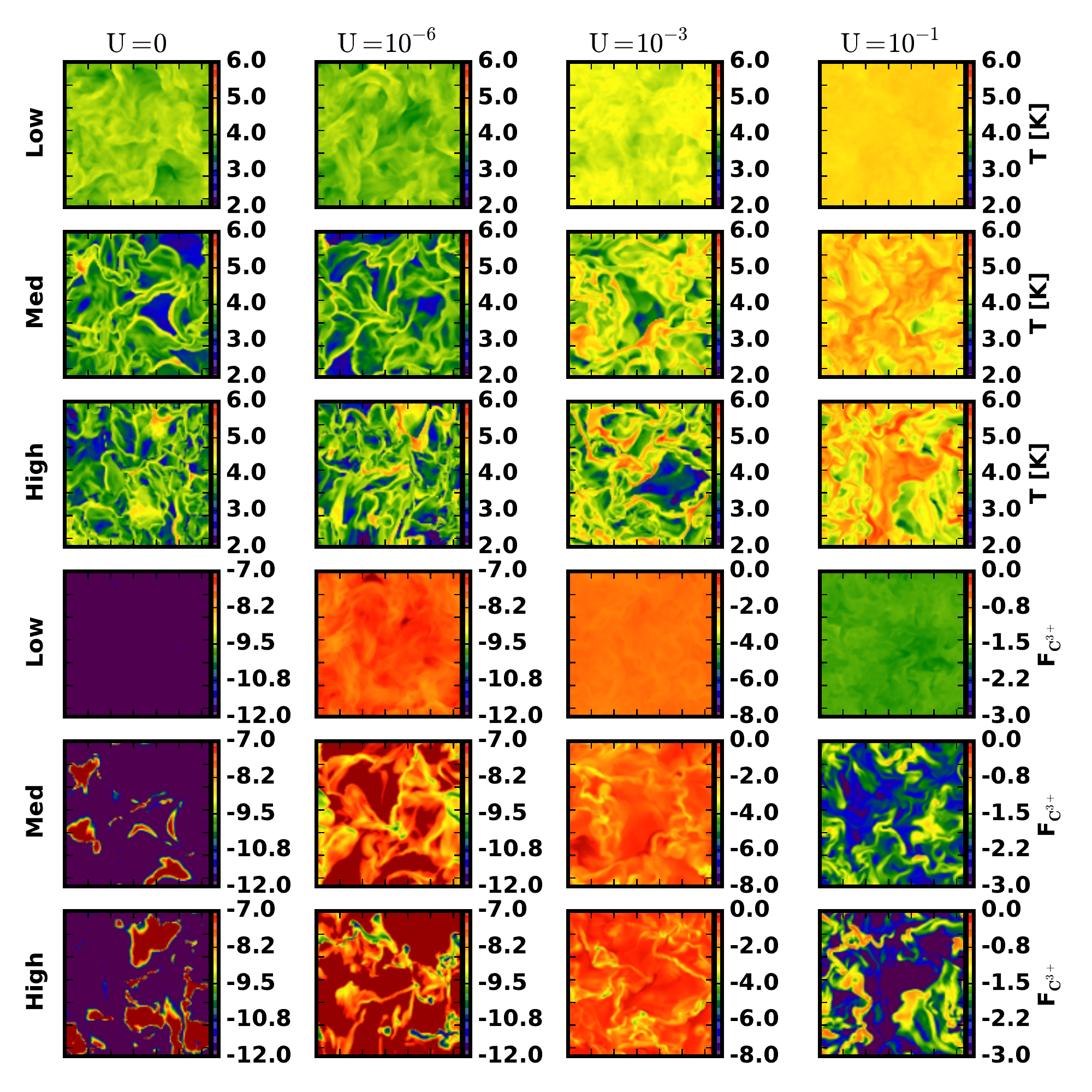}
\caption{Slices of Temperature (Top Three Rows) and F$_{\rm C^{3+}}$ (Bottom Three Rows) for our representative runs for Low, Medium, and High. Each column gives the slices for the given ionization parameter.}
\label{fig:slice}
\end{center}
\end{figure*}

Figure~\ref{fig:slice} shows slices of temperature and C$^{3+}$ for the Low, Medium, and High runs. Each column gives the results for a different ionization parameter, with a zero background in the first column, to our highest ionization parameter in the fourth column. In case of subsonic or transonic turbulence, as shown in the Low run, the temperature is largely uniform with only slight variations in terms of temperature and F$_{\rm C^{3+}}$. At high ionization parameter these differences are even less apparent since the turbulence has transitioned from transsonic to subsonic. This trend is also seen in the slices of F$_{\rm C^{3+}}$. In the case of Medium and High strong features are seen in both temperature and F$_{\rm C^{3+}}$ and are found at all ionization parameters as the Mach number stays supersonic. However, as the background increases the difference between the minimum temperature and the maximum temperature become more pronounced. F$_{\rm C^{3+}}$ also keeps its filamentary appearance at all ionization potentials. Finally, as mentioned in Paper 1, the peaks in F$_{\rm C^{3+}}$ and temperature do not overlap but show substantial variations from each other. This again shows that the gas is far from local ionization equilibrium, and that only simulations such as those described here can provide accurate results.

\begin{table*}
\centering
\resizebox{0.90\textwidth}{!}{%
\begin{threeparttable}
\caption{Summary of models run.}
\label{tab:simruns}
\begin{tabular}{|l|rrrrrrrrrrr|}
\hline
Name           & $\bar{\rho}$   & Column         & $\sigma_{1D}$  & T$_{\rm MW}$   & T$_{\rm VW}$   & M$_{\rm MW}$   & M$_{\rm VW}$   & $\bar{s}$      & $\sigma_{s}$   & s$_{skew}$     & s$_{kurt}$     \\
\hline
N1E17\_S3      & 7e-28          & 1.0E+17        & 3.5            & 0.98           & 0.98           & 0.55           & 0.56           & 0.01           & 0.11           & -0.58          & 1.25           \\
N2E16\_S12     & 2e-28          & 2.9E+16        & 11.5           & 9.52           & 9.49           & 0.47           & 0.48           & 0.00           & 0.08           & -0.74          & 1.23           \\
N1E17\_S12     & 7e-28          & 1.0E+17        & 11.5           & 5.20           & 5.14           & 0.64           & 0.65           & 0.00           & 0.16           & -0.88          & 2.06           \\
N2E17\_S12(Low)& 2e-27          & 2.9E+17        & 11.5           & 1.30           & 1.19           & 1.23           & 1.28           & -0.07          & 0.48           & -0.38          & 0.33           \\
N1E18\_S12     & 7e-27          & 1.0E+18        & 11.5           & 0.88           & 0.77           & 1.60           & 1.67           & -0.17          & 0.68           & -0.39          & 0.45           \\
N1E17\_S20     & 1e-27          & 1.4E+17        & 20.2           & 12.37          & 12.19          & 0.73           & 0.74           & 0.00           & 0.19           & -0.62          & 1.45           \\
N7E17\_S20     & 5e-27          & 7.1E+17        & 20.2           & 1.13           & 0.93           & 2.53           & 2.60           & -0.39          & 0.99           & -0.35          & 0.22           \\
N1E18\_S20     & 1e-26          & 1.4E+18        & 20.2           & 0.93           & 0.78           & 2.68           & 2.79           & -0.50          & 1.19           & -0.68          & 0.84           \\
N7E18\_S20     & 5e-26          & 7.1E+18        & 20.2           & 0.62           & 0.53           & 3.36           & 3.58           & -0.53          & 1.15           & -0.25          & 0.09           \\
N1E18\_S35     & 1e-26          & 1.4E+18        & 34.6           & 3.91           & 3.81           & 2.26           & 2.37           & -0.42          & 1.04           & -0.34          & 0.26           \\
N4E18\_S35(Med)& 3e-26          & 4.3E+18        & 34.6           & 1.01           & 0.83           & 4.29           & 4.52           & -0.84          & 1.45           & -0.24          & 0.24           \\
N1E19\_S35     & 1e-25          & 1.4E+19        & 34.6           & 0.71           & 0.60           & 5.63           & 5.81           & -1.02          & 1.59           & -0.16          & -0.25          \\
N4E19\_S35     & 3e-25          & 4.3E+19        & 34.6           & 0.55           & 0.48           & 6.74           & 6.97           & -1.07          & 1.70           & -0.46          & 0.51           \\
N5E18\_S46     & 4e-26          & 5.7E+18        & 46.2           & 1.47           & 1.87           & 4.87           & 5.20           & -0.93          & 1.51           & -0.26          & 0.25           \\
N1E19\_S46     & 1e-25          & 1.4E+19        & 46.2           & 0.88           & 0.91           & 7.40           & 7.57           & -1.22          & 1.83           & -0.42          & -0.01          \\
N5E19\_S46     & 4e-25          & 5.7E+19        & 46.2           & 0.63           & 0.61           & 8.20           & 8.79           & -1.46          & 2.01           & -0.37          & -0.08          \\
N1E20\_S46     & 1e-24          & 1.4E+20        & 46.2           & 0.51           & 0.51           & 9.09           & 9.86           & -1.62          & 2.22           & -0.53          & 0.16           \\
N1E19\_S58     & 7e-26          & 1.0E+19        & 57.7           & 1.73           & 2.84           & 5.71           & 5.66           & -1.22          & 1.70           & -0.03          & -0.28          \\
N2E19\_S58(High)& 2e-25          & 2.9E+19        & 57.7           & 0.87           & 1.05           & 8.76           & 8.85           & -1.28          & 1.82           & -0.24          & -0.22          \\
N1E20\_S58     & 7e-25          & 1.0E+20        & 57.7           & 0.61           & 0.60           & 10.55          & 10.87          & -1.45          & 1.99           & -0.30          & -0.08          \\
N2E20\_S58     & 2e-24          & 2.9E+20        & 57.7           & 0.50           & 0.51           & 11.70          & 12.00          & -1.58          & 2.17           & -0.51          & 0.15 		  \\
\hline
\end{tabular}
\begin{tablenotes}{
\item \textbf{Notes.} $\bar \rho$ is the mean density in units of  gm cm$^{-3}.$  Column is the column density in units of cm$^{-2}.$   $\sigma_{1D}$ is the 1-D velocity dispersion in units of km/s.  $T_{\rm MW}$ and $T_{\rm VW}$ are the mass-weighted and volume-weighted temperatures in units of $10^4$ K.  $M_{\rm MW}$ and $M_{\rm VW}$ are the mass-weighted and volume-weighted turbulent Mach numbers.  $\bar s$ is the volumed averaged value of $s \equiv \ln \rho /\bar \rho.$  $\sigma_{\rm s},$ $s_{\rm skew}$ and $s_{\rm kurt}$ are the rms, skewness and kurtosis excess of the volume-weighted probably density function of $s$.  Temperatures, Mach numbers, and $s$ statistics are all reported in the case without an ionizing background.\\}
\end{tablenotes}
\end{threeparttable}
}
\end{table*}

\subsection{Probability Density Functions}

\begin{figure*}
\begin{center}
\setlength{\tabcolsep}{0mm}
\begin{tabular}{cc}
\includegraphics[trim=0.0mm 0.0mm 0.0mm 0.0mm, clip, scale=0.4]{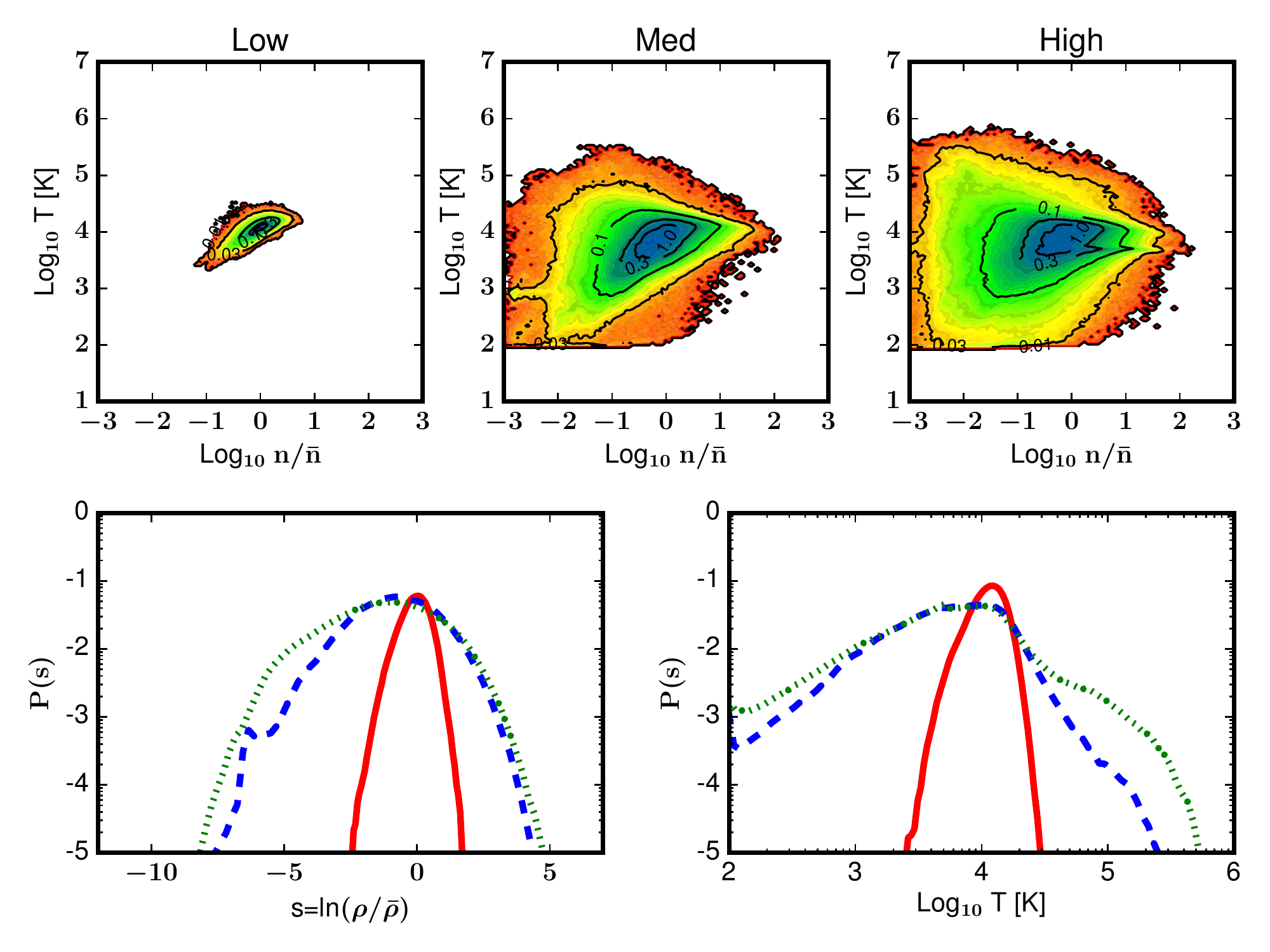} 
\includegraphics[trim=0.0mm 0.0mm 0.0mm 0.0mm, clip, scale=0.4]{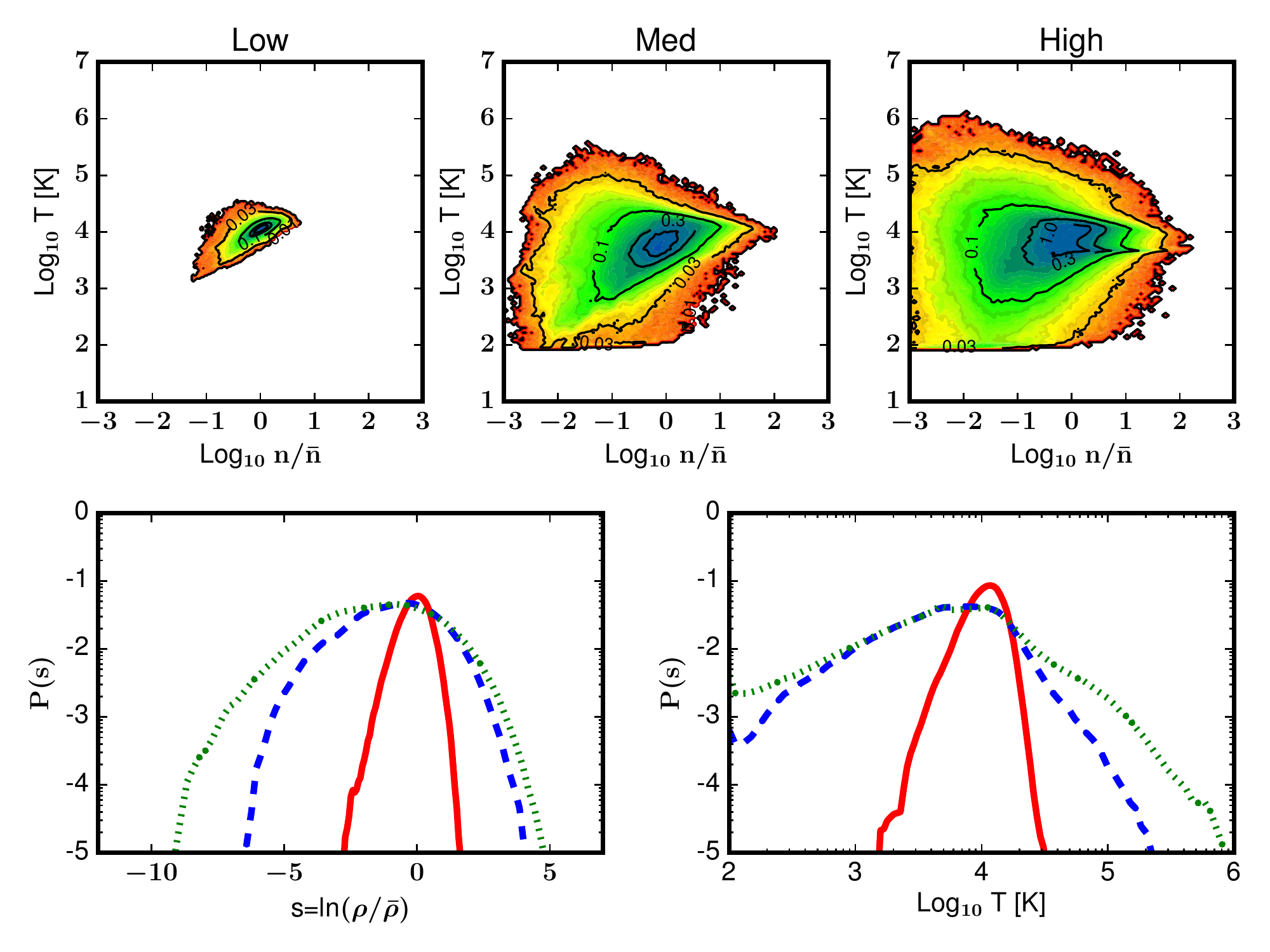} \\
\includegraphics[trim=0.0mm 0.0mm 0.0mm 0.0mm, clip, scale=0.4]{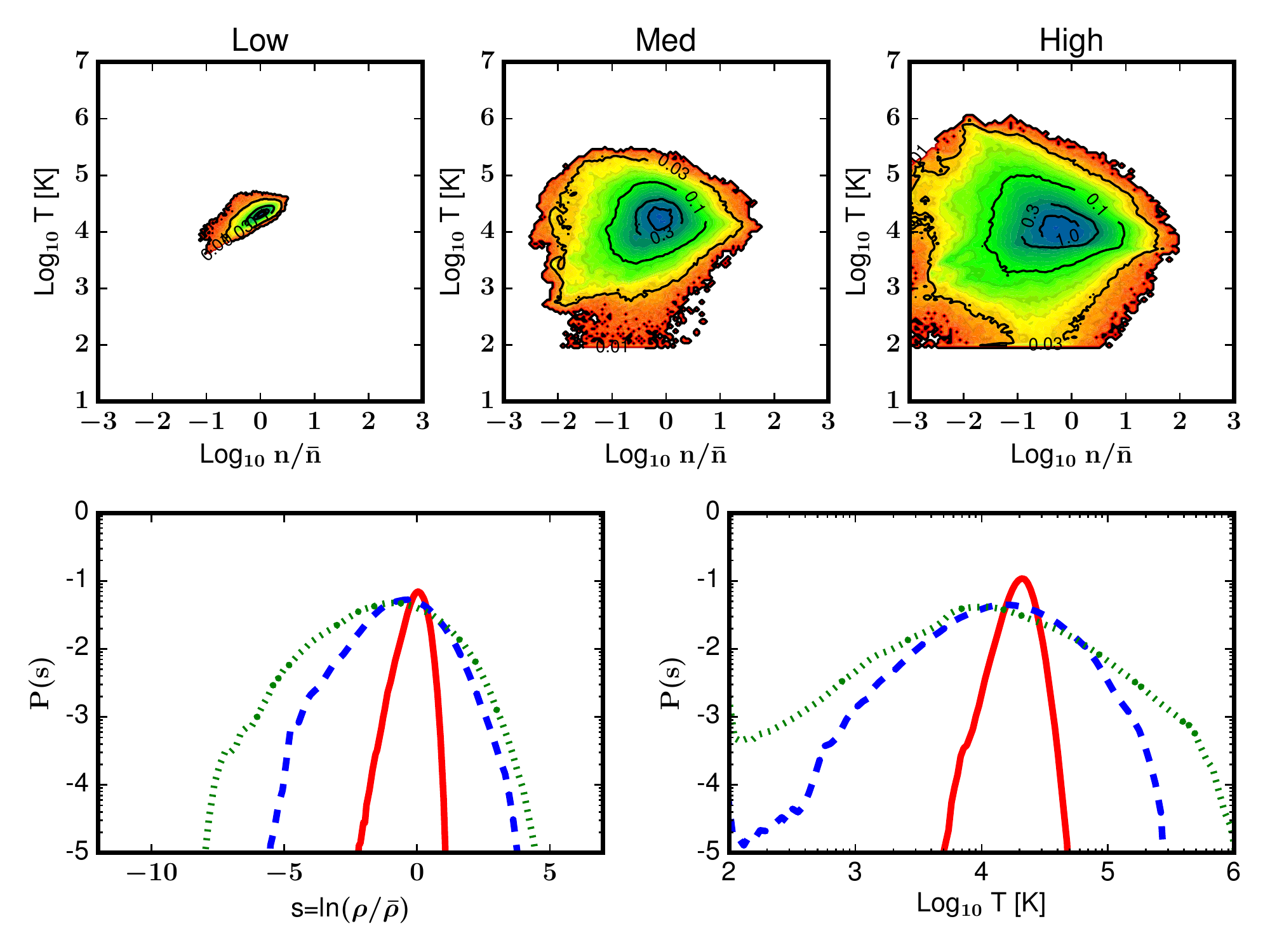} 
\includegraphics[trim=0.0mm 0.0mm 0.0mm 0.0mm, clip, scale=0.4]{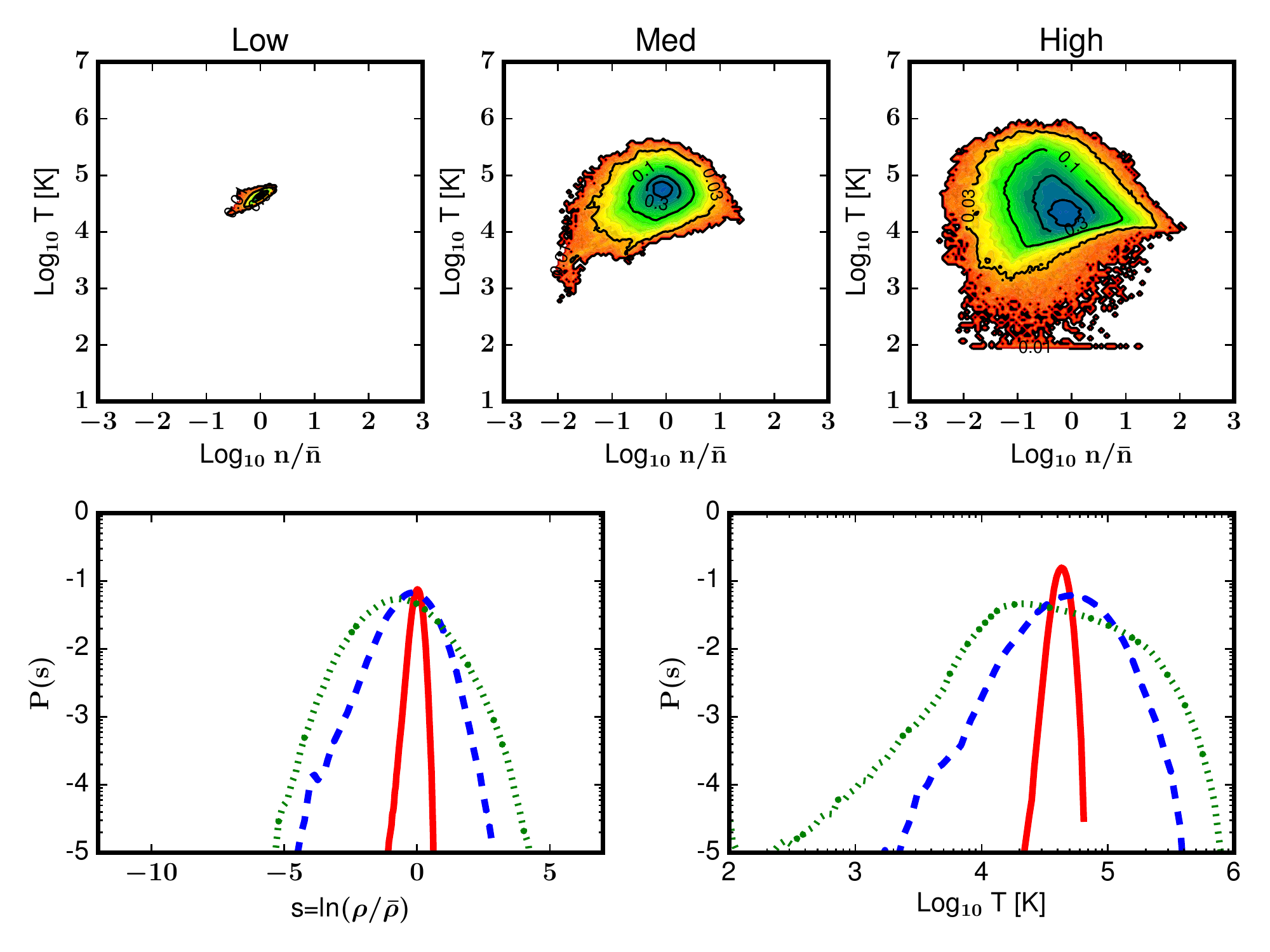} 
\end{tabular}
\caption{Set of 2D and 1D probability density functions for Low, Medium, and High over a range of ionization parameters. Each set of panels shows the following. The top set of panels shows the two dimensional probability density function for {\it Left Panel}: Low ,  {\it Middle Panel}: Medium and {\it Right Panel}: High. Temperature is given along the $y$-axis and number density is given along the $x$-axis. All contours are labeled by their values relative to the PDF bin with the most mass. The bottom set of panels show the one dimensional where the bottom left panel shows the logarithmic density PDF and the bottom right panel shows the temperature PDF. Low is shown by the (red) solid line, Medium by the (blue) dashed line, and High by the (dotted) green line. The top left set of panels shows the results for no background, the top right for a low ionization parameter, ($U=10^{-6}$), the bottom left panel shows a moderate ionization parameter, ($U=10^{-3}$), and the bottom right shows the PDFs for a high ionization parameter ($U=10^{-1}$). }
\label{fig:pdfs}
\end{center}
\end{figure*}

Figure~\ref{fig:pdfs} shows several probability density functions, which quantify the fraction of the volume in the simulation that is located at various temperatures and densities for Low, Medium, and High at several ionizations parameters. As expected from Figure~\ref{fig:slice}, the gas in Low starts as a slightly supersonic turbulent gas with a relatively small spread in term of temperature, number density, and the logarithmic density. At low and intermediate backgrounds, the spread in the PDFs remains largely the same, with only a slight decrease in the width in the temperature and logarithmic density PDFs. At large ionization parameters, however, the steady state temperature increases to nearly 10$^{5}$ K. This has caused the turbulent motions to become subsonic. As a result, the spread in the PDFs have become very small. Low also shows a rough correlation between $T$ and $n$, the total number density, where $T$ is roughly proportional to $n^{-1}$, as is expected in a constant pressure medium. In Medium and High, however, the density and temperature span a much larger range, over roughly five orders of magnitude in density and up to four orders of magnitude in temperature. The inclusion of a background does not radically change the resulting PDFs with only slight variations in the peak of the two dimensional PDFs. Furthermore, these quantities remain largely uncorrelated with each other. 

Although simulations that include full rate equations, photoionization, and cooling for a large number of atomic species are new \citep[\eg][]{Gray2015}, several previous studies of isothermal, supersonic turbulence have found that the gas approximates a lognormal distribution \citep[][]{Vazquez1994,Padoan1997,Klessen2000,Ostriker2001,Li2003,Kritsuk2007,Federrath2008,Federrath2010,Lemaster2008,Schmidt2009,Glover2010,Collins2011,Padoan2011,Price2011,Molina2012}, defined as,
\be
p(s) = \frac{1}{\sqrt{2 \pi \sigma_{\rm s}^2}} \exp\left[-\frac{ (s-s_0)^2 }{ 2\sigma_s^2} \right],
\label{eqn:lognormal}
\ee
where $s$ is the logarithmic density, $s\equiv$ln$(\rho/\bar{\rho})$ and the mean logarithmic density, $s_0$, is related to the standard deviation as $s_0=-\sigma_s^2/2.$ The lognormal distribution is a result of a constant barrage of shocks, at a particular location within the gas, with Mach numbers that are independent of the local density which cause density variations. For an isothermal gas, the resulting variance of the lognormal distribution is given by
\be
\sigma_s^2 = \ln \left( 1 + b^2 M^2 \right),
\label{eqn:sigmam}
\ee
where $b$ is a constant that depends on the forcing that drives the turbulence and $M$ is the Mach number. \cite{Federrath2008} showed that in the isothermal case, $b=1$ for purely compressive forcing and $b=1/3$ for purely solenoidal forcing. 

\begin{figure}[t]
\begin{center}
\includegraphics[trim=0.0mm 0.0mm 0.0mm 0.0mm, clip, scale=0.4]{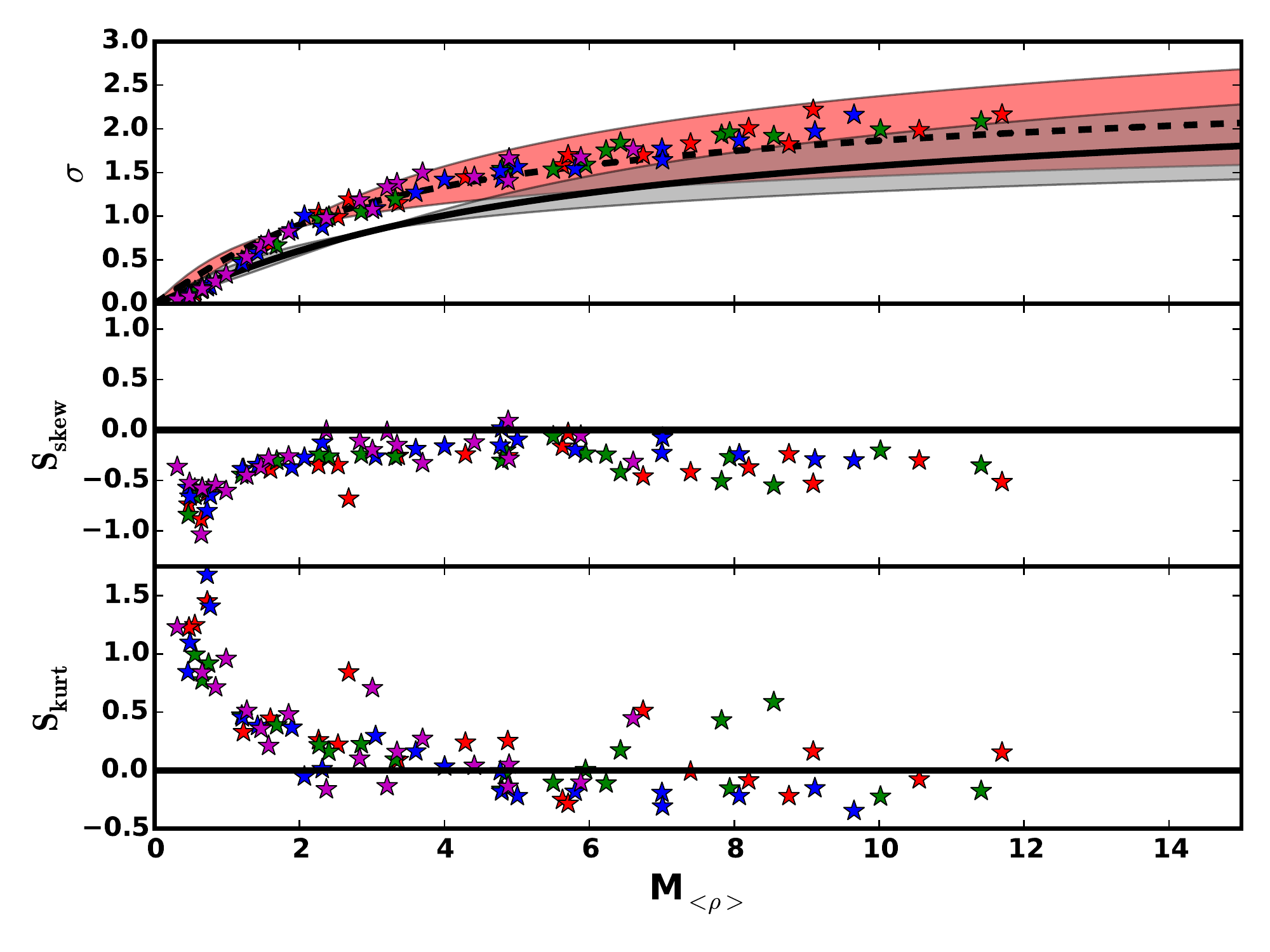}
\caption{Statistical measures of the s=ln($\rho/\bar{\rho}$) PDF for each models. The $x$-axis is the density weighted Mach number. {\it Top}: The variance in $s$ for each model. The solid (black) line shows the expected $\sigma_s-M$ relation for $b=1/3$ isothermal turbulence while the black shaded region give the expected values for polytropic turbulence with $\gamma$=0.5 to 2. The dashed (black) line and red shaded region shows the same relation for $b=0.56$. The stars represent values from each of our models. The red stars correspond to U=0, the green to $U=10^{-6}$, blue to $U=10^{-3}$, and magenta stars to $U=10^{-1}$. }
\label{fig:statfit}
\end{center}
\end{figure}

We show the logarithmic density PDFs for Low, Medium, and High for a range of ionization parameters in Figure~\ref{fig:pdfs} as the bottom right panel of each set of figures. A similar trend is seen if logarithmic density PDFs as in Paper 1. With zero background the width of the PDF scales with the strength of the driving, with Low having the smallest width and increasing with Medium and High. This trend is also seen as the strength of the background increases. In Table~\ref{tab:simruns} we give several important statistical quantities in terms of the logarithmic density for each run once it has reached steady state and before any background is added. In particular, we calculate the first four moments of the logarithmic density distribution, the mean, variance, skewness, and kurtosis. 

As mentioned in Paper 1, the mean and variance have the usual definitions. The skewness, $\langle (s-s_0)^3\rangle / \sigma^{3/2}$, measures the symmetry of the distribution. The kurtosis, $\langle (s-s_0)^4\rangle / \sigma^{2}$, gauges the peakedness of the distribution and measures the importance of the distribution tails verus the peak. A small value for the kurtosis represents a narrower distribution while a large kurtosis denotes s flatter distribution. A Gaussian distribution will have a kurtosis equal to 3. In Table~\ref{tab:simruns} we give the kurtosis excess where we subtract off the factor of 3. 

As discussed by \cite{Federrath2010}, numerical resolution plays an important role in defining the probability density, especially in tails of the distribution. In Paper 1, we showed that we obtain similar values for $s$ and $\sigma_s$ for models in our resolution but find scatter in values for the skewness and kurtosis. It is likely that this scatter would remain even as the resolution increases.  
 
Figure~\ref{fig:statfit} we show the variance, skewness, and kurtosis excess as a function of Mach number for each model at the four ionization parameters. As mentioned above, a simple relation was found between the Mach number and the density variance for isothermal turbulence. As shown in Paper 1, this relation matched the results of our solenoidally-driven gamma-law gas as long as the stirring parameter was allowed to vary. This difference is attributed to the non-isothermality of the gas. Recently, \cite{Federrath2015} studied turbulent motions using a polytropic equation of state (P=P$_{0} \left( \rho/\rho_0\right)^{\Gamma}$) over a range of polytropic exponents $\Gamma$ between 0.7 and $5/3$. Detailed simulations of molecular clouds found that $\Gamma<$1 in regions with a density range of 10 cm$^{\rm -3}$ $\lesssim$ $n$ $\gtrsim$ 10$^{4}$ cm$^{\rm -3}$ due to efficient radiative cooling, from both molecular and fine structure cooling, \citep{Glover2007a,Glover2007b}. As a result, analytic functions for the density contrast for a set of polytropic exponents are given as,
\begin{align}
\frac{\rho}{\rho_0} &= \frac{1}{8} \left(4 b^2 M^2 + b^4 M^4 + b^3 M^3 \sqrt{8+b^2M^2} \right) &\Gamma&=1/2 \label{eqn:hsigma} \\
\frac{\rho}{\rho_0} &= b^2 M^2   &\Gamma&=1 \label{eqn:isosigma} \\
\frac{\rho}{\rho_0} &= \frac{1}{2} \left(-1.0 + \sqrt{1+ 8b^2+M^2}\right)  &\Gamma&=2 \label{eqn:tsigma},
\end{align}
where $\Gamma$ is the polytropic index, $b$ is the forcing constant, and $M$ is the Mach number. The density variance is then given as $\sigma^2_s$=ln(1+$\rho/\rho_0$). In the top panel of Figure~\ref{fig:statfit} the points represent the variance found in our models over our range of ionization parameters. The solid black line shows the expected isothermal $\sigma_s-M$ relation for a stirring parameter of $b$=1/3 given by eqn.~(\ref{eqn:isosigma}), consistent with solenoidal driving. The shaded black region shows the range of $\sigma_s$ bounded by polytropic indexes between 0.5 and 2.0, given by eqns.~(\ref{eqn:hsigma}) and (\ref{eqn:tsigma}). The dashed black line and the red shaded region give the expected range in $\sigma_s$ for a stirring parameter of 0.56, determined by a least-square fit to eqn.~(\ref{eqn:isosigma}). For supersonic flows, eqn.~(\ref{eqn:isosigma})  matches the data well but overestimated $\sigma_s$ for subsonic flows. In a range of moderate Mach numbers (2$<M<$6) there is a slight systematic effect of increasing the variance as the ionization parameter increases.

\subsection{Effect of the UV Background}

\begin{figure*}
\begin{center}
\setlength{\tabcolsep}{0mm}
\includegraphics[trim=0.0mm 0.0mm 0.0mm 0.0mm, clip, scale=0.70]{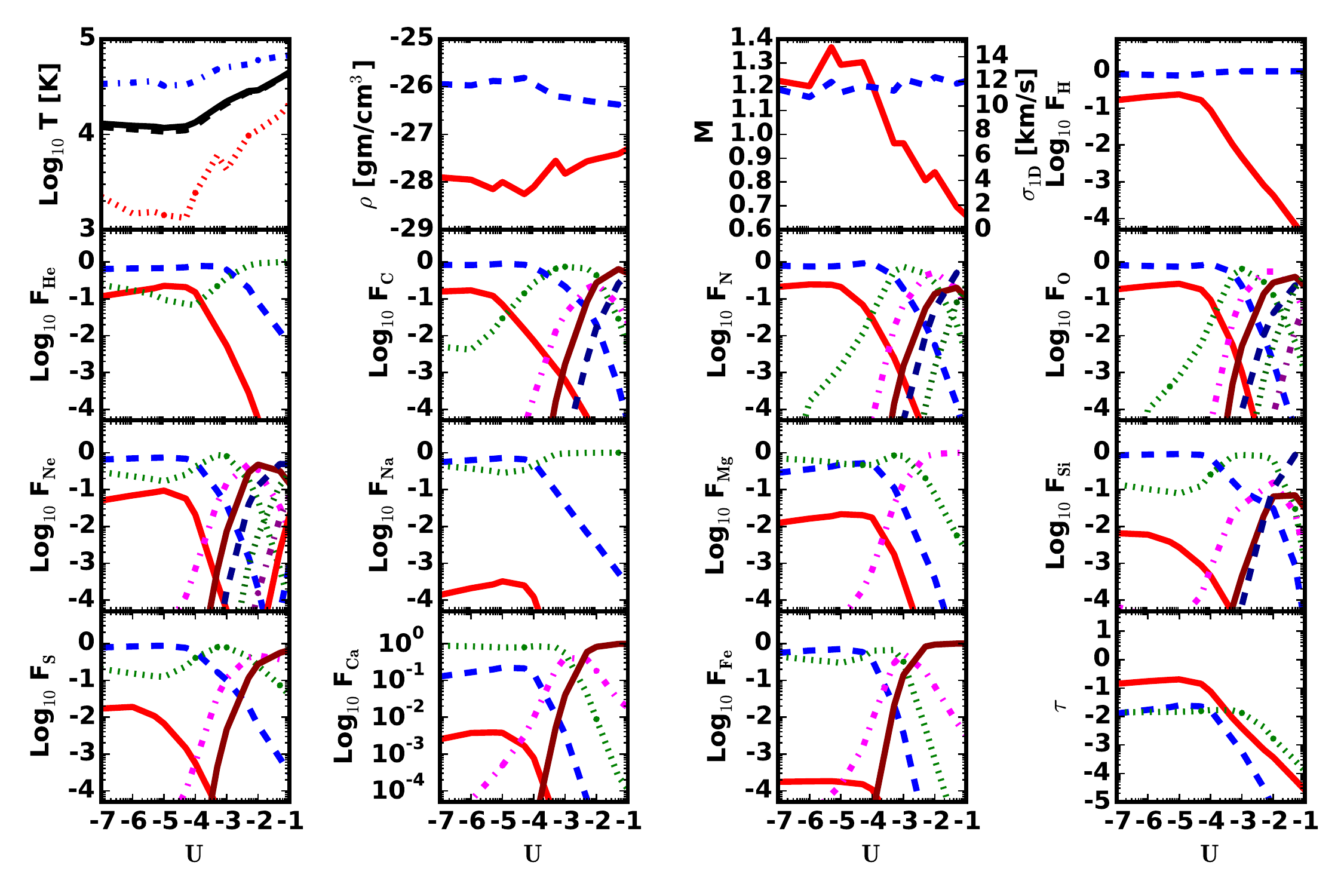}
\caption{Hydrodynamical and Chemical evolution of N2E17\_S12 (Low). {\it Top Left}: The dash-dotted (blue) line shows the maximum temperature, the dotted (red) line shows the minimum temperature, the solid (black) line shows the density weighted average temperature, and the (dotted) black line shows the volume weighted average temperature. {\it Top Left Middle}: The solid (red) line shows the minimum density while the dashed (blue) line shows the maximum density. {\it Top Right Middle}: The solid (red) line shows Mach number while the dashed (blue) line shows the 1D velocity dispersion. {\it Top Right - Bottom Middle Right}: The fractional abundance for each of the species given by the label. In each panel, each line has the same legend as Fig.~\ref{fig:chemtest}. {\it Bottom Right }: The (solid) red line shows the neutral hydrogen optical depth, the (dashed) blue line shows the neutral helium optical depth, and the (dotted) green shows the singly ionized helium optical depth. The $x$-axis in each panel is given in terms of the ionization parameter. Therefore each line shows the final steady state value at each ionization parameter. An ionization parameter less than U=$10^{-7}$ corresponds to a zero background. }
\label{fig:evos}
\end{center}
\end{figure*}

Figure~\ref{fig:evos} shows the chemical and hydrodynamical evolution of the Low run as a function of the ionization parameter.  Here the values for each quantity correspond to the steady state value at each ionization parameter.  When $U=0,$ the gas is largely either neutral or singly ionized, and the changes in the steady state values at low and moderate value of the ionization parameter are small. In fact, unlike in a purely photoionized medium, the ionization state of all the ions are essentially unchanged from $U=0$ up to an ionization parameter of $U=10^{-4}$. At higher ionization parameters,  however, the average temperature begins to rise and highly ionized species begin to appear. In particular, moderately ionized carbon, nitrogen, oxygen, neon, sulfur, and calcium become the dominant ions.

\begin{figure*}
\begin{center}
\setlength{\tabcolsep}{0mm}
\begin{tabular}{cc}
\includegraphics[trim=0.0mm 0.0mm 0.0mm 0.0mm, clip, scale=0.70]{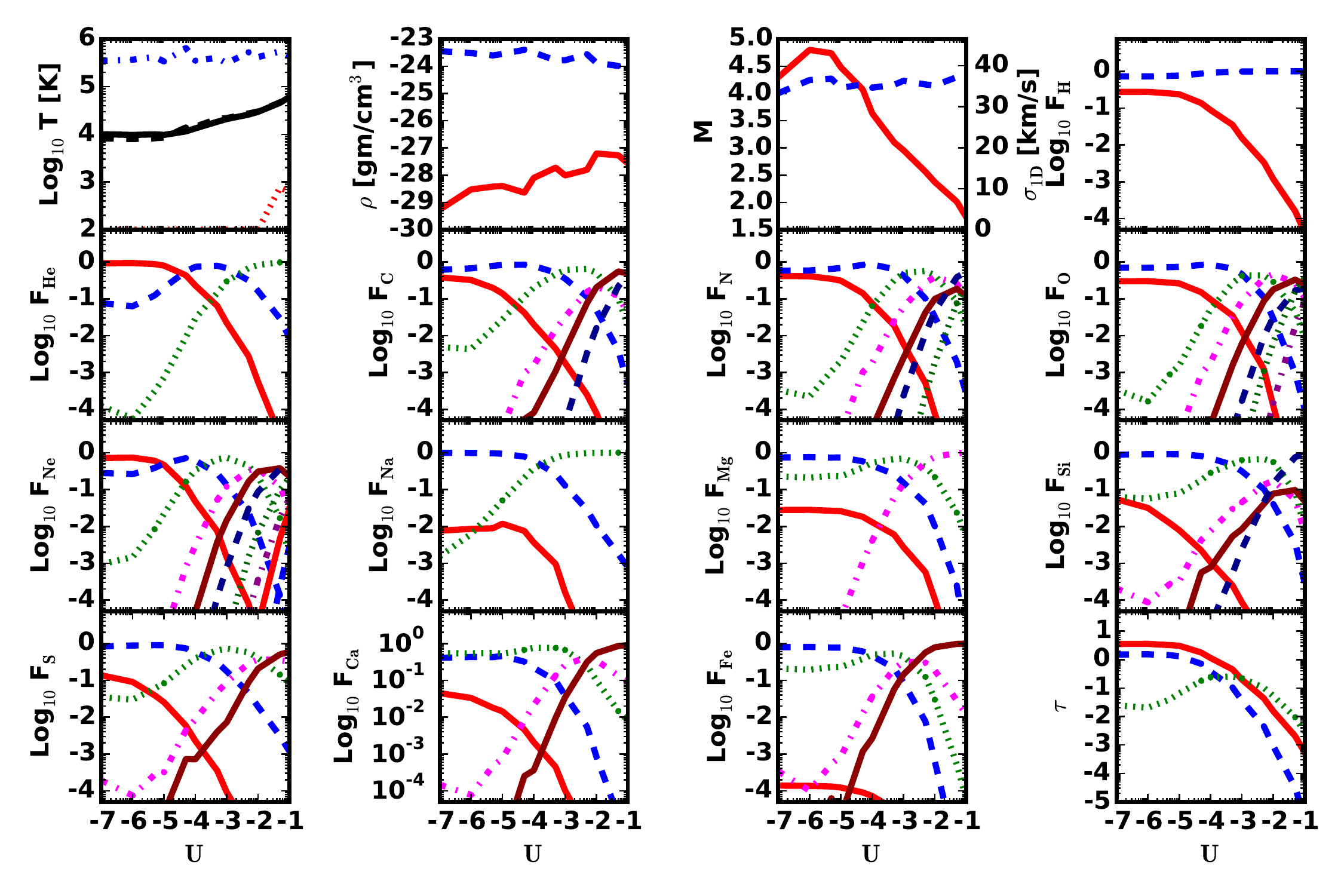} 
\end{tabular}
\caption{Hydrodynamical and Chemical evolution of N4E18\_S35 (Med). Panels and axes are the same as Fig.~\ref{fig:evos}. }
\label{fig:evos2}
\end{center}
\end{figure*}

Figure~\ref{fig:evos2} shows the evolution of the Medium run.  As in the Low case, when $U \leq 10^{-4},$ the gas is mostly neutral or singly ionized.  In fact, the gas is slightly more neutral than in the Low run, because the Medium run not only has a  higher velocity dispersion, but also a higher density, increasing the cooling rate. For example, in the Medium run, neutral neon is dominant up to U$=10^{-6}$ whereas, in the Low run, singly ionized neon is dominant in the range.  Above $U = 10^{-4},$ the increase in the temperature is the Medium run is similar to that found in the Low run, with a steady increase in temperature as a function of ionization parameter, up to 10$^{5}$K. Thus the spread in density does not show the decrease seen in the Low run, and the gas remains strongly shocked with the Mach number remaining above 1. 

\begin{figure*}
\begin{center}
\setlength{\tabcolsep}{0mm}
\begin{tabular}{cc}
\includegraphics[trim=0.0mm 0.0mm 0.0mm 0.0mm, clip, scale=0.70]{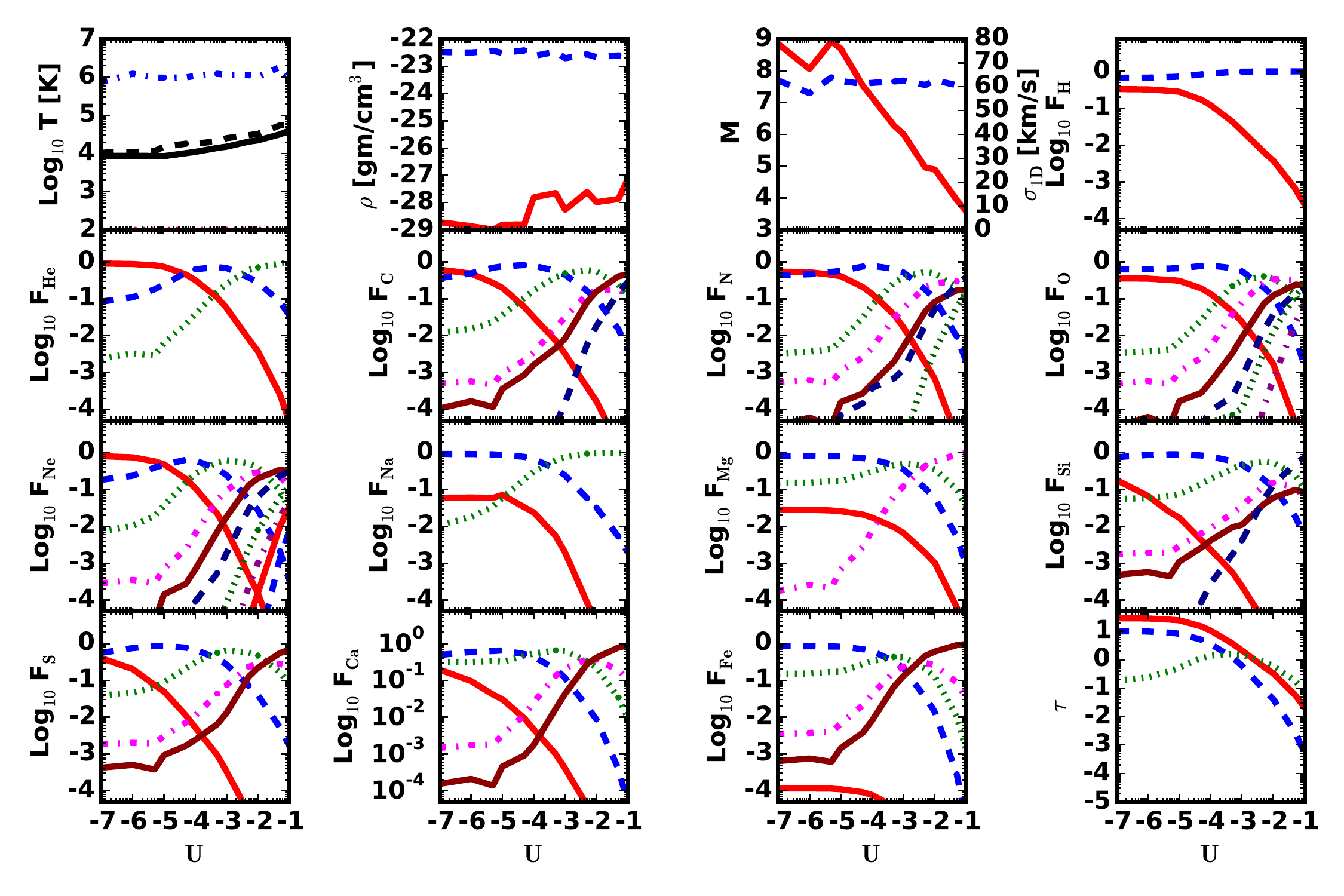} 
\end{tabular}
\caption{Hydrodynamical and Chemical evolution of N2E19\_S58 (High). Panels and axes are the same as Fig.~\ref{fig:evos}.}
\label{fig:evos3}
\end{center}
\end{figure*}

Figure~\ref{fig:evos3} shows the evolution for the High run, which has both the highest velocity dispersion and the highest density.  This follows the same general trends seen in the Low and Medium run. The gas starts mostly neutral or singly ionized and begins to transition to higher ionization states once the ionization parameter is $\geq 10^{-5}$. Compared to the Low and Medium runs,  there is even less of a variation in the hydrodynamic variables, such that the spread in density remains essentially constant as a function of ionization parameter. Finally, the High run shows a large spread in abundances for the chemical species, with higher ionization states being found at all ionization parameters. Thus, for example, \sifourp\ and \sifivep\ are found at low ionization parameters in the High run although they are almost completely absent in the Medium and Low runs.

\subsection{Ionization States}
\label{subsec_is}

The large spread in abundances for the chemical species is due to the large spread in temperature and density, most notably in the supersonic cases. In fact, the instantaneous chemical makeup in a particular location is a strong function of its thermodynamic history as well as its current temperature, density, and chemical makeup. 

\begin{figure*}
\begin{center}
\includegraphics[trim=7.0mm 7.0mm 0.0mm 10.0mm, clip, scale=0.65]{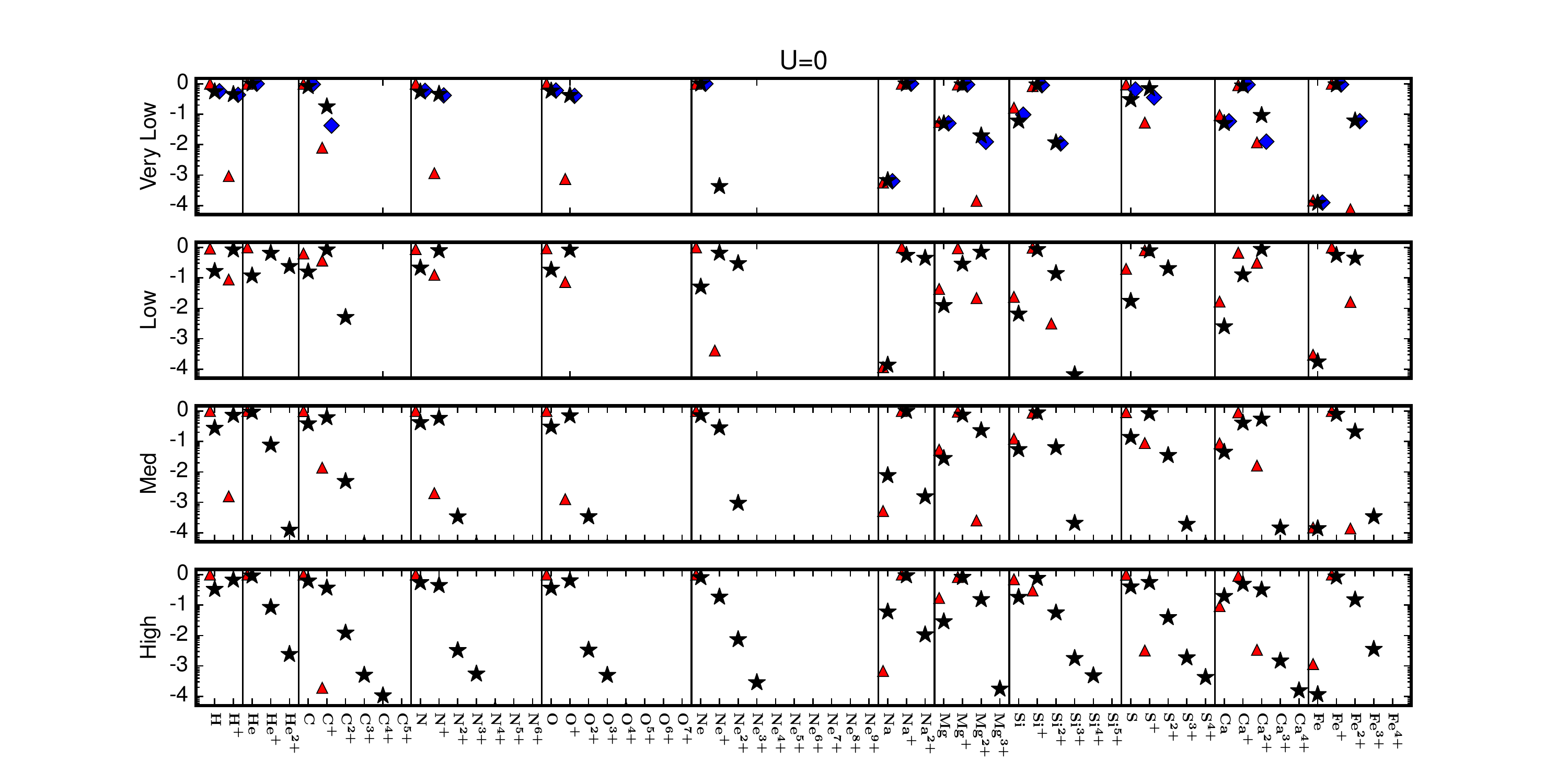} 
\caption{Steady state fractional species abundances for Very Low, Low, Medium, and High with zero UV background. The species is given along the $x$-axis while the fractional species abundance is given on the $y$-axis. The (red) triangles show F$_{\rm mean}$, the expected species abundances assuming collisional ionization equilibrium at the mass weighted mean temperature, and the (black) stars show F$_{\rm Turb}$, the final global steady state values.  In the Very Low case the blue diamonds show $F_{\rm mean}$ as calculated using the Glover \& Abel (2008) and Glover \& Savin (2009)  rates.}
\label{fig:chemcompa}
\end{center}
\end{figure*}

\begin{figure*}
\begin{center}
\includegraphics[trim=7.0mm 7.0mm 0.0mm 10.0mm, clip, scale=0.65]{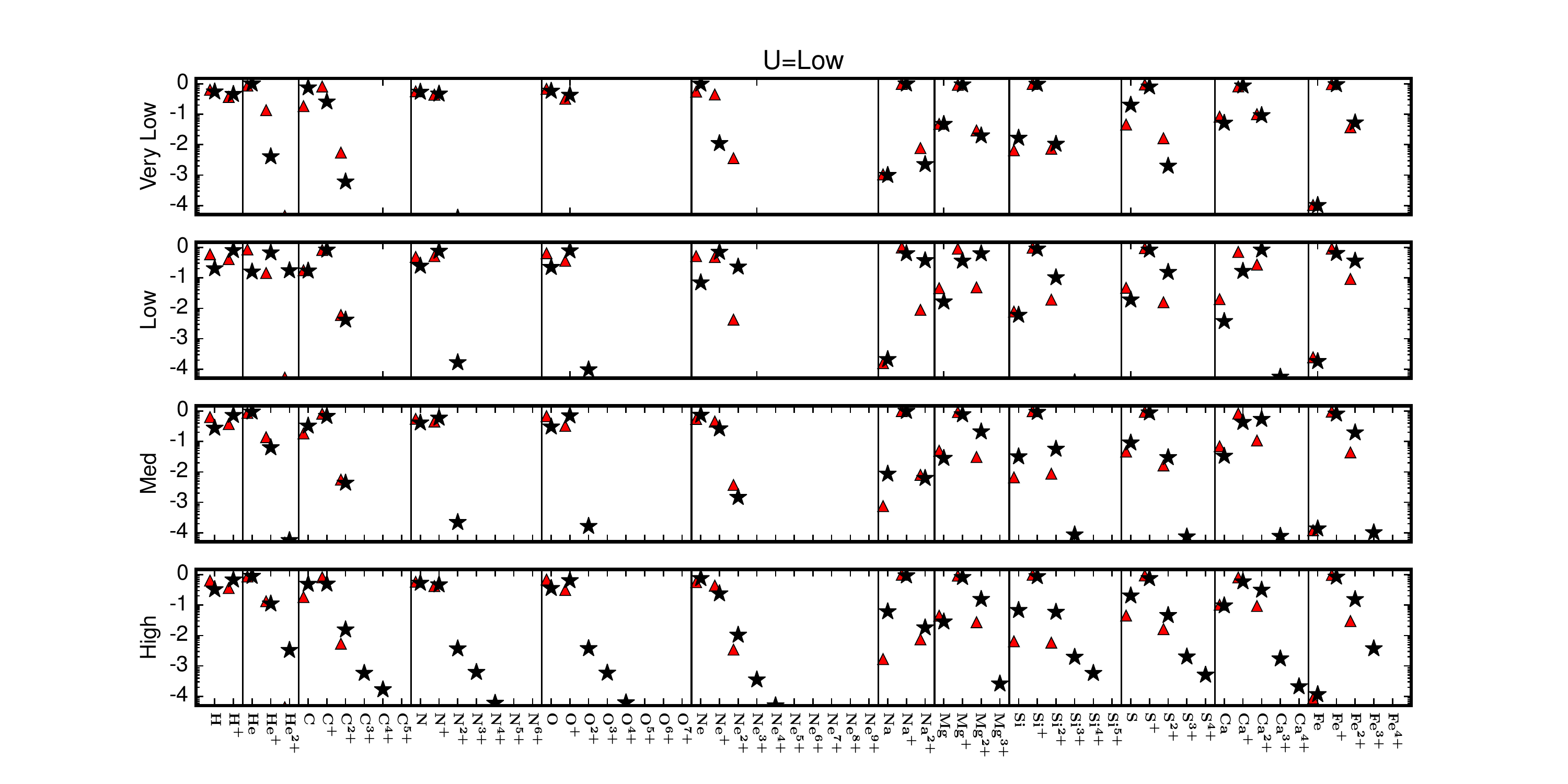}
\caption{Steady state fractional species abundances for Very Low, Low, Medium, and High with an ionization parameter of U=10$^{-6}$. Symbols are the same as Figure~\ref{fig:chemcompa}. }
\label{fig:chemcompb}
\end{center}
\end{figure*}

\begin{figure*}
\begin{center}
\includegraphics[trim=7.0mm 7.0mm 0.0mm 10.0mm, clip, scale=0.65]{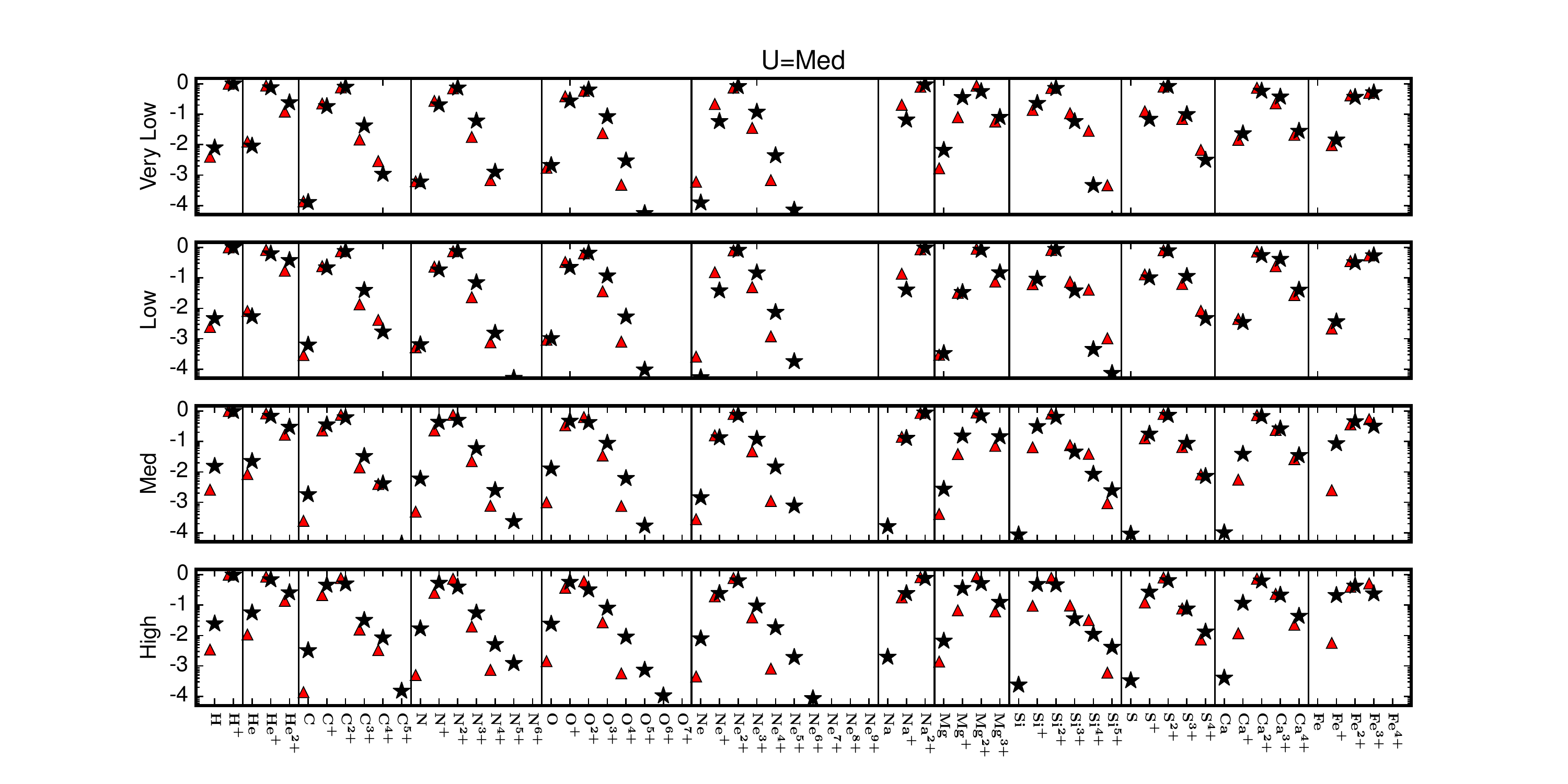} 
\caption{Steady state fractional species abundances for Very Low, Low, Medium, and High with an ionization parameter of U=10$^{-3}$. Symbols are the same as Figure~\ref{fig:chemcompa}.}
\label{fig:chemcompc}
\end{center}
\end{figure*}

\begin{figure*}
\begin{center}
\includegraphics[trim=7.0mm 7.0mm 0.0mm 10.0mm, clip, scale=0.65]{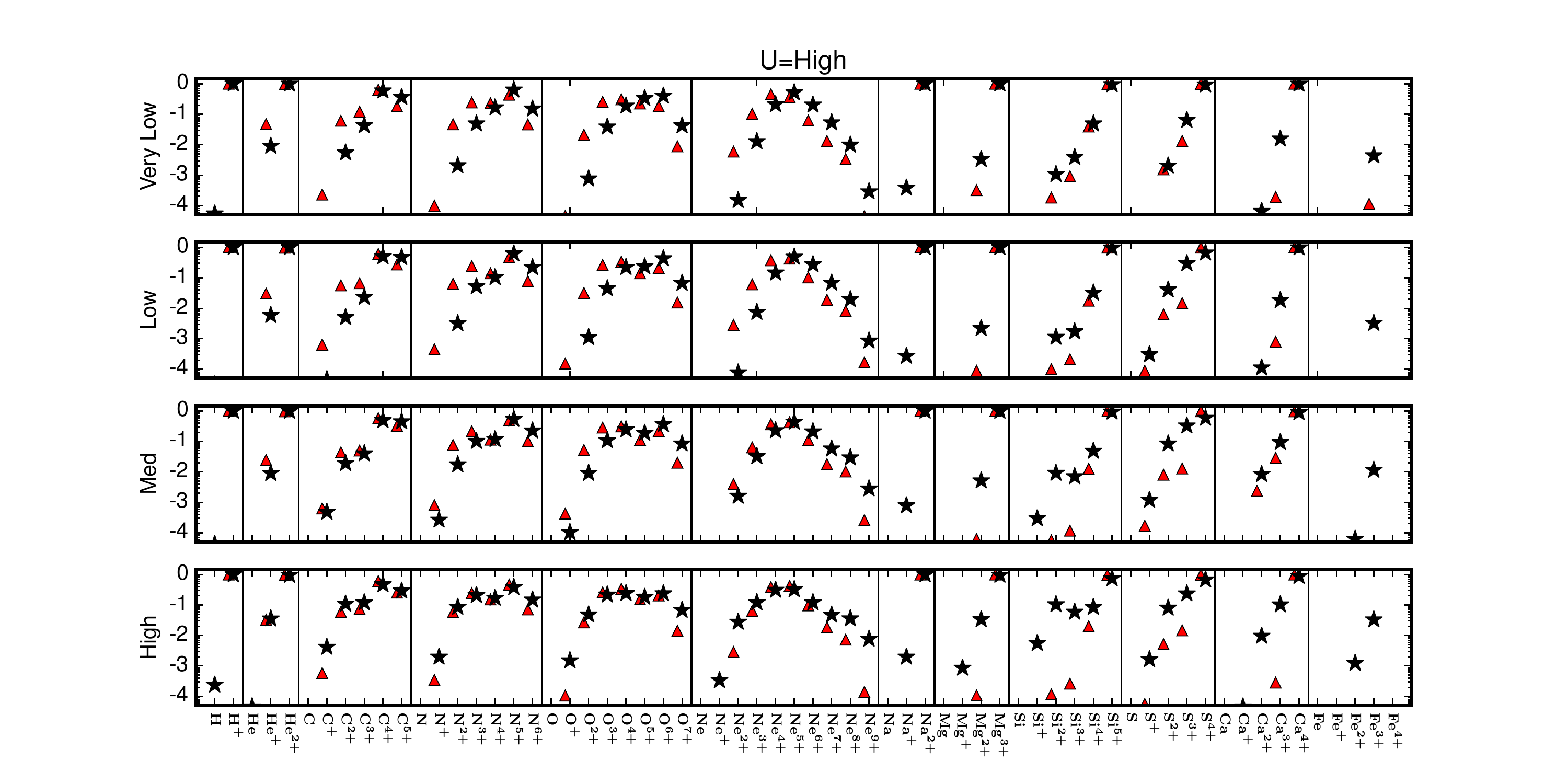} 
\caption{Steady state fractional species abundances for Very Low, Low, Medium, and High with an ionization parameter of U=10$^{-1}$. Symbols are the same as Figure~\ref{fig:chemcompa}.}
\label{fig:chemcompd}
\end{center}
\end{figure*}

To quantify the error involved with assuming constant temperature rather than using  our full treatment of turbulence, cooling, and chemical reactions we calculated approximate abundances assuming the full medium was in local ionization equilibrium at the mass weighted average temperature.  Figures~\ref{fig:chemcompa}-\ref{fig:chemcompd} show the comparison of this estimates to the steady state abundances for the Low, Medium, and High run, as well as an additional run N1E16\_S3, with \soned = 3 km/s.  This run, referred to as Very Low below is chosen to represent a case in which turbulence is highly-subsonic, such that density and temperature fluctuations are 10-20\%. 

Figure~\ref{fig:chemcompa} shows the comparison between the constant temperature estimate and our exact results in the case with no UV background. The name of each species is given along the $x$-axis while the fractional abundance of each species is given along the $y$-axis. The black stars show the exact results while the red diamonds show the results from the single temperature estimate. 

In the case of the Very Low run, which has a Mach number $0.51,$ the single temperature estimates from Cloudy provide a reasonable match to our final results for several ions however, there are several ions which the match is poor, H and \hp\ for example. To show that the difference between our final results and Cloudy are due to altering rates for reactions 1,2 and 3, we have rerun our collisional ionization equilibrium test with the \cite{Glover2008} and \cite{Glover2009}  rates rather than the  \cite{Badnell2006RR} and \cite{Badnell2006H} rates used in Cloudy. These values are shown as the blue diamonds in Figure~\ref{fig:chemcompa}. We find that these values match very well with the Very Low results for nearly all our ions, although turbulent heating has a noticeable effect on a handful of species, such as C$^+$ and Ca$^{2+}$, even in the $M=0.51$ case.

On the other hand, the Low case with zero background and a mass weighed Mach number of 1.23 produces a  much larger spread in density and temperature, which leads to a large difference in fractional species abundances. In particular, the simulations find both H and \hp\  in nearly equal amount whereas a single temperature approximation without turbulence would  predict almost pure neutral hydrogen. Other singly ionized species, show similar large differences between the true values and the simple estimates.

For the Medium ($M=4.3$, $T\approx$10$^{4}$K) and High ($M=8.7$, $T\approx$10$^{4}$ K) runs, the differences between the exact and approximate case is even more pronounced. For example, for the High case, the mean temperature approximation produces estimates for neutral carbon, nitrogen, oxygen, neon, sulfur, and calcium that are much larger than the true values while producing nearly zero ionized ions. The steady state values for this figure, as well as those for Figures~\ref{fig:chemcompb}-\ref{fig:chemcompd} are given in Table.~\ref{tab:ion}.

Figure~\ref{fig:chemcompb} shows the same comparions for a low, U=10$^{-6}$ ionization parameter. For the Low run, the mean temperature approximation starts to match better than with no UV background although there remains a sizable difference between these values and the true values. Only a handful of species, specifically \nthreep\ and Ne, are off by a factor of $\approx$10. For Medium and High, the difference is more pronounced. As mentioned in \cite{Gray2015} this  is due to the fact that the ionization timescale is much longer than the recombination timescale, which means that even though a parcel of gas is shocked to high temperatures, the ions do not have enough time to fully ionize.

Figure~\ref{fig:chemcompc} shows the ion comparisons for a moderate, U=10$^{-3}$, ionization parameter. For the Low run, the mean temperature approximation matches the real values for most ions. However, there is a factor of a few difference for many neutral ions. For example, hydrogen, helium, nitrogen, and oxygen have true values that are significantly higher than predicted. The simple approximation also does a poor job matching higher ionization states such as \sithreep and \cfour. The Medium and High runs also follow the same trend seen in the lower ionization parameter.

Finally, Figure~\ref{fig:chemcompd} shows the comparison for a large, U=10$^{-1}$, ionization parameter. The approximations for Low predict much higher neutral and lower ionized ion fractions than seen in our simulations. For example, N, \np, O, \oone, Ne, and \neone. However, more abundant ions, \eg\ \ofive and \nefive, are well matched by these approximations. The Medium and High run again follow the similar trends as seen at lower ionization parameters, with large differences being found between the approximate and exact cases for many ions.

\section{Parameter Dependencies} 

Above, we have explored the detailed evolution of three representative cases over a range of ionization parameters. Here we turn our attention to the full suite of models and ionization parameters. Table~\ref{tab:simruns} gives a summary of the models run, and the key hydrodynamical quantities (evaluated when $U=0$), which span a large range in column density, temperature, and Mach numbers. In most cases, heating from stirring and from photoionization is quickly balanced by cooling, as seen in Figures~\ref{fig:evos}-\ref{fig:evos3}. However, if the stirring is sufficiently strong or if the photoionization heating becomes strong enough that the average temperature of the medium exceeds $\approx$ 10$^{5}$ K, the model will undergo thermal runaway. This is explained by the fact that most elements have peaks in their cooling functions at $\approx$10$^{5}$ K \citep[\eg][]{Gnat2012,Oppenheimer2013}. Therefore, once the heating has crossed this barrier, cooling can no longer balance the heating and no meaningful steady state can be achieved. 

A summary of our results is given in Appendix~\ref{summ_results}, which shows the final abundances for all models that were able to reach a steady state. Each table shows the results for a different ionization parameter.  For example, Table~\ref{tab:uminus6} shows the results for an ionization parameter of U=10$^{-6}$ while Table~\ref{tab:uminus1} shows the results for U=10$^{-1}$. These abundances reveal many interesting trends affecting commonly observed species. For example, \hep\ is found in substantial quantities even when the average temperature $\approx$10$^{4}$K, and in many cases the \hp\ fraction is greater than 50\%. At moderate to high ionization parameters, this trend is even more pronounced.

Under uniform collisional ionization equilibrium conditions, certain species should not be found at temperatures below 10$^{5}$ K, see Figure~\ref{fig:chemtest}, \cthree, and \othree, for example. However, In certain low column density, high \soned, conditions these species can be found with substantial abundances (as seen for \soned=20 km s$^{-1}$ M$_{\rm MW}$=1.5 and \soned=35 km s$^{-1}$ M$_{\rm MW}$=1.9). 

The effect of the ionizing background can also be quite severe, although it is harder to quantify. As shown in Figure~\ref{fig:chemtest}, for even a small background, the abundances can be substantially different. However one general trend is that for moderate to high ionization potentials, our turbulent models produce much more neutral ions than would be expected in a simple single temperature approximation. For example, with an ionization parameter of only 10$^{-6}$, the simple approximation predicts that hydrogen should be roughly 60\% neutral at 10$^{4}$ K. However, the true abundances for H and \hp depend greatly on the turbulence. Both models N2E17\_S12 and N1E18\_S12 have mean temperatures of $\approx$10$^{4}$K and similar Mach numbers (1.23 and 1.6 respectively) but have very different abundances for hydrogen, where nearly all the hydrogen is neutral in N1E18\_S12. This is likely due to more efficient collisional recombination from the denser environment. 

In addition to Appendix~\ref{summ_results} we make available online\footnote{http://zofia.sese.asu.edu/\~{}evan/turbspecies/} the data files from each run. Each file presents a variety of information, including the true abundances and the simple abundance estimates presented in Section~\ref{subsec_is}. We also give Doppler parameters for each species, $i$, defined as,
\be
b^2=\sigma_{i,{\rm 1D}}^2 +2k_{\rm b}T_{\rm MW}/A_i m_{\rm H},
\ee
where $\sigma_{i,{\rm 1D}}$ is the 1D velocity dispersion, $k_{\rm B}$ is the Boltzmann constant, and $A_i$ is the ion atomic mass. While the masses of hydrogen and helium are small enough that the temperature term makes a substantial contribution, the Doppler parameters of the heavier elements are very close to $\sigma_{i,{\rm 1D}}$ in general, providing a good measurement of the local velocity dispersion.

\section{Conclusions}
Turbulence permeates the universe, often moving at supersonic speeds due to the high efficiency of radiative cooling. These random motions provide overall support against gravity, but also concentrate a portion of the material to very high densities, giving rise to a multiphase distribution with unique thermodynamic properties. This can have a dramatic effect on the chemical makeup of the medium. Specifically, if the recombination time for a given species is long compared to the eddy turnover time, it cannot reach an equilibrium state before it is further acted upon by the turbulence. This creates a situation in which the final ionization state is not only a function of the temperature and density, but also a function of the rate at which parcels of gas move through these conditions.

To study how these abundances are altered by turbulence, we have expanded the nonequilibrium atomic chemistry package from \cite{Gray2015} to include several new elements and ions and their associated cooling functions within FLASH. This package tracks the evolution of twelve elements and 65 separate species. In addition, we have used the method of \cite{Gnat2012} to derive ion-by-ion cooling curves for each of the ions under consideration that allow us to follow the thermodynamic evolution of the gas. We have also implemented a variable photoionizing background based on the spectra from \cite{Haardt2012}. The result is a very fast and efficient package, which we are able to run for many eddy turnover times for many cases.

Using this updated package, we have performed a suite of direct numerical simulations of solenoidally-driven turbulence over a range from 1D velocity dispersion of \soned= 3.5-58 km s$^{-1}$ and the product of the mean density and turbulent length scale from 10$^{16}$ to 10$^{20}$ cm$^{-2}$ for solar metallicity gas, concentrating on three representative models and four representative backgrounds.  

As found in isothermal models of driven turbulence, the gas approximates a lognormal distribution, whose logarithmic density variance in the supersonic case is well approximated by $\sigma^2$=ln$(1 + b^2 M^2)$. On the other hand, this expression overestimates the variance at subsonic Mach numbers, and the $b \approx$0.50 value that best fits our data is significantly different from the $b$ = 1/3 value measured in solenoidally-driven, isothermal turbulence. However, all our data is well described over the range of astronomically valid polytropic exponents. 

We compare the final steady state abundances in our simulations to those obtained assuming the gas is in collisional ionization equilibrium, using both the mean temperature and the full temperature PDF. We find that at low Mach numbers and in the absence of an ionizing background,
the estimates agree with within a factor of $\approx$2 for most species, save for \hetwo\ and \ctwo, which show large deviations due to their long recombination times. At intermediate Mach numbers, several species such as \hep, \ntwop, \netwo, and \sitwop, begin to differ by order of magnitude from the simple estimates. At high Mach numbers the abundances can vary by many orders of magnitude from simple estimates. The additional effect of the photoionizing background further effects these estimates. With a low background, the estimates begin to match the true value well for low Mach number models, but many of the same discrepancies are seen at moderate to high Mach numbers. At moderate to high ionization parameters, many of the estimates are within a factor of ten for the most abundant species. However, in high Mach number flows, they fail to reproduce the high abundances of some neutral ions, \eg neutral hydrogen, carbon, and oxygen. 

These results underscore the fact that transsonic and supersonic turbulence can drastically alter the abundances, and that only nonequilibrium calculations can predict these changes accurately. Thus we make all of the derived properties from our models available online. In particular, we present the logarithmic density statistics and other hydrodynamic quantities, such as Mach number and average temperature. We also give the final abundances for each species, the abundance values from the two simple estimates, and species by species Doppler parameters.

In future work we plan on running additional models that vary both the metallicity of the gas as well as the shape of the UV background. This will result in a large set of tables, useful for a variety of theoretical and observational applications.

\begin{table*}
\resizebox{0.95\textwidth}{!}{%
\centering
\begin{threeparttable}
\caption{}
\label{tab:ion}
\begin{tabular}{|l||rrrr|rrrr|rrrr|}
%\begin{table*}
%\begin{ThreePartTable}
%\caption{Species fractions in our representative runs.}
%\label{tab:ion}
%\resizebox{\linewidth}{!}{%
%\begin{TableNotes}
%\item {\bf Notes:} The top row gives the name of the representative model. Low corresponds to N2E17\_S12, Med to N4E18\_S35, and High to N2E19\_S58 respectively. The subscript in the second row coincides with the ionization parameter, where $_{,Z}$ is no background, $_{,L}$ corresponds to a U=1.0$\times$10$^{-6}$,  $_{,M}$ corresponds to a U=1.0$\times$10$^{-4}$, and $_{,H}$ corresponds to a U=1.0$\times$10$^{-2}$ respectively. 
%\end{TableNotes}
%\begin{tabular}{l|rrrrrrrrrrrr}
\hline
& Low                  & Low                  & Low                  & Low                  & Medium               & Medium               & Medium               & Medium               & High                 & High                 & High                 & High                 \\
\hline
                     & F$_{\rm Turb, Z}$    & F$_{\rm Turb, L}$    & F$_{\rm Turb, M}$    & F$_{\rm Turb, H}$    & F$_{\rm Turb, Z}$    & F$_{\rm Turb, L}$    & F$_{\rm Turb, M}$    & F$_{\rm Turb, H}$    & F$_{\rm Turb, Z}$    & F$_{\rm Turb, L}$    & F$_{\rm Turb, M}$    & F$_{\rm Turb, H}$    \\
${\rm H}$            & -0.78                & -0.69                & -2.34                & -4.51                & -0.56                & -0.56                & -1.81                & -4.37                & -0.48                & -0.49                & -1.61                & -3.62                \\
${\rm H^+}$          & -0.08                & -0.10                & -0.00                & -0.00                & -0.14                & -0.14                & -0.01                & -0.00                & -0.17                & -0.17                & -0.01                & -0.00                \\
${\rm He}$           & -0.93                & -0.80                & -2.27                & -6.50                & -0.03                & -0.03                & -1.65                & -5.79                & -0.04                & -0.05                & -1.25                & -4.34                \\
${\rm He^{+}}$       & -0.19                & -0.17                & -0.21                & -2.23                & -1.12                & -1.20                & -0.17                & -2.05                & -1.07                & -0.96                & -0.16                & -1.45                \\
${\rm He^{2+}}$      & -0.62                & -0.76                & -0.43                & -0.00                & -3.91                & -4.25                & -0.53                & -0.00                & -2.61                & -2.48                & -0.59                & -0.02                \\
${\rm C}$            & -0.80                & -0.77                & -3.20                & -8.24                & -0.42                & -0.49                & -2.74                & -6.66                & -0.21                & -0.31                & -2.49                & -5.66                \\
${\rm C^{+}}$        & -0.08                & -0.08                & -0.67                & -4.34                & -0.21                & -0.17                & -0.45                & -3.32                & -0.44                & -0.30                & -0.34                & -2.38                \\
${\rm C^{2+}}$       & -2.30                & -2.38                & -0.13                & -2.30                & -2.31                & -2.36                & -0.22                & -1.72                & -1.91                & -1.81                & -0.30                & -0.96                \\
${\rm C^{3+}}$       & -13.32               & -7.78                & -1.41                & -1.63                & -4.40                & -5.21                & -1.49                & -1.40                & -3.30                & -3.24                & -1.50                & -0.92                \\
${\rm C^{4+}}$       & -21.93               & -13.47               & -2.77                & -0.30                & -5.85                & -7.38                & -2.39                & -0.31                & -3.97                & -3.78                & -2.07                & -0.32                \\
${\rm C^{5+}}$       & -21.62               & -18.70               & -5.02                & -0.33                & -13.22               & -12.07               & -4.37                & -0.35                & -9.78                & -7.96                & -3.82                & -0.53                \\
${\rm N}$            & -0.67                & -0.61                & -3.19                & -8.71                & -0.38                & -0.39                & -2.23                & -6.87                & -0.26                & -0.28                & -1.77                & -5.74                \\
${\rm N^{+}}$        & -0.10                & -0.12                & -0.73                & -4.85                & -0.23                & -0.23                & -0.37                & -3.58                & -0.35                & -0.33                & -0.28                & -2.70                \\
${\rm N^{2+}}$       & -7.02                & -3.78                & -0.13                & -2.50                & -3.47                & -3.66                & -0.30                & -1.76                & -2.49                & -2.43                & -0.40                & -1.07                \\
${\rm N^{3+}}$       & -10.55               & -8.90                & -1.15                & -1.28                & -4.42                & -5.37                & -1.23                & -0.99                & -3.26                & -3.21                & -1.25                & -0.68                \\
${\rm N^{4+}}$       & -10.42               & -11.38               & -2.81                & -0.98                & -6.02                & -7.52                & -2.60                & -0.93                & -4.45                & -4.22                & -2.29                & -0.77                \\
${\rm N^{5+}}$       & -8.81                & -9.69                & -4.29                & -0.21                & -7.92                & -9.25                & -3.63                & -0.27                & -4.96                & -4.56                & -2.91                & -0.42                \\
${\rm N^{6+}}$       & -23.38               & -15.12               & -6.95                & -0.66                & -16.75               & -14.44               & -5.92                & -0.65                & -12.29               & -9.39                & -4.90                & -0.83                \\
${\rm O}$            & -0.74                & -0.65                & -2.99                & -9.53                & -0.52                & -0.52                & -1.90                & -7.42                & -0.44                & -0.45                & -1.62                & -5.54                \\
${\rm O^{+}}$        & -0.09                & -0.11                & -0.65                & -5.55                & -0.16                & -0.16                & -0.33                & -3.99                & -0.20                & -0.19                & -0.25                & -2.82                \\
${\rm O^{2+}}$       & -8.04                & -4.02                & -0.18                & -2.95                & -3.47                & -3.78                & -0.37                & -2.04                & -2.48                & -2.43                & -0.50                & -1.32                \\
${\rm O^{3+}}$       & -16.82               & -9.07                & -0.93                & -1.35                & -4.48                & -5.54                & -1.05                & -0.97                & -3.30                & -3.23                & -1.10                & -0.66                \\
${\rm O^{4+}}$       & -19.68               & -13.75               & -2.28                & -0.65                & -6.10                & -7.65                & -2.21                & -0.61                & -4.46                & -4.20                & -2.04                & -0.62                \\
${\rm O^{5+}}$       & -19.07               & -18.25               & -4.02                & -0.63                & -8.33                & -10.21               & -3.77                & -0.72                & -5.72                & -5.17                & -3.13                & -0.74                \\
${\rm O^{6+}}$       & -20.07               & -21.40               & -5.97                & -0.37                & -10.63               & -13.16               & -5.12                & -0.44                & -6.69                & -5.74                & -3.97                & -0.62                \\
${\rm O^{7+}}$       & -26.53               & -26.54               & -8.95                & -1.17                & -20.86               & -18.52               & -7.64                & -1.08                & -15.09               & -10.82               & -6.13                & -1.17                \\
${\rm Ne}$           & -1.30                & -1.16                & -4.27                & -11.94               & -0.14                & -0.13                & -2.84                & -8.81                & -0.09                & -0.12                & -2.10                & -6.55                \\
${\rm Ne^{+}}$       & -0.19                & -0.15                & -1.42                & -7.34                & -0.55                & -0.58                & -0.87                & -5.14                & -0.73                & -0.63                & -0.61                & -3.47                \\
${\rm Ne^{2+}}$      & -0.53                & -0.64                & -0.09                & -4.11                & -3.03                & -2.83                & -0.14                & -2.79                & -2.13                & -1.98                & -0.20                & -1.56                \\
${\rm Ne^{3+}}$      & -15.15               & -5.73                & -0.83                & -2.13                & -4.90                & -6.21                & -0.92                & -1.49                & -3.54                & -3.45                & -1.03                & -0.92                \\
${\rm Ne^{4+}}$      & -25.92               & -11.25               & -2.13                & -0.83                & -6.54                & -8.43                & -1.83                & -0.64                & -4.62                & -4.30                & -1.74                & -0.50                \\
${\rm Ne^{5+}}$      & -25.43               & -15.87               & -3.75                & -0.31                & -8.76                & -11.08               & -3.11                & -0.36                & -5.88                & -5.26                & -2.72                & -0.49                \\
${\rm Ne^{6+}}$      & -25.85               & -21.31               & -6.14                & -0.56                & -11.76               & -15.13               & -4.99                & -0.68                & -7.76                & -6.68                & -4.07                & -0.92                \\
${\rm Ne^{7+}}$      & -25.80               & -25.77               & -8.70                & -1.17                & -15.40               & -19.49               & -7.09                & -1.24                & -10.06               & -8.03                & -5.56                & -1.32                \\
${\rm Ne^{8+}}$      & -25.67               & -25.69               & -11.41               & -1.71                & -18.23               & -23.14               & -9.02                & -1.53                & -11.79               & -9.05                & -7.07                & -1.44                \\
${\rm Ne^{9+}}$      & -25.85               & -25.85               & -14.91               & -3.07                & -25.89               & -25.88               & -11.87               & -2.55                & -21.83               & -14.45               & -9.49                & -2.11                \\
${\rm Na}$           & -3.86                & -3.68                & -6.26                & -9.58                & -2.12                & -2.07                & -3.79                & -8.74                & -1.22                & -1.21                & -2.71                & -6.98                \\
${\rm Na^{+}}$       & -0.26                & -0.20                & -1.39                & -3.57                & -0.00                & -0.01                & -0.89                & -3.11                & -0.03                & -0.04                & -0.62                & -2.70                \\
${\rm Na^{2+}}$      & -0.35                & -0.43                & -0.02                & -0.00                & -2.81                & -2.21                & -0.06                & -0.00                & -1.97                & -1.74                & -0.12                & -0.00                \\
${\rm Mg}$           & -1.91                & -1.79                & -3.48                & -9.39                & -1.55                & -1.55                & -2.56                & -7.60                & -1.54                & -1.55                & -2.18                & -5.13                \\
${\rm Mg^{+}}$       & -0.54                & -0.45                & -1.48                & -6.09                & -0.13                & -0.12                & -0.82                & -5.10                & -0.09                & -0.09                & -0.45                & -3.07                \\
${\rm Mg^{2+}}$      & -0.15                & -0.20                & -0.09                & -2.66                & -0.64                & -0.68                & -0.15                & -2.29                & -0.81                & -0.81                & -0.29                & -1.47                \\
${\rm Mg^{3+}}$      & -18.90               & -5.39                & -0.83                & -0.00                & -5.18                & -5.65                & -0.84                & -0.00                & -3.75                & -3.58                & -0.91                & -0.02                \\
${\rm Si}$           & -2.18                & -2.22                & -4.70                & -9.59                & -1.26                & -1.50                & -4.06                & -8.01                & -0.74                & -1.17                & -3.62                & -6.74                \\
${\rm Si^{+}}$       & -0.07                & -0.05                & -1.04                & -4.62                & -0.05                & -0.04                & -0.51                & -3.53                & -0.12                & -0.06                & -0.32                & -2.25                \\
${\rm Si^{2+}}$      & -0.85                & -0.99                & -0.06                & -2.95                & -1.20                & -1.25                & -0.20                & -2.04                & -1.25                & -1.23                & -0.33                & -0.98                \\
${\rm Si^{3+}}$      & -4.18                & -4.44                & -1.42                & -2.77                & -3.68                & -4.06                & -1.34                & -2.15                & -2.75                & -2.71                & -1.44                & -1.23                \\
${\rm Si^{4+}}$      & -13.47               & -10.94               & -3.35                & -1.49                & -4.71                & -5.81                & -2.07                & -1.32                & -3.32                & -3.24                & -1.96                & -1.07                \\
${\rm Si^{5+}}$      & -11.45               & -12.39               & -4.13                & -0.02                & -8.19                & -9.27                & -2.61                & -0.03                & -5.69                & -5.06                & -2.38                & -0.13                \\
${\rm S}$            & -1.77                & -1.72                & -4.80                & -8.36                & -0.86                & -1.04                & -4.04                & -7.26                & -0.40                & -0.69                & -3.48                & -7.04                \\
${\rm S^{+}}$        & -0.11                & -0.08                & -1.00                & -3.52                & -0.08                & -0.06                & -0.75                & -2.92                & -0.25                & -0.13                & -0.57                & -2.79                \\
${\rm S^{2+}}$       & -0.69                & -0.82                & -0.11                & -1.39                & -1.46                & -1.53                & -0.14                & -1.08                & -1.41                & -1.34                & -0.19                & -1.10                \\
${\rm S^{3+}}$       & -8.69                & -5.96                & -0.95                & -0.53                & -3.72                & -4.13                & -1.06                & -0.49                & -2.73                & -2.70                & -1.12                & -0.63                \\
${\rm S^{4+}}$       & -16.21               & -11.71               & -2.33                & -0.18                & -4.40                & -5.21                & -2.14                & -0.23                & -3.37                & -3.30                & -1.88                & -0.17                \\
${\rm Ca}$           & -2.60                & -2.43                & -5.52                & -12.50               & -1.35                & -1.48                & -3.99                & -9.95                & -0.71                & -1.02                & -3.39                & -7.71                \\
${\rm Ca^{+}}$       & -0.90                & -0.78                & -2.46                & -7.62                & -0.39                & -0.37                & -1.42                & -5.06                & -0.31                & -0.23                & -0.93                & -4.34                \\
${\rm Ca^{2+}}$      & -0.06                & -0.08                & -0.26                & -3.95                & -0.26                & -0.27                & -0.18                & -2.08                & -0.50                & -0.50                & -0.20                & -2.01                \\
${\rm Ca^{3+}}$      & -10.80               & -4.26                & -0.39                & -1.74                & -3.84                & -4.11                & -0.58                & -1.03                & -2.83                & -2.76                & -0.67                & -0.99                \\
${\rm Ca^{4+}}$      & -19.91               & -9.53                & -1.40                & -0.01                & -5.02                & -6.13                & -1.46                & -0.05                & -3.81                & -3.68                & -1.36                & -0.05                \\
${\rm Fe}$           & -3.76                & -3.74                & -5.71                & -13.43               & -3.86                & -3.87                & -4.82                & -11.30               & -3.94                & -3.94                & -4.66                & -8.50                \\
${\rm Fe^{+}}$       & -0.26                & -0.19                & -2.44                & -8.98                & -0.10                & -0.09                & -1.07                & -7.14                & -0.07                & -0.07                & -0.68                & -4.86                \\
${\rm Fe^{2+}}$      & -0.35                & -0.44                & -0.50                & -5.67                & -0.68                & -0.71                & -0.35                & -4.20                & -0.82                & -0.81                & -0.37                & -2.91                \\
${\rm Fe^{3+}}$      & -8.24                & -4.68                & -0.27                & -2.49                & -3.47                & -3.99                & -0.49                & -1.94                & -2.45                & -2.43                & -0.63                & -1.48                \\
${\rm Fe^{4+}}$      & -17.58               & -9.58                & -0.85                & -0.00                & -4.41                & -5.39                & -0.84                & -0.01                & -3.19                & -3.12                & -0.88                & -0.02                \\
\hline
\hline
%\insertTableNotes
%\end{tabular}
%}
%\end{ThreePartTable}
%\end{table*}
\end{tabular}%
\begin{tablenotes}{
\item {\bf Notes:} The top row gives the name of the representative model. Low corresponds to N2E17\_S12, Med to N4E18\_S35, and High to N2E19\_S58 respectively. The subscript in the second row coincides with the ionization parameter, where F$_{{\rm Turb,Z}}$ is no background, F$_{{\rm Turb,L}}$ corresponds to a U=1.0$\times$10$^{-6}$,  F$_{{\rm Turb,M}}$ corresponds to a U=1.0$\times$10$^{-3}$, and F$_{{\rm Turb,H}}$ corresponds to a U=1.0$\times$10$^{-1}$ respectively.}
\end{tablenotes}
\end{threeparttable}
}
\end{table*}

\acknowledgements

We would like to thank Daniel Kasen, Christopher Matzner, Cody Raskin, Eve Ostriker, Robert J. Thacker, and David Williamson for helpful discussions. We would also like to thank the anonymous referee for their comments which helped to improve this paper. The software used in this work was in part developed by the DOE NNSA-ASC OASCR Flash Center at the University of Chicago. This work was performed under the auspices of the U.S. Department of Energy by Lawrence Livermore National Laboratory under Contract DE-AC52-07NA27344. ES was supported by NSF grant AST11-03608 and AST14-07835, and NASA theory grant NNX15AK82G. The figures and analysis presented here were created using the {\bf yt} analysis package \citep{Turk2011}. The authors also acknowledge the Texas Advanced Computing Center (TACC) at The University of Texas at Austin, the San Diego Supercomputer Center (SDSC) at the University of California, San Diego, and the Extreme Science and Engineering Discovery Environment (XSEDE) for providing HPC resources that have contributed to the results reported within this paper.

\bibliographystyle{apj}
\bibliography{ms.bib}

\appendix

\section{List of Atomic Reactions}
\label{chemtable}
\resizebox{0.90\columnwidth}{!}{
\begin{threeparttable}
		\label{tab:rxns}
\begin{tabular}{|lll|lll|lll|}
	\hline
	Number & Reaction & Source & Number & Reaction & Source &  Number & Reaction & Source\\
	\hline
	\hline
R001  &  H$^{+}$    +  e$^{-}$    \ra  H                                        &12   & R002  &  He$^{+}$   +  e$^{-}$    \ra  He                                       &12/23& R003  &  He$^{2+}$  +  e$^{-}$    \ra  He$^{+}$                                 & 12  \\ 
R004  &  C$^{+}$    +  e$^{-}$    \ra  C                                        &  5  & R005  &  C$^{2+}$   +  e$^{-}$    \ra  C$^{+}$                                  &  4  & R006  &  C$^{3+}$   +  e$^{-}$    \ra  C$^{2+}$                                 &  3  \\
R007  &  C$^{4+}$   +  e$^{-}$    \ra  C$^{3+}$                                 &  2  & R008  &  C$^{5+}$   +  e$^{-}$    \ra  C$^{4+}$                                 &  1  & R009  &  N$^{+}$    +  e$^{-}$    \ra  N                                        &  6  \\
R010  &  N$^{2+}$   +  e$^{-}$    \ra  N$^{+}$                                  &  5  & R011  &  N$^{3+}$   +  e$^{-}$    \ra  N$^{2+}$                                 &  4  & R012  &  N$^{4+}$   +  e$^{-}$    \ra  N$^{3+}$                                 &  3  \\ 
R013  &  N$^{5+}$   +  e$^{-}$    \ra  N$^{4+}$                                 &  2  & R014  &  N$^{6+}$   +  e$^{-}$    \ra  N$^{5+}$                                 &  1  & R015  &  O$^{+}$    +  e$^{-}$    \ra  O                                        &  7  \\
R016  &  O$^{2+}$   +  e$^{-}$    \ra  O$^{+}$                                  &  6  & R017  &  O$^{3+}$   +  e$^{-}$    \ra  O$^{2+}$                                 &  5  & R018  &  O$^{4+}$   +  e$^{-}$    \ra  O$^{3+}$                                 &  4  \\
R019  &  O$^{5+}$   +  e$^{-}$    \ra  O$^{4+}$                                 &  3  & R020  &  O$^{6+}$   +  e$^{-}$    \ra  O$^{5+}$                                 &  2  & R021  &  O$^{7+}$   +  e$^{-}$    \ra  O$^{6+}$                                 &  1  \\
R022  &  Ne$^{+}$   +  e$^{-}$    \ra  Ne                                       &  19 & R023  &  Ne$^{2+}$  +  e$^{-}$    \ra  Ne$^{+}$                                 &  20 & R024  &  Ne$^{3+}$  +  e$^{-}$    \ra  Ne$^{2+}$                                &  7  \\
R025  &  Ne$^{4+}$  +  e$^{-}$    \ra  Ne$^{3+}$                                &  6  & R026  &  Ne$^{5+}$  +  e$^{-}$    \ra  Ne$^{4+}$                                &  5  & R027  &  Ne$^{6+}$  +  e$^{-}$    \ra  Ne$^{5+}$                                &  4  \\
R028  &  Ne$^{7+}$  +  e$^{-}$    \ra  Ne$^{6+}$                                &  3  & R029  &  Ne$^{8+}$  +  e$^{-}$    \ra  Ne$^{7+}$                                &  2  & R030  &  Ne$^{9+}$  +  e$^{-}$    \ra  Ne$^{8+}$                                &  1  \\
R031  &  Na$^{+}$   +  e$^{-}$    \ra  Na                                       &  8  & R032  &  Na$^{2+}$  +  e$^{-}$    \ra  Na$^{+}$                                 &  19 & R033  &  Mg$^{+}$   +  e$^{-}$    \ra  Mg                                       &  9  \\
R034  &  Mg$^{2+}$  +  e$^{-}$    \ra  Mg$^{+}$                                 &  8  & R035  &  Mg$^{3+}$  +  e$^{-}$    \ra  Mg$^{2+}$                                &  19 & R036  &  Si$^{+}$   +  e$^{-}$    \ra  Si                                       &  21 \\
R037  &  Si$^{2+}$  +  e$^{-}$    \ra  Si$^{+}$                                 &  2  & R038  &  Si$^{3+}$  +  e$^{-}$    \ra  Si$^{2+}$                                &  9  & R039  &  Si$^{4+}$  +  e$^{-}$    \ra  Si$^{3+}$                                &  8  \\
R040  &  Si$^{5+}$  +  e$^{-}$    \ra  Si$^{4+}$                                &  19 & R041  &  S$^{+}$    +  e$^{-}$    \ra  S                                        &13,14& R042  &  S$^{2+}$   +  e$^{-}$    \ra  S$^{+}$                                  &13,14\\
R043  &  S$^{3+}$   +  e$^{-}$    \ra  S$^{2+}$                                 &13,14& R044  &  S$^{4+}$   +  e$^{-}$    \ra  S$^{3+}$                                 &13,14& R045  &  Ca$^{+}$   +  e$^{-}$    \ra  Ca                                       &13,14\\
R046  &  Ca$^{2+}$  +  e$^{-}$    \ra  Ca$^{+}$                                 &13,14& R047  &  Ca$^{3+}$  +  e$^{-}$    \ra  Ca$^{2+}$                                &13,14& R048  &  Ca$^{4+}$  +  e$^{-}$    \ra  Ca$^{3+}$                                &13,14\\
R049  &  Fe$^{+}$   +  e$^{-}$    \ra  Fe                                       &13,14& R050  &  Fe$^{2+}$  +  e$^{-}$    \ra  Fe$^{+}$                                 &13,14& R051  &  Fe$^{3+}$  +  e$^{-}$    \ra  Fe$^{2+}$                                &13,14\\
R052  &  Fe$^{4+}$  +  e$^{-}$    \ra  Fe$^{3+}$                                &13,14& R053  &  H          +  e$^{-}$    \ra  H$^{+}$    +  e$^{-}$    +  e$^{-}$      &11,23& R054  &  He         +  e$^{-}$    \ra  He$^{+}$   +  e$^{-}$    +  e$^{-}$      &  11 \\
R055  &  He$^{+}$   +  e$^{-}$    \ra  He$^{2+}$  +  e$^{-}$    +  e$^{-}$      &  11 & R056  &  C          +  e$^{-}$    \ra  C$^{+}$    +  e$^{-}$    +  e$^{-}$      &  11 & R057  &  C$^{+}$    +  e$^{-}$    \ra  C$^{2+}$   +  e$^{-}$    +  e$^{-}$      &  11 \\
R058  &  C$^{2+}$   +  e$^{-}$    \ra  C$^{3+}$   +  e$^{-}$    +  e$^{-}$      &  11 & R059  &  C$^{3+}$   +  e$^{-}$    \ra  C$^{4+}$   +  e$^{-}$    +  e$^{-}$      &  11 & R060  &  C$^{4+}$   +  e$^{-}$    \ra  C$^{5+}$   +  e$^{-}$    +  e$^{-}$      &  11 \\
R061  &  N          +  e$^{-}$    \ra  N$^{+}$    +  e$^{-}$    +  e$^{-}$      &  11 & R062  &  N$^{+}$    +  e$^{-}$    \ra  N$^{2+}$   +  e$^{-}$    +  e$^{-}$      &  11 & R063  &  N$^{2+}$   +  e$^{-}$    \ra  N$^{3+}$   +  e$^{-}$    +  e$^{-}$      &  11 \\
R064  &  N$^{3+}$   +  e$^{-}$    \ra  N$^{4+}$   +  e$^{-}$    +  e$^{-}$      &  11 & R065  &  N$^{4+}$   +  e$^{-}$    \ra  N$^{5+}$   +  e$^{-}$    +  e$^{-}$      &  11 & R066  &  N$^{5+}$   +  e$^{-}$    \ra  N$^{6+}$   +  e$^{-}$    +  e$^{-}$      &  11 \\
R067  &  O          +  e$^{-}$    \ra  O$^{+}$    +  e$^{-}$    +  e$^{-}$      &  11 & R068  &  O$^{+}$    +  e$^{-}$    \ra  O$^{2+}$   +  e$^{-}$    +  e$^{-}$      &  11 & R069  &  O$^{2+}$   +  e$^{-}$    \ra  O$^{3+}$   +  e$^{-}$    +  e$^{-}$      &  11 \\
R070  &  O$^{3+}$   +  e$^{-}$    \ra  O$^{4+}$   +  e$^{-}$    +  e$^{-}$      &  11 & R071  &  O$^{4+}$   +  e$^{-}$    \ra  O$^{5+}$   +  e$^{-}$    +  e$^{-}$      &  11 & R072  &  O$^{5+}$   +  e$^{-}$    \ra  O$^{6+}$   +  e$^{-}$    +  e$^{-}$      &  11 \\
R073  &  O$^{6+}$   +  e$^{-}$    \ra  O$^{7+}$   +  e$^{-}$    +  e$^{-}$      &  11 & R074  &  Ne         +  e$^{-}$    \ra  Ne$^{+}$   +  e$^{-}$    +  e$^{-}$      &  11 & R075  &  Ne$^{+}$   +  e$^{-}$    \ra  Ne$^{2+}$  +  e$^{-}$    +  e$^{-}$      &  11 \\
R076  &  Ne$^{2+}$  +  e$^{-}$    \ra  Ne$^{3+}$  +  e$^{-}$    +  e$^{-}$      &  11 & R077  &  Ne$^{3+}$  +  e$^{-}$    \ra  Ne$^{4+}$  +  e$^{-}$    +  e$^{-}$      &  11 & R078  &  Ne$^{4+}$  +  e$^{-}$    \ra  Ne$^{5+}$  +  e$^{-}$    +  e$^{-}$      &  11 \\
R079  &  Ne$^{5+}$  +  e$^{-}$    \ra  Ne$^{6+}$  +  e$^{-}$    +  e$^{-}$      &  11 & R080  &  Ne$^{6+}$  +  e$^{-}$    \ra  Ne$^{7+}$  +  e$^{-}$    +  e$^{-}$      &  11 & R081  &  Ne$^{7+}$  +  e$^{-}$    \ra  Ne$^{8+}$  +  e$^{-}$    +  e$^{-}$      &  11 \\
R082  &  Ne$^{8+}$  +  e$^{-}$    \ra  Ne$^{9+}$  +  e$^{-}$    +  e$^{-}$      &  11 & R083  &  Na         +  e$^{-}$    \ra  Na$^{+}$   +  e$^{-}$    +  e$^{-}$      &  11 & R084  &  Na$^{+}$   +  e$^{-}$    \ra  Na$^{2+}$  +  e$^{-}$    +  e$^{-}$      &  11 \\
R085  &  Mg         +  e$^{-}$    \ra  Mg$^{+}$   +  e$^{-}$    +  e$^{-}$      &  11 & R086  &  Mg$^{+}$   +  e$^{-}$    \ra  Mg$^{2+}$  +  e$^{-}$    +  e$^{-}$      &  11 & R087  &  Mg$^{2+}$  +  e$^{-}$    \ra  Mg$^{3+}$  +  e$^{-}$    +  e$^{-}$      &  11 \\
R088  &  Si         +  e$^{-}$    \ra  Si$^{+}$   +  e$^{-}$    +  e$^{-}$      &  11 & R089  &  Si$^{+}$   +  e$^{-}$    \ra  Si$^{2+}$  +  e$^{-}$    +  e$^{-}$      &  11 & R090  &  Si$^{2+}$  +  e$^{-}$    \ra  Si$^{3+}$  +  e$^{-}$    +  e$^{-}$      &  11 \\
R091  &  Si$^{3+}$  +  e$^{-}$    \ra  Si$^{4+}$  +  e$^{-}$    +  e$^{-}$      &  11 & R092  &  Si$^{4+}$  +  e$^{-}$    \ra  Si$^{5+}$  +  e$^{-}$    +  e$^{-}$      &  11 & R093  &  S          +  e$^{-}$    \ra  S$^{+}$    +  e$^{-}$    +  e$^{-}$      &  11 \\
R094  &  S$^{+}$    +  e$^{-}$    \ra  S$^{2+}$   +  e$^{-}$    +  e$^{-}$      &  11 & R095  &  S$^{2+}$   +  e$^{-}$    \ra  S$^{3+}$   +  e$^{-}$    +  e$^{-}$      &  11 & R096  &  S$^{3+}$   +  e$^{-}$    \ra  S$^{4+}$   +  e$^{-}$    +  e$^{-}$      &  11 \\
R097  &  Ca         +  e$^{-}$    \ra  Ca$^{+}$   +  e$^{-}$    +  e$^{-}$      &  11 & R098  &  Ca$^{+}$   +  e$^{-}$    \ra  Ca$^{2+}$  +  e$^{-}$    +  e$^{-}$      &  11 & R099  &  Ca$^{2+}$  +  e$^{-}$    \ra  Ca$^{3+}$  +  e$^{-}$    +  e$^{-}$      &  11 \\
R100  &  Ca$^{3+}$  +  e$^{-}$    \ra  Ca$^{4+}$  +  e$^{-}$    +  e$^{-}$      &  11 & R101  &  Fe         +  e$^{-}$    \ra  Fe$^{+}$   +  e$^{-}$    +  e$^{-}$      &  11 & R102  &  Fe$^{+}$   +  e$^{-}$    \ra  Fe$^{2+}$  +  e$^{-}$    +  e$^{-}$      &  11 \\
R103  &  Fe$^{2+}$  +  e$^{-}$    \ra  Fe$^{3+}$  +  e$^{-}$    +  e$^{-}$      &  11 & R104  &  Fe$^{3+}$  +  e$^{-}$    \ra  Fe$^{4+}$  +  e$^{-}$    +  e$^{-}$      &  11 & R105  &  He$^{+}$   +  H          \ra  He         +  H$^{+}$                    &  12 \\
R106  &  He         +  H$^{+}$    \ra  He$^{+}$   +  H                          &  12 & R107  &  H          +  $\gamma$   \ra  H$^{+}$    +  e$^{-}$                    &  15 & R108  &  He         +  $\gamma$   \ra  He$^{+}$   +  e$^{-}$                    &  15 \\
R109  &  He$^{+}$   +  $\gamma$   \ra  He$^{2+}$  +  e$^{-}$                    &  15 & R110  &  C          +  $\gamma$   \ra  C$^{+}$    +  e$^{-}$                    &  15 & R111  &  C$^{+}$    +  $\gamma$   \ra  C$^{2+}$   +  e$^{-}$                    &  15 \\
R112  &  C$^{2+}$   +  $\gamma$   \ra  C$^{3+}$   +  e$^{-}$                    &  15 & R113  &  C$^{3+}$   +  $\gamma$   \ra  C$^{4+}$   +  e$^{-}$                    &  15 & R114  &  C$^{4+}$   +  $\gamma$   \ra  C$^{5+}$   +  e$^{-}$                    &  15 \\
R115  &  N          +  $\gamma$   \ra  N$^{+}$    +  e$^{-}$                    &  15 & R116  &  N$^{+}$    +  $\gamma$   \ra  N$^{2+}$   +  e$^{-}$                    &  15 & R117  &  N$^{2+}$   +  $\gamma$   \ra  N$^{3+}$   +  e$^{-}$                    &  15 \\
R118  &  N$^{3+}$   +  $\gamma$   \ra  N$^{4+}$   +  e$^{-}$                    &  15 & R119  &  N$^{4+}$   +  $\gamma$   \ra  N$^{5+}$   +  e$^{-}$                    &  15 & R120  &  N$^{5+}$   +  $\gamma$   \ra  N$^{6+}$   +  e$^{-}$                    &  15 \\
R121  &  O          +  $\gamma$   \ra  O$^{+}$    +  e$^{-}$                    &  15 & R122  &  O$^{+}$    +  $\gamma$   \ra  O$^{2+}$   +  e$^{-}$                    &  15 & R123  &  O$^{2+}$   +  $\gamma$   \ra  O$^{3+}$   +  e$^{-}$                    &  15 \\
R124  &  O$^{3+}$   +  $\gamma$   \ra  O$^{4+}$   +  e$^{-}$                    &  15 & R125  &  O$^{4+}$   +  $\gamma$   \ra  O$^{5+}$   +  e$^{-}$                    &  15 & R126  &  O$^{5+}$   +  $\gamma$   \ra  O$^{6+}$   +  e$^{-}$                    &  15 \\
R127  &  O$^{6+}$   +  $\gamma$   \ra  O$^{7+}$   +  e$^{-}$                    &  15 & R128  &  Ne         +  $\gamma$   \ra  Ne$^{+}$   +  e$^{-}$                    &  15 & R129  &  Ne$^{+}$   +  $\gamma$   \ra  Ne$^{2+}$  +  e$^{-}$                    &  15 \\
R130  &  Ne$^{2+}$  +  $\gamma$   \ra  Ne$^{3+}$  +  e$^{-}$                    &  15 & R131  &  Ne$^{3+}$  +  $\gamma$   \ra  Ne$^{4+}$  +  e$^{-}$                    &  15 & R132  &  Ne$^{4+}$  +  $\gamma$   \ra  Ne$^{5+}$  +  e$^{-}$                    &  15 \\
R133  &  Ne$^{5+}$  +  $\gamma$   \ra  Ne$^{6+}$  +  e$^{-}$                    &  15 & R134  &  Ne$^{6+}$  +  $\gamma$   \ra  Ne$^{7+}$  +  e$^{-}$                    &  15 & R135  &  Ne$^{7+}$  +  $\gamma$   \ra  Ne$^{8+}$  +  e$^{-}$                    &  15 \\
R136  &  Ne$^{8+}$  +  $\gamma$   \ra  Ne$^{9+}$  +  e$^{-}$                    &  15 & R137  &  Na         +  $\gamma$   \ra  Na$^{+}$   +  e$^{-}$                    &  15 & R138  &  Na$^{+}$   +  $\gamma$   \ra  Na$^{2+}$  +  e$^{-}$                    &  15 \\
R139  &  Mg         +  $\gamma$   \ra  Mg$^{+}$   +  e$^{-}$                    &  15 & R140  &  Mg$^{+}$   +  $\gamma$   \ra  Mg$^{2+}$  +  e$^{-}$                    &  15 & R141  &  Mg$^{2+}$  +  $\gamma$   \ra  Mg$^{3+}$  +  e$^{-}$                    &  15 \\
R142  &  Si         +  $\gamma$   \ra  Si$^{+}$   +  e$^{-}$                    &  15 & R143  &  Si$^{+}$   +  $\gamma$   \ra  Si$^{2+}$  +  e$^{-}$                    &  15 & R144  &  Si$^{2+}$  +  $\gamma$   \ra  Si$^{3+}$  +  e$^{-}$                    &  15 \\
R145  &  Si$^{3+}$  +  $\gamma$   \ra  Si$^{4+}$  +  e$^{-}$                    &  15 & R146  &  Si$^{4+}$  +  $\gamma$   \ra  Si$^{5+}$  +  e$^{-}$                    &  15 & R147  &  S          +  $\gamma$   \ra  S$^{+}$    +  e$^{-}$                    &  15 \\
R148  &  S$^{+}$    +  $\gamma$   \ra  S$^{2+}$   +  e$^{-}$                    &  15 & R149  &  S$^{2+}$   +  $\gamma$   \ra  S$^{3+}$   +  e$^{-}$                    &  15 & R150  &  S$^{3+}$   +  $\gamma$   \ra  S$^{4+}$   +  e$^{-}$                    &  15 \\
R151  &  Ca         +  $\gamma$   \ra  Ca$^{+}$   +  e$^{-}$                    &  15 & R152  &  Ca$^{+}$   +  $\gamma$   \ra  Ca$^{2+}$  +  e$^{-}$                    &  15 & R153  &  Ca$^{2+}$  +  $\gamma$   \ra  Ca$^{3+}$  +  e$^{-}$                    &  15 \\ 
R154  &  Ca$^{3+}$  +  $\gamma$   \ra  Ca$^{4+}$  +  e$^{-}$                    &  15 & R155  &  Fe         +  $\gamma$   \ra  Fe$^{+}$   +  e$^{-}$                    &  15 & R156  &  Fe$^{+}$   +  $\gamma$   \ra  Fe$^{2+}$  +  e$^{-}$                    &  15 \\ 
R157  &  Fe$^{2+}$  +  $\gamma$   \ra  Fe$^{3+}$  +  e$^{-}$                    &  15 & R158  &  Fe$^{3+}$  +  $\gamma$   \ra  Fe$^{4+}$  +  e$^{-}$                    &  15 & R159  &  He$^{2+}$  +  H          \ra  He$^{+}$   +  H$^{+}$                    &  17 \\
R160  &  C$^{+}$    +  H          \ra  C          +  H$^{+}$                    &  17 & R161  &  C$^{2+}$   +  H          \ra  C$^{+}$    +  H$^{+}$                    &  17 & R162  &  C$^{3+}$   +  H          \ra  C$^{2+}$   +  H$^{+}$                    &  17 \\
R163  &  C$^{4+}$   +  H          \ra  C$^{3+}$   +  H$^{+}$                    &  17 & R164  &  N$^{+}$    +  H          \ra  N          +  H$^{+}$                    &  17 & R165  &  N$^{2+}$   +  H          \ra  N$^{+}$    +  H$^{+}$                    &  17 \\
R166  &  N$^{3+}$   +  H          \ra  N$^{2+}$   +  H$^{+}$                    &  17 & R167  &  N$^{4+}$   +  H          \ra  N$^{3+}$   +  H$^{+}$                    &  17 & R168  &  O$^{+}$    +  H          \ra  O          +  H$^{+}$                    &  17 \\
R169  &  O$^{2+}$   +  H          \ra  O$^{+}$    +  H$^{+}$                    &  17 & R170  &  O$^{3+}$   +  H          \ra  O$^{2+}$   +  H$^{+}$                    &  17 & R171  &  O$^{4+}$   +  H          \ra  O$^{3+}$   +  H$^{+}$                    &  17 \\
R172  &  Ne$^{2+}$  +  H          \ra  Ne$^{+}$   +  H$^{+}$                    &  17 & R173  &  Ne$^{3+}$  +  H          \ra  Ne$^{2+}$  +  H$^{+}$                    &  17 & R174  &  Ne$^{4+}$  +  H          \ra  Ne$^{3+}$  +  H$^{+}$                    &  17 \\
R175  &  Na$^{2+}$  +  H          \ra  Na$^{+}$   +  H$^{+}$                    &  17 & R176  &  Mg$^{2+}$  +  H          \ra  Mg$^{+}$   +  H$^{+}$                    &  17 & R177  &  Mg$^{3+}$  +  H          \ra  Mg$^{2+}$  +  H$^{+}$                    &  17 \\
R178  &  Si$^{2+}$  +  H          \ra  Si$^{+}$   +  H$^{+}$                    &  17 & R179  &  Si$^{3+}$  +  H          \ra  Si$^{2+}$  +  H$^{+}$                    &  17 & R180  &  Si$^{4+}$  +  H          \ra  Si$^{3+}$  +  H$^{+}$                    &  17 \\
R181  &  S$^{+}$    +  H          \ra  S          +  H$^{+}$                    &  17 & R182  &  S$^{2+}$   +  H          \ra  S$^{+}$    +  H$^{+}$                    &  17 & R183  &  S$^{3+}$   +  H          \ra  S$^{2+}$   +  H$^{+}$                    &  17 \\
R184  &  S$^{4+}$   +  H          \ra  S$^{3+}$   +  H$^{+}$                    &  17 & R185  &  Ca$^{3+}$  +  H          \ra  Ca$^{2+}$  +  H$^{+}$                    &  17 & R186  &  Ca$^{4+}$  +  H          \ra  Ca$^{3+}$  +  H$^{+}$                    &  17 \\
R187  &  Fe$^{2+}$  +  H          \ra  Fe$^{+}$   +  H$^{+}$                    &  17 & R188  &  Fe$^{3+}$  +  H          \ra  Fe$^{2+}$  +  H$^{+}$                    &  17 & R189  &  Fe$^{4+}$  +  H          \ra  Fe$^{3+}$  +  H$^{+}$                    &  17 \\
R190  &  C          +  H$^{+}$    \ra  C$^{+}$    +  H                          &  17 & R191  &  N          +  H$^{+}$    \ra  N$^{+}$    +  H                          &  17 & R192  &  O          +  H$^{+}$    \ra  O$^{+}$    +  H                          &  17 \\
R193  &  Na         +  H$^{+}$    \ra  Na$^{+}$   +  H                          &  17 & R194  &  Mg         +  H$^{+}$    \ra  Mg$^{+}$   +  H                          &  17 & R195  &  Mg$^{+}$   +  H$^{+}$    \ra  Mg$^{2+}$  +  H                          &  17 \\
R196  &  Si         +  H$^{+}$    \ra  Si$^{+}$   +  H                          &  17 & R197  &  Si$^{+}$   +  H$^{+}$    \ra  Si$^{2+}$  +  H                          &  17 & R198  &  S          +  H$^{+}$    \ra  S$^{+}$    +  H                          &  17 \\
R199  &  Fe         +  H$^{+}$    \ra  Fe$^{+}$   +  H                          &  17 & R200  &  Fe$^{+}$   +  H$^{+}$    \ra  Fe$^{2+}$  +  H                          &  17 & R201  &  C$^{3+}$   +  He         \ra  C$^{2+}$   +  He$^{+}$                   &  17 \\
R202  &  C$^{4+}$   +  He         \ra  C$^{3+}$   +  He$^{+}$                   &  17 & R203  &  N$^{2+}$   +  He         \ra  N$^{+}$    +  He$^{+}$                   &  17 & R204  &  N$^{3+}$   +  He         \ra  N$^{2+}$   +  He$^{+}$                   &  17 \\
R205  &  N$^{4+}$   +  He         \ra  N$^{3+}$   +  He$^{+}$                   &  17 & R206  &  O$^{2+}$   +  He         \ra  O$^{+}$    +  He$^{+}$                   &  17 & R207  &  O$^{3+}$   +  He         \ra  O$^{2+}$   +  He$^{+}$                   &  17 \\
R208  &  O$^{4+}$   +  He         \ra  O$^{3+}$   +  He$^{+}$                   &  17 & R209  &  Ne$^{2+}$  +  He         \ra  Ne$^{+}$   +  He$^{+}$                   &  17 & R210  &  Ne$^{3+}$  +  He         \ra  Ne$^{2+}$  +  He$^{+}$                   &  17 \\
R211  &  Ne$^{4+}$  +  He         \ra  Ne$^{3+}$  +  He$^{+}$                   &  17 & R212  &  Na$^{2+}$  +  He         \ra  Na$^{+}$   +  He$^{+}$                   &  17 & R213  &  Mg$^{3+}$  +  He         \ra  Mg$^{2+}$  +  He$^{+}$                   &  17 \\
R214  &  S$^{3+}$   +  He         \ra  S$^{2+}$   +  He$^{+}$                   &  17 & R215  &  S$^{4+}$   +  He         \ra  S$^{3+}$   +  He$^{+}$                   &  17 & R216  &  Ca$^{3+}$  +  He         \ra  Ca$^{2+}$  +  He$^{+}$                   &  17 \\
R217  &  Ca$^{4+}$  +  He         \ra  Ca$^{3+}$  +  He$^{+}$                   &  17 & R218  &  Si$^{3+}$  +  He         \ra  Si$^{2+}$  +  He$^{+}$                   &  17 & R219  &  Si$^{4+}$  +  He         \ra  Si$^{3+}$  +  He$^{+}$                   &  17 \\
R220  &  Fe$^{3+}$  +  He         \ra  Fe$^{2+}$  +  He$^{+}$                   &  17 & R221  &  Fe$^{4+}$  +  He         \ra  Fe$^{3+}$  +  He$^{+}$                   &  17 & R222  &  Si         +  He$^{+}$   \ra  Si$^{+}$   +  He                         &  17 \\
R223  &  Fe         +  C$^{+}$    \ra  C          +  Fe$^{+}$                   &  18 & R224  &  Mg         +  C$^{+}$    \ra  C          +  Mg$^{+}$                   &  18 & R225  &  Na         +  C$^{+}$    \ra  C          +  Na$^{+}$                   &  18 \\
R226  &  S          +  C$^{+}$    \ra  C          +  S$^{+}$                    &  18 & R227  &  Si         +  C$^{+}$    \ra  C          +  Si$^{+}$                   &  18 & R228  &  Na         +  Fe$^{+}$   \ra  Fe         +  Na$^{+}$                   &  18 \\ 
R229  &  Na         +  Mg$^{+}$   \ra  Mg         +  Na$^{+}$                   &  18 & R230  &  Fe         +  N$^{+}$    \ra  Fe$^{+}$   +  N                          &  18 & R231  &  Mg         +  N$^{+}$    \ra  Mg$^{+}$   +  N                          &  18 \\
R232  &  Fe         +  O$^{+}$    \ra  O          +  Fe$^{+}$                   &  18 & R233  &  Fe         +  S$^{+}$    \ra  S          +  Fe$^{+}$                   &  18 & R234  &  Mg         +  S$^{+}$    \ra  S          +  Mg$^{+}$                   &  18 \\
R235  &  Na         +  S$^{+}$    \ra  S          +  Na$^{+}$                   &  18 & R236  &  Si         +  S$^{+}$    \ra  S          +  Si$^{+}$                   &  18 & R237  &  Fe         +  Si$^{+}$   \ra  Si         +  Fe$^{+}$                   &  18 \\ 
R238  &  Mg         +  Si$^{+}$   \ra  Si         +  Mg$^{+}$                   &  18 & R239  &  Na         +  Si$^{+}$   \ra  Si         +  Na$^{+}$                   &  18 & R240  &  Si$^{2+}$  +  He$^{+}$   \ra  Si$^{3+}$  +  He                         &  18 \\
\hline
\end{tabular}
\begin{tablenotes}
		\item {\bf Notes:} List of the reactions considered and their source. 
		Radiative Recombination rates for all species are taken from \cite{Badnell2006RR}. 
		(1) \cite{Badnell2006H}, (2) \cite{Bautista2007}, (3) \cite{Colgan2004}, 
		(4) \cite{Colgan2003}, (5) \cite{Altun2004}, (6) \cite{Zatsarinny2004a}, 
		(7) \cite{Mitnik2004}, (8) \cite{Zatsarinny2004b}, (9) \cite{Altun2006}, 
		(10) \cite{Verner1996A}, (11) \cite{Voronov1997}, (12) \cite{Glover2008}, 
		(13) \cite{Shull1982}, (14) \cite{Mazzotta1998}, (15) \cite{Haardt2012}, 
		(16) \cite{BadnellWebsite}, (17) \cite{Kingdon1996}, (18) \cite{Wakelam2012}, 
		(19) \cite{Zatsarinny2006}, (20) \cite{Zatsarinny2003}, (21) \cite{AbdelNaby2012},
		(22) \cite{Altun2007}, and (23) \cite{Glover2009}.
\end{tablenotes}
\end{threeparttable}
}

\section{Summary of Results}
\label{summ_results}

\subsection{Table 3: U=0}
\resizebox{0.95\textwidth}{!}{%
\begin{threeparttable}
	%\caption{Results for U=0.}
	%\caption{}
	\label{tab:uzero}
	\begin{tabular}{|l||r|rrrr|rrrr|rrrr|rrrr|rrrr|}
	\hline
	
	Col. (cm$^{-2}$)              & 1.0E17    & 2.9E17    & 1.0E17    & 2.9E17    & 1.0E18    & 1.4E17    & 7.1E17    & 1.4E18    & 7.1E18    & 1.4E18    & 4.3E18    & 1.4E19    & 4.3E19    & 5.7E18    & 1.4E19    & 5.7E19    & 1.4E20    & 1.0E19    & 2.9E19    & 1.0E20    & 2.9E20     \\
	$\sigma_{1D}$ (km s$^{-1}$)   & 3.5       & 11.5      & 11.5      & 11.5      & 11.5      & 20.2      & 20.2      & 20.2      & 20.2      & 34.6      & 34.6      & 34.6      & 34.6      & 46.2      & 46.2      & 46.2      & 46.2      & 57.7      & 57.7      & 57.7      & 57.7       \\
	T$_{DW}$ (10$^{4}$ K)         & 1.0       & 1.3       & 5.2       & 1.3       & 0.9       & 12.4      & 1.1       & 0.9       & 0.6       & 3.9       & 1.0       & 0.7       & 0.6       & 1.5       & 0.9       & 0.6       & 0.5       & 1.7       & 0.9       & 0.6       & 0.5        \\
	M$_{DW}$                      & 0.6       & 1.2       & 0.6       & 1.2       & 1.6       & 0.7       & 2.5       & 2.7       & 3.4       & 2.3       & 4.3       & 5.6       & 6.7       & 4.9       & 7.4       & 8.2       & 9.1       & 5.7       & 8.8       & 10.6      & 11.7       \\
	\hline
	\hline
	${\rm H}$                     & -0.3      & -0.8      & -3.6      & -0.8      & -0.4      & -4.9      & -0.6      & -0.5      & -0.4      & -1.8      & -0.6      & -0.4      & -0.4      & -0.8      & -0.5      & -0.4      & -0.4      & -0.8      & -0.5      & -0.4      & -0.4       \\
	${\rm H^+}$                   & -0.3      & -0.1      & -0.0      & -0.1      & -0.2      & -0.0      & -0.1      & -0.2      & -0.2      & -0.0      & -0.1      & -0.2      & -0.2      & -0.1      & -0.2      & -0.2      & -0.2      & -0.1      & -0.2      & -0.2      & -0.2       \\
	\hline
	${\rm He}$                    & -0.0      & -0.9      & -3.1      & -0.9      & -0.2      & -4.6      & -0.2      & -0.2      & -0.0      & -1.4      & -0.0      & -0.0      & -0.0      & -0.3      & -0.0      & -0.0      & -0.0      & -0.3      & -0.0      & -0.0      & -0.0       \\
	${\rm He^{+}}$                & -4.8      & -0.2      & -0.7      & -0.2      & -0.5      & -1.3      & -0.4      & -0.5      & -2.6      & -0.2      & -1.1      & -2.9      & -3.9      & -0.3      & -1.3      & -2.3      & -3.2      & -0.3      & -1.1      & -2.1      & -2.7       \\
	${\rm He^{2+}}$               & -21.6     & -0.6      & -0.1      & -0.6      & -2.8      & -0.0      & -2.6      & -2.6      & -8.7      & -0.6      & -3.9      & -6.2      & -7.8      & -1.7      & -3.4      & -4.2      & -5.6      & -1.2      & -2.6      & -4.1      & -4.7       \\
	\hline
	${\rm C}$                     & -0.1      & -0.8      & -3.0      & -0.8      & -0.2      & -5.5      & -0.7      & -0.4      & -0.1      & -1.7      & -0.4      & -0.1      & -0.0      & -0.6      & -0.2      & -0.1      & -0.0      & -0.5      & -0.2      & -0.1      & -0.0       \\
	${\rm C^{+}}$                 & -0.8      & -0.1      & -0.6      & -0.1      & -0.4      & -2.4      & -0.1      & -0.2      & -0.9      & -0.4      & -0.2      & -0.6      & -1.1      & -0.2      & -0.4      & -0.8      & -1.2      & -0.2      & -0.4      & -0.8      & -1.2       \\
	${\rm C^{2+}}$                & -8.1      & -2.3      & -0.1      & -2.3      & -5.8      & -0.6      & -2.8      & -3.6      & -5.0      & -0.3      & -2.3      & -3.8      & -4.4      & -1.1      & -2.1      & -3.0      & -3.9      & -1.0      & -1.9      & -3.0      & -3.5       \\
	${\rm C^{3+}}$                & -20.3     & -13.3     & -2.2      & -13.3     & -17.7     & -0.3      & -8.8      & -8.3      & -12.0     & -1.5      & -4.4      & -7.7      & -9.1      & -2.5      & -4.3      & -5.2      & -6.7      & -2.0      & -3.3      & -4.7      & -5.4       \\
	${\rm C^{4+}}$                & -20.1     & -21.9     & -1.2      & -21.9     & -26.5     & -0.7      & -14.2     & -12.8     & -18.6     & -2.0      & -5.8      & -9.9      & -13.2     & -3.1      & -5.7      & -6.5      & -8.8      & -2.1      & -4.0      & -5.3      & -6.4       \\
	${\rm C^{5+}}$                & -17.9     & -21.6     & -17.7     & -21.6     & -26.2     & -12.3     & -23.6     & -26.2     & -26.3     & -9.5      & -13.2     & -25.7     & -20.3     & -8.7      & -12.1     & -13.7     & -18.2     & -6.0      & -9.8      & -11.3     & -13.7      \\
	\hline
	${\rm N}$                     & -0.3      & -0.7      & -2.8      & -0.7      & -0.3      & -5.3      & -0.5      & -0.3      & -0.2      & -1.6      & -0.4      & -0.2      & -0.1      & -0.6      & -0.3      & -0.2      & -0.1      & -0.5      & -0.3      & -0.2      & -0.1       \\
	${\rm N^{+}}$                 & -0.3      & -0.1      & -0.4      & -0.1      & -0.3      & -2.1      & -0.2      & -0.3      & -0.5      & -0.2      & -0.2      & -0.4      & -0.6      & -0.2      & -0.3      & -0.5      & -0.6      & -0.2      & -0.4      & -0.5      & -0.6       \\
	${\rm N^{2+}}$                & -18.8     & -7.0      & -0.3      & -7.0      & -9.8      & -0.5      & -5.2      & -5.5      & -7.7      & -0.5      & -3.5      & -5.7      & -6.6      & -1.6      & -3.2      & -4.0      & -5.3      & -1.3      & -2.5      & -3.8      & -4.2       \\
	${\rm N^{3+}}$                & -19.5     & -10.5     & -2.0      & -10.5     & -19.3     & -0.2      & -9.7      & -9.1      & -13.3     & -1.3      & -4.4      & -8.0      & -9.6      & -2.4      & -4.4      & -5.3      & -6.8      & -1.9      & -3.3      & -4.7      & -5.3       \\
	${\rm N^{4+}}$                & -19.4     & -10.4     & -6.2      & -10.4     & -22.5     & -1.4      & -15.3     & -13.8     & -20.2     & -2.7      & -6.0      & -11.1     & -14.2     & -3.5      & -6.0      & -6.8      & -9.1      & -2.7      & -4.5      & -5.6      & -6.8       \\
	${\rm N^{5+}}$                & -17.1     & -8.8      & -5.4      & -8.8      & -20.6     & -3.3      & -14.3     & -15.6     & -24.1     & -3.6      & -7.9      & -13.4     & -15.8     & -4.2      & -7.0      & -7.8      & -10.4     & -2.7      & -5.0      & -6.2      & -7.8       \\
	${\rm N^{6+}}$                & -17.8     & -23.4     & -17.5     & -23.4     & -25.8     & -17.3     & -25.8     & -25.8     & -25.8     & -12.9     & -16.7     & -25.8     & -19.9     & -11.1     & -15.1     & -17.5     & -19.2     & -7.4      & -12.3     & -14.2     & -17.0      \\
	\hline
	${\rm O}$                     & -0.2      & -0.7      & -3.2      & -0.7      & -0.4      & -5.3      & -0.6      & -0.4      & -0.3      & -1.8      & -0.5      & -0.4      & -0.3      & -0.8      & -0.5      & -0.4      & -0.3      & -0.7      & -0.4      & -0.4      & -0.3       \\
	${\rm O^{+}}$                 & -0.4      & -0.1      & -0.4      & -0.1      & -0.3      & -2.0      & -0.1      & -0.2      & -0.3      & -0.2      & -0.2      & -0.2      & -0.3      & -0.1      & -0.2      & -0.3      & -0.3      & -0.1      & -0.2      & -0.3      & -0.3       \\
	${\rm O^{2+}}$                & -19.8     & -8.0      & -0.2      & -8.0      & -10.7     & -0.5      & -5.7      & -5.9      & -8.2      & -0.5      & -3.5      & -5.8      & -6.9      & -1.6      & -3.2      & -4.0      & -5.3      & -1.3      & -2.5      & -3.7      & -4.2       \\
	${\rm O^{3+}}$                & -20.4     & -16.8     & -1.7      & -16.8     & -26.9     & -0.2      & -10.3     & -9.5      & -13.9     & -1.3      & -4.5      & -8.0      & -10.0     & -2.4      & -4.4      & -5.3      & -6.9      & -1.9      & -3.3      & -4.7      & -5.4       \\
	${\rm O^{4+}}$                & -19.4     & -19.7     & -5.7      & -19.7     & -26.6     & -1.5      & -15.8     & -14.1     & -20.9     & -2.8      & -6.1      & -11.2     & -14.7     & -3.6      & -6.0      & -6.7      & -9.2      & -2.7      & -4.5      & -5.6      & -6.8       \\
	${\rm O^{5+}}$                & -18.0     & -19.1     & -11.8     & -19.1     & -26.3     & -4.3      & -17.2     & -20.6     & -25.7     & -4.5      & -8.3      & -15.2     & -17.7     & -4.8      & -7.8      & -8.5      & -12.0     & -3.5      & -5.7      & -6.6      & -8.2       \\
	${\rm O^{6+}}$                & -18.4     & -20.1     & -15.0     & -20.1     & -26.3     & -7.6      & -20.6     & -25.9     & -26.4     & -5.9      & -10.6     & -20.0     & -20.5     & -5.7      & -9.3      & -10.5     & -14.6     & -3.7      & -6.7      & -8.0      & -9.8       \\
	${\rm O^{7+}}$                & -19.7     & -26.5     & -18.6     & -26.5     & -26.5     & -18.3     & -26.5     & -26.5     & -26.6     & -17.0     & -20.9     & -26.6     & -20.4     & -14.0     & -18.5     & -21.4     & -19.8     & -8.8      & -15.1     & -17.4     & -19.4      \\
	\hline
	${\rm Ne}$                    & -0.0      & -1.3      & -3.4      & -1.3      & -0.4      & -5.0      & -0.5      & -0.4      & -0.0      & -2.1      & -0.1      & -0.0      & -0.0      & -0.6      & -0.1      & -0.0      & -0.0      & -0.5      & -0.1      & -0.0      & -0.0       \\
	${\rm Ne^{+}}$                & -3.4      & -0.2      & -0.5      & -0.2      & -0.2      & -1.7      & -0.2      & -0.2      & -1.6      & -0.4      & -0.5      & -1.9      & -2.7      & -0.2      & -0.8      & -1.7      & -2.4      & -0.2      & -0.7      & -1.5      & -2.1       \\
	${\rm Ne^{2+}}$               & -16.5     & -0.5      & -0.2      & -0.5      & -2.6      & -0.4      & -2.5      & -2.5      & -8.1      & -0.3      & -3.0      & -5.0      & -6.4      & -1.2      & -2.7      & -3.5      & -4.8      & -1.0      & -2.1      & -3.5      & -4.0       \\
	${\rm Ne^{3+}}$               & -20.2     & -15.2     & -1.6      & -15.2     & -24.1     & -0.3      & -11.2     & -10.3     & -16.8     & -1.5      & -4.9      & -9.0      & -11.5     & -2.7      & -4.9      & -5.8      & -7.6      & -2.0      & -3.5      & -5.0      & -5.8       \\
	${\rm Ne^{4+}}$               & -20.3     & -25.9     & -4.3      & -25.9     & -26.4     & -1.6      & -17.6     & -15.7     & -17.8     & -2.8      & -6.5      & -12.9     & -16.9     & -3.6      & -6.4      & -7.1      & -10.1     & -2.6      & -4.6      & -5.7      & -7.0       \\
	${\rm Ne^{5+}}$               & -18.3     & -25.4     & -10.8     & -25.4     & -25.4     & -4.6      & -23.9     & -22.4     & -25.4     & -4.6      & -8.8      & -16.7     & -19.1     & -4.8      & -8.3      & -8.8      & -12.9     & -3.5      & -5.9      & -6.7      & -8.4       \\
	${\rm Ne^{6+}}$               & -19.0     & -25.9     & -18.2     & -25.9     & -25.9     & -9.3      & -25.9     & -25.9     & -26.1     & -7.1      & -11.8     & -21.1     & -18.6     & -6.5      & -11.1     & -11.7     & -17.3     & -4.7      & -7.8      & -8.6      & -11.0      \\
	${\rm Ne^{7+}}$               & -18.5     & -25.8     & -18.2     & -25.8     & -25.7     & -15.0     & -25.8     & -25.7     & -25.7     & -10.1     & -15.4     & -25.7     & -19.4     & -8.5      & -13.3     & -15.2     & -18.8     & -5.8      & -10.1     & -11.2     & -14.7      \\
	${\rm Ne^{8+}}$               & -19.1     & -25.7     & -17.6     & -25.7     & -25.7     & -17.4     & -25.7     & -25.7     & -25.7     & -13.1     & -18.2     & -25.8     & -19.8     & -10.9     & -15.9     & -18.8     & -19.0     & -6.2      & -11.8     & -14.2     & -16.1      \\
	${\rm Ne^{9+}}$               & -21.9     & -25.9     & -18.6     & -25.9     & -25.9     & -18.1     & -25.9     & -25.9     & -25.9     & -25.9     & -25.9     & -26.0     & -19.2     & -22.3     & -26.0     & -19.9     & -18.5     & -12.3     & -21.8     & -19.1     & -18.2      \\
	\hline
	${\rm Na}$                    & -3.2      & -3.9      & -6.8      & -3.9      & -2.5      & -7.2      & -2.9      & -2.3      & -1.2      & -4.1      & -2.1      & -1.2      & -0.7      & -2.3      & -1.5      & -0.8      & -0.5      & -2.0      & -1.2      & -0.7      & -0.5       \\
	${\rm Na^{+}}$                & -0.0      & -0.3      & -0.9      & -0.3      & -0.0      & -1.7      & -0.0      & -0.0      & -0.0      & -0.5      & -0.0      & -0.0      & -0.1      & -0.0      & -0.0      & -0.1      & -0.1      & -0.1      & -0.0      & -0.1      & -0.2       \\
	${\rm Na^{2+}}$               & -15.8     & -0.4      & -0.1      & -0.4      & -1.8      & -0.0      & -1.6      & -1.7      & -6.7      & -0.2      & -2.8      & -4.8      & -6.1      & -1.0      & -2.5      & -3.3      & -4.6      & -0.8      & -2.0      & -3.3      & -3.8       \\
	\hline
	${\rm Mg}$                    & -1.3      & -1.9      & -6.5      & -1.9      & -1.4      & -8.1      & -1.7      & -1.5      & -1.5      & -3.1      & -1.6      & -1.5      & -1.5      & -1.8      & -1.5      & -1.5      & -1.6      & -1.7      & -1.5      & -1.5      & -1.6       \\
	${\rm Mg^{+}}$                & -0.0      & -0.5      & -3.4      & -0.5      & -0.1      & -4.2      & -0.3      & -0.1      & -0.0      & -1.6      & -0.1      & -0.0      & -0.0      & -0.4      & -0.1      & -0.0      & -0.0      & -0.3      & -0.1      & -0.0      & -0.0       \\
	${\rm Mg^{2+}}$               & -1.7      & -0.2      & -0.0      & -0.2      & -1.1      & -0.2      & -0.4      & -0.8      & -2.0      & -0.0      & -0.6      & -1.4      & -2.0      & -0.3      & -0.8      & -1.4      & -1.8      & -0.3      & -0.8      & -1.4      & -1.8       \\
	${\rm Mg^{3+}}$               & -23.6     & -18.9     & -1.0      & -18.9     & -22.7     & -0.4      & -10.4     & -9.9      & -14.4     & -1.8      & -5.2      & -9.0      & -11.0     & -2.9      & -5.1      & -6.0      & -8.0      & -2.0      & -3.8      & -5.1      & -6.0       \\
	\hline
	${\rm Si}$                    & -1.2      & -2.2      & -5.0      & -2.2      & -1.8      & -8.3      & -1.9      & -1.7      & -0.4      & -3.0      & -1.3      & -0.4      & -0.2      & -1.9      & -0.8      & -0.3      & -0.2      & -1.7      & -0.7      & -0.3      & -0.2       \\
	${\rm Si^{+}}$                & -0.0      & -0.1      & -1.9      & -0.1      & -0.0      & -4.7      & -0.0      & -0.0      & -0.2      & -0.8      & -0.1      & -0.2      & -0.4      & -0.1      & -0.1      & -0.3      & -0.5      & -0.2      & -0.1      & -0.3      & -0.5       \\
	${\rm Si^{2+}}$               & -1.9      & -0.9      & -0.4      & -0.9      & -1.7      & -2.3      & -1.1      & -1.5      & -2.1      & -0.2      & -1.2      & -1.8      & -2.2      & -0.6      & -1.3      & -1.8      & -2.2      & -0.6      & -1.3      & -1.8      & -2.2       \\
	${\rm Si^{3+}}$               & -11.6     & -4.2      & -1.2      & -4.2      & -6.3      & -1.4      & -4.7      & -5.3      & -8.0      & -1.0      & -3.7      & -6.1      & -7.0      & -2.0      & -3.5      & -4.3      & -5.5      & -1.7      & -2.7      & -4.0      & -4.5       \\
	${\rm Si^{4+}}$               & -20.1     & -13.5     & -0.3      & -13.5     & -20.0     & -0.0      & -10.7     & -9.8      & -14.6     & -1.2      & -4.7      & -8.7      & -10.5     & -2.4      & -4.7      & -5.6      & -7.3      & -1.8      & -3.3      & -4.8      & -5.5       \\
	${\rm Si^{5+}}$               & -17.4     & -11.5     & -8.0      & -11.5     & -24.0     & -4.0      & -14.5     & -18.1     & -23.3     & -4.3      & -8.2      & -14.0     & -15.8     & -4.7      & -7.4      & -8.3      & -11.0     & -3.2      & -5.7      & -6.9      & -8.6       \\
	\hline
	${\rm S}$                     & -0.5      & -1.8      & -4.4      & -1.8      & -0.7      & -6.9      & -1.4      & -1.0      & -0.2      & -2.7      & -0.9      & -0.3      & -0.1      & -1.1      & -0.5      & -0.2      & -0.1      & -0.9      & -0.4      & -0.2      & -0.1       \\
	${\rm S^{+}}$                 & -0.2      & -0.1      & -1.3      & -0.1      & -0.1      & -3.4      & -0.0      & -0.1      & -0.4      & -0.7      & -0.1      & -0.3      & -0.7      & -0.2      & -0.2      & -0.5      & -0.8      & -0.2      & -0.3      & -0.5      & -0.9       \\
	${\rm S^{2+}}$                & -5.8      & -0.7      & -0.1      & -0.7      & -2.5      & -1.2      & -1.3      & -2.0      & -3.7      & -0.2      & -1.5      & -2.7      & -3.3      & -0.6      & -1.5      & -2.3      & -3.0      & -0.6      & -1.4      & -2.3      & -2.8       \\
	${\rm S^{3+}}$                & -24.8     & -8.7      & -0.8      & -8.7      & -12.8     & -0.3      & -6.3      & -6.3      & -9.2      & -0.8      & -3.7      & -6.2      & -7.3      & -1.9      & -3.5      & -4.3      & -5.6      & -1.6      & -2.7      & -4.0      & -4.5       \\
	${\rm S^{4+}}$                & -25.3     & -16.2     & -3.1      & -16.2     & -19.7     & -0.3      & -10.0     & -9.1      & -13.9     & -1.6      & -4.4      & -7.8      & -9.5      & -2.6      & -4.3      & -5.3      & -6.8      & -2.0      & -3.4      & -4.7      & -5.5       \\
	\hline
	${\rm Ca}$                    & -1.3      & -2.6      & -7.2      & -2.6      & -1.3      & -8.1      & -2.1      & -1.5      & -0.6      & -3.5      & -1.3      & -0.6      & -0.3      & -1.5      & -0.9      & -0.4      & -0.2      & -1.3      & -0.7      & -0.3      & -0.2       \\
	${\rm Ca^{+}}$                & -0.1      & -0.9      & -3.4      & -0.9      & -0.2      & -3.5      & -0.7      & -0.4      & -0.2      & -1.9      & -0.4      & -0.2      & -0.3      & -0.6      & -0.3      & -0.3      & -0.5      & -0.6      & -0.3      & -0.4      & -0.5       \\
	${\rm Ca^{2+}}$               & -1.0      & -0.1      & -0.1      & -0.1      & -0.6      & -0.6      & -0.1      & -0.3      & -1.2      & -0.1      & -0.3      & -0.8      & -1.4      & -0.1      & -0.4      & -1.0      & -1.3      & -0.2      & -0.5      & -0.9      & -1.3       \\
	${\rm Ca^{3+}}$               & -21.6     & -10.8     & -0.8      & -10.8     & -20.7     & -0.3      & -7.0      & -6.8      & -9.8      & -0.8      & -3.8      & -6.4      & -7.9      & -1.9      & -3.6      & -4.5      & -5.9      & -1.6      & -2.8      & -4.1      & -4.7       \\
	${\rm Ca^{4+}}$               & -24.3     & -19.9     & -2.6      & -19.9     & -24.5     & -0.7      & -12.0     & -10.9     & -16.3     & -1.9      & -5.0      & -9.6      & -11.8     & -2.9      & -5.0      & -5.9      & -7.9      & -2.2      & -3.8      & -5.1      & -6.1       \\
	\hline
	${\rm Fe}$                    & -3.9      & -3.8      & -6.9      & -3.8      & -3.9      & -9.3      & -3.8      & -3.9      & -4.0      & -4.7      & -3.9      & -4.0      & -4.1      & -4.0      & -3.9      & -4.0      & -4.1      & -4.0      & -3.9      & -4.0      & -4.1       \\
	${\rm Fe^{+}}$                & -0.0      & -0.3      & -3.3      & -0.3      & -0.0      & -5.0      & -0.1      & -0.1      & -0.0      & -1.2      & -0.1      & -0.0      & -0.0      & -0.3      & -0.1      & -0.0      & -0.0      & -0.3      & -0.1      & -0.0      & -0.0       \\
	${\rm Fe^{2+}}$               & -1.2      & -0.4      & -0.8      & -0.4      & -1.1      & -2.1      & -0.6      & -0.9      & -1.4      & -0.3      & -0.7      & -1.1      & -1.5      & -0.4      & -0.8      & -1.2      & -1.5      & -0.4      & -0.8      & -1.2      & -1.5       \\
	${\rm Fe^{3+}}$               & -22.8     & -8.2      & -0.2      & -8.2      & -10.9     & -0.6      & -5.9      & -6.1      & -8.5      & -0.5      & -3.5      & -5.9      & -7.0      & -1.6      & -3.2      & -3.9      & -5.3      & -1.3      & -2.4      & -3.7      & -4.1       \\
	${\rm Fe^{4+}}$               & -25.4     & -17.6     & -0.7      & -17.6     & -20.2     & -0.1      & -10.4     & -9.5      & -14.3     & -1.1      & -4.4      & -8.2      & -10.0     & -2.3      & -4.4      & -5.3      & -6.9      & -1.7      & -3.2      & -4.6      & -5.3       \\	
\hline
\hline
\end{tabular}
\begin{tablenotes}
\item	{\bf Notes:} Abundances values are given as log$_{10}(F_i/F_t)$. 
\end{tablenotes}
\end{threeparttable}
}
\subsection{Table 4: U$=$10$^{-6}$}
\resizebox{0.95\textwidth}{!}{%
\begin{threeparttable}
	%\caption{Results for U=1.0$\times$10$^{-6}$.}
	\label{tab:uminus6}
	\begin{tabular}{|l||r|rrrr|rrrr|rrrr|rrrr|rrrr|}
	\hline
	
	Col. (cm$^{-2}$)              & 1.0E17    & 2.9E17    & 1.0E17    & 2.9E17    & 1.0E18    & 1.4E17    & 7.1E17    & 1.4E18    & 7.1E18    & 1.4E18    & 4.3E18    & 1.4E19    & 4.3E19    & 5.7E18    & 1.4E19    & 5.7E19    & 1.4E20    & 1.0E19    & 2.9E19    & 1.0E20    & 2.9E20     \\
	$\sigma_{1D}$ (km s$^{-1}$)   & 3.5       & 11.5      & 11.5      & 11.5      & 11.5      & 20.2      & 20.2      & 20.2      & 20.2      & 34.6      & 34.6      & 34.6      & 34.6      & 46.2      & 46.2      & 46.2      & 46.2      & 57.7      & 57.7      & 57.7      & 57.7       \\
	T$_{DW}$ (10$^{4}$ K)         & 0.9       & 1.2       & 4.9       & 1.2       & 0.8       & 12.2      & 1.1       & 0.9       & 0.6       & 3.7       & 1.0       & 0.7       & 0.5       & 1.5       & 0.9       & 0.6       & 0.5       & 1.8       & 0.9       & 0.6       & 0.5        \\
	M$_{DW}$                      & 0.6       & 1.2       & 0.7       & 1.2       & 1.7       & 0.7       & 2.4       & 2.9       & 3.3       & 2.3       & 4.8       & 5.9       & 6.2       & 4.8       & 6.4       & 7.8       & 8.5       & 5.5       & 7.9       & 10.0      & 11.4       \\
	\hline
	\hline
	${\rm H}$                     & -0.3      & -0.7      & -3.5      & -0.7      & -0.4      & -4.8      & -0.6      & -0.5      & -0.4      & -1.8      & -0.6      & -0.4      & -0.4      & -0.8      & -0.5      & -0.4      & -0.4      & -0.8      & -0.5      & -0.4      & -0.4       \\
	${\rm H^+}$                   & -0.3      & -0.1      & -0.0      & -0.1      & -0.2      & -0.0      & -0.1      & -0.2      & -0.2      & -0.0      & -0.1      & -0.2      & -0.2      & -0.1      & -0.2      & -0.2      & -0.2      & -0.1      & -0.2      & -0.2      & -0.2       \\
	\hline
	${\rm He}$                    & -0.0      & -0.8      & -2.9      & -0.8      & -0.1      & -4.7      & -0.2      & -0.1      & -0.0      & -1.3      & -0.0      & -0.0      & -0.0      & -0.3      & -0.0      & -0.0      & -0.0      & -0.3      & -0.1      & -0.0      & -0.0       \\
	${\rm He^{+}}$                & -2.4      & -0.2      & -0.6      & -0.2      & -0.6      & -1.4      & -0.5      & -0.6      & -1.5      & -0.1      & -1.2      & -1.5      & -1.5      & -0.3      & -1.1      & -1.5      & -1.5      & -0.4      & -1.0      & -1.4      & -1.4       \\
	${\rm He^{2+}}$               & -6.4      & -0.8      & -0.1      & -0.8      & -3.2      & -0.0      & -2.9      & -3.2      & -4.8      & -0.7      & -4.2      & -4.8      & -4.7      & -1.7      & -3.3      & -4.3      & -4.5      & -1.2      & -2.5      & -3.2      & -4.1       \\
	\hline
	${\rm C}$                     & -0.1      & -0.8      & -2.8      & -0.8      & -0.4      & -5.4      & -0.8      & -0.5      & -0.3      & -1.7      & -0.5      & -0.3      & -0.2      & -0.6      & -0.3      & -0.2      & -0.2      & -0.6      & -0.3      & -0.2      & -0.2       \\
	${\rm C^{+}}$                 & -0.6      & -0.1      & -0.6      & -0.1      & -0.2      & -2.3      & -0.1      & -0.2      & -0.3      & -0.3      & -0.2      & -0.3      & -0.4      & -0.2      & -0.3      & -0.4      & -0.4      & -0.2      & -0.3      & -0.4      & -0.4       \\
	${\rm C^{2+}}$                & -3.2      & -2.4      & -0.2      & -2.4      & -2.6      & -0.6      & -2.4      & -2.6      & -2.7      & -0.3      & -2.4      & -2.7      & -2.7      & -1.1      & -2.1      & -2.6      & -2.7      & -1.0      & -1.8      & -2.3      & -2.6       \\
	${\rm C^{3+}}$                & -9.0      & -7.8      & -2.3      & -7.8      & -8.3      & -0.3      & -7.6      & -8.1      & -8.1      & -1.6      & -5.2      & -6.3      & -7.4      & -2.4      & -4.0      & -5.3      & -6.0      & -2.0      & -3.2      & -3.8      & -5.0       \\
	${\rm C^{4+}}$                & -15.2     & -13.5     & -1.3      & -13.5     & -14.1     & -0.6      & -11.5     & -13.6     & -13.2     & -2.2      & -7.4      & -8.2      & -10.7     & -3.0      & -5.0      & -6.7      & -7.9      & -2.0      & -3.8      & -4.3      & -5.8       \\
	${\rm C^{5+}}$                & -18.3     & -18.7     & -6.8      & -18.7     & -19.2     & -6.2      & -16.1     & -17.5     & -17.9     & -7.3      & -12.1     & -12.5     & -14.3     & -7.7      & -9.6      & -11.1     & -12.7     & -5.7      & -8.0      & -8.9      & -10.9      \\
	\hline
	${\rm N}$                     & -0.3      & -0.6      & -2.7      & -0.6      & -0.3      & -5.2      & -0.5      & -0.4      & -0.2      & -1.6      & -0.4      & -0.2      & -0.2      & -0.6      & -0.3      & -0.2      & -0.1      & -0.5      & -0.3      & -0.2      & -0.1       \\
	${\rm N^{+}}$                 & -0.3      & -0.1      & -0.3      & -0.1      & -0.3      & -2.0      & -0.2      & -0.3      & -0.4      & -0.2      & -0.2      & -0.4      & -0.5      & -0.2      & -0.3      & -0.5      & -0.5      & -0.2      & -0.3      & -0.5      & -0.5       \\
	${\rm N^{2+}}$                & -4.4      & -3.8      & -0.3      & -3.8      & -4.2      & -0.5      & -3.9      & -4.1      & -4.3      & -0.5      & -3.7      & -4.1      & -4.2      & -1.6      & -3.1      & -3.9      & -4.1      & -1.4      & -2.4      & -3.0      & -3.7       \\
	${\rm N^{3+}}$                & -9.9      & -8.9      & -2.4      & -8.9      & -9.3      & -0.2      & -8.6      & -9.0      & -8.7      & -1.4      & -5.4      & -6.4      & -7.8      & -2.4      & -4.0      & -5.3      & -6.0      & -2.0      & -3.2      & -3.7      & -4.9       \\
	${\rm N^{4+}}$                & -15.9     & -11.4     & -6.4      & -11.4     & -14.6     & -1.4      & -13.0     & -14.0     & -13.2     & -3.0      & -7.5      & -8.4      & -11.4     & -3.5      & -5.4      & -7.2      & -8.5      & -2.6      & -4.2      & -4.8      & -6.5       \\
	${\rm N^{5+}}$                & -17.3     & -9.7      & -5.5      & -9.7      & -18.9     & -3.2      & -15.0     & -16.5     & -17.1     & -3.7      & -9.3      & -9.6      & -13.0     & -4.1      & -6.1      & -8.3      & -10.8     & -2.4      & -4.6      & -5.1      & -7.8       \\
	${\rm N^{6+}}$                & -18.4     & -15.1     & -11.3     & -15.1     & -24.4     & -9.1      & -20.9     & -21.7     & -22.1     & -9.2      & -14.4     & -14.8     & -18.0     & -9.3      & -11.1     & -13.2     & -15.2     & -7.3      & -9.4      & -10.1     & -12.8      \\
	\hline
	${\rm O}$                     & -0.2      & -0.7      & -3.1      & -0.7      & -0.4      & -5.3      & -0.6      & -0.4      & -0.4      & -1.8      & -0.5      & -0.4      & -0.3      & -0.7      & -0.5      & -0.4      & -0.3      & -0.8      & -0.4      & -0.4      & -0.4       \\
	${\rm O^{+}}$                 & -0.4      & -0.1      & -0.3      & -0.1      & -0.3      & -1.9      & -0.1      & -0.2      & -0.2      & -0.2      & -0.2      & -0.2      & -0.3      & -0.1      & -0.2      & -0.2      & -0.3      & -0.1      & -0.2      & -0.2      & -0.3       \\
	${\rm O^{2+}}$                & -4.7      & -4.0      & -0.3      & -4.0      & -4.4      & -0.5      & -4.1      & -4.3      & -4.4      & -0.5      & -3.8      & -4.3      & -4.4      & -1.5      & -3.1      & -3.9      & -4.3      & -1.3      & -2.4      & -3.0      & -3.8       \\
	${\rm O^{3+}}$                & -10.2     & -9.1      & -2.0      & -9.1      & -9.6      & -0.2      & -8.9      & -9.2      & -9.2      & -1.5      & -5.5      & -6.5      & -8.3      & -2.4      & -4.1      & -5.4      & -6.2      & -1.9      & -3.2      & -3.8      & -4.9       \\
	${\rm O^{4+}}$                & -15.2     & -13.7     & -6.0      & -13.7     & -14.3     & -1.6      & -12.5     & -13.7     & -13.3     & -3.0      & -7.7      & -8.4      & -11.2     & -3.5      & -5.4      & -7.2      & -8.6      & -2.6      & -4.2      & -4.8      & -6.6       \\
	${\rm O^{5+}}$                & -18.2     & -18.2     & -11.0     & -18.2     & -18.5     & -4.4      & -16.2     & -17.7     & -17.1     & -4.8      & -10.2     & -10.5     & -14.2     & -4.6      & -6.8      & -9.4      & -11.7     & -3.1      & -5.2      & -5.9      & -8.8       \\
	${\rm O^{6+}}$                & -18.9     & -21.4     & -15.1     & -21.4     & -23.4     & -7.5      & -21.2     & -22.1     & -21.4     & -5.8      & -13.2     & -13.2     & -17.9     & -5.5      & -7.8      & -11.4     & -14.1     & -3.0      & -5.7      & -6.5      & -10.9      \\
	${\rm O^{7+}}$                & -20.6     & -26.5     & -18.8     & -26.5     & -26.5     & -13.6     & -26.5     & -26.5     & -26.4     & -11.5     & -18.5     & -18.6     & -21.5     & -11.0     & -13.0     & -16.2     & -18.4     & -8.7      & -10.8     & -11.4     & -15.6      \\
	\hline
	${\rm Ne}$                    & -0.0      & -1.2      & -3.3      & -1.2      & -0.3      & -5.0      & -0.5      & -0.3      & -0.1      & -2.0      & -0.1      & -0.1      & -0.0      & -0.5      & -0.1      & -0.1      & -0.1      & -0.5      & -0.1      & -0.1      & -0.1       \\
	${\rm Ne^{+}}$                & -2.0      & -0.2      & -0.5      & -0.2      & -0.3      & -1.7      & -0.2      & -0.3      & -1.0      & -0.3      & -0.6      & -1.0      & -1.0      & -0.2      & -0.7      & -0.9      & -0.9      & -0.2      & -0.6      & -0.9      & -0.9       \\
	${\rm Ne^{2+}}$               & -4.8      & -0.6      & -0.2      & -0.6      & -2.3      & -0.4      & -2.4      & -2.5      & -3.1      & -0.3      & -2.8      & -3.2      & -3.1      & -1.2      & -2.5      & -3.0      & -2.9      & -1.0      & -2.0      & -2.6      & -2.8       \\
	${\rm Ne^{3+}}$               & -10.3     & -5.7      & -1.8      & -5.7      & -7.6      & -0.3      & -7.5      & -7.6      & -8.2      & -1.7      & -6.2      & -7.1      & -7.7      & -2.6      & -4.4      & -5.9      & -6.7      & -2.0      & -3.5      & -4.0      & -5.2       \\
	${\rm Ne^{4+}}$               & -16.3     & -11.2     & -4.6      & -11.2     & -13.2     & -1.6      & -12.8     & -12.9     & -13.3     & -3.1      & -8.4      & -9.3      & -11.9     & -3.6      & -5.7      & -7.8      & -9.4      & -2.5      & -4.3      & -4.9      & -6.7       \\
	${\rm Ne^{5+}}$               & -19.2     & -15.9     & -9.1      & -15.9     & -17.7     & -4.7      & -17.2     & -17.2     & -17.1     & -4.9      & -11.1     & -11.5     & -14.8     & -4.7      & -7.0      & -9.8      & -12.3     & -3.1      & -5.3      & -6.1      & -9.2       \\
	${\rm Ne^{6+}}$               & -19.9     & -21.3     & -14.3     & -21.3     & -22.9     & -9.4      & -22.5     & -22.2     & -21.6     & -7.4      & -15.1     & -15.6     & -18.3     & -6.2      & -9.1      & -13.1     & -16.0     & -4.0      & -6.7      & -7.9      & -12.9      \\
	${\rm Ne^{7+}}$               & -19.3     & -25.8     & -18.4     & -25.8     & -25.7     & -14.9     & -25.4     & -25.5     & -25.5     & -10.2     & -19.5     & -19.6     & -20.3     & -7.7      & -11.3     & -16.6     & -18.9     & -5.0      & -8.0      & -9.3      & -16.1      \\
	${\rm Ne^{8+}}$               & -20.1     & -25.7     & -17.9     & -25.7     & -25.7     & -17.5     & -25.5     & -25.6     & -25.7     & -11.8     & -23.1     & -23.9     & -20.4     & -9.3      & -13.3     & -17.8     & -19.2     & -5.8      & -9.1      & -10.5     & -18.6      \\
	${\rm Ne^{9+}}$               & -23.4     & -25.9     & -19.1     & -25.9     & -25.9     & -18.3     & -25.9     & -25.9     & -25.9     & -17.8     & -25.9     & -26.0     & -19.8     & -15.2     & -18.9     & -19.4     & -18.7     & -12.0     & -14.5     & -15.5     & -18.3      \\
	\hline
	${\rm Na}$                    & -3.0      & -3.7      & -6.7      & -3.7      & -2.4      & -7.2      & -2.9      & -2.3      & -1.2      & -4.3      & -2.1      & -1.1      & -0.7      & -2.1      & -1.4      & -0.8      & -0.6      & -2.0      & -1.2      & -0.7      & -0.5       \\
	${\rm Na^{+}}$                & -0.0      & -0.2      & -0.8      & -0.2      & -0.0      & -1.6      & -0.0      & -0.0      & -0.0      & -0.4      & -0.0      & -0.0      & -0.1      & -0.0      & -0.0      & -0.1      & -0.1      & -0.1      & -0.0      & -0.1      & -0.2       \\
	${\rm Na^{2+}}$               & -2.6      & -0.4      & -0.1      & -0.4      & -1.8      & -0.0      & -1.7      & -1.9      & -2.1      & -0.2      & -2.2      & -2.2      & -2.3      & -1.1      & -2.1      & -2.3      & -2.3      & -0.8      & -1.7      & -2.1      & -2.3       \\
	\hline
	${\rm Mg}$                    & -1.3      & -1.8      & -6.2      & -1.8      & -1.4      & -8.2      & -1.7      & -1.5      & -1.5      & -3.0      & -1.5      & -1.5      & -1.5      & -1.8      & -1.5      & -1.5      & -1.6      & -1.8      & -1.5      & -1.5      & -1.6       \\
	${\rm Mg^{+}}$                & -0.0      & -0.4      & -3.3      & -0.4      & -0.0      & -4.3      & -0.3      & -0.1      & -0.0      & -1.5      & -0.1      & -0.0      & -0.0      & -0.3      & -0.1      & -0.0      & -0.0      & -0.3      & -0.1      & -0.0      & -0.0       \\
	${\rm Mg^{2+}}$               & -1.7      & -0.2      & -0.0      & -0.2      & -1.3      & -0.2      & -0.4      & -0.8      & -1.9      & -0.0      & -0.7      & -1.4      & -1.9      & -0.3      & -0.8      & -1.5      & -1.8      & -0.3      & -0.8      & -1.3      & -1.7       \\
	${\rm Mg^{3+}}$               & -7.3      & -5.4      & -1.0      & -5.4      & -6.7      & -0.4      & -5.6      & -6.1      & -7.2      & -1.9      & -5.7      & -6.4      & -7.0      & -2.8      & -4.6      & -5.9      & -6.6      & -1.9      & -3.6      & -4.1      & -5.4       \\
	\hline
	${\rm Si}$                    & -1.8      & -2.2      & -4.8      & -2.2      & -1.9      & -8.2      & -2.0      & -1.8      & -1.2      & -3.0      & -1.5      & -1.1      & -1.0      & -1.9      & -1.2      & -1.0      & -0.9      & -1.8      & -1.2      & -0.9      & -0.9       \\
	${\rm Si^{+}}$                & -0.0      & -0.0      & -1.7      & -0.0      & -0.0      & -4.7      & -0.0      & -0.0      & -0.0      & -0.8      & -0.0      & -0.0      & -0.1      & -0.1      & -0.0      & -0.1      & -0.1      & -0.1      & -0.1      & -0.1      & -0.1       \\
	${\rm Si^{2+}}$               & -2.0      & -1.0      & -0.3      & -1.0      & -1.8      & -2.4      & -1.1      & -1.5      & -2.0      & -0.2      & -1.2      & -1.8      & -2.1      & -0.7      & -1.3      & -1.8      & -2.1      & -0.6      & -1.2      & -1.7      & -2.0       \\
	${\rm Si^{3+}}$               & -7.3      & -4.4      & -1.2      & -4.4      & -6.4      & -1.4      & -4.7      & -5.6      & -7.0      & -1.0      & -4.1      & -5.2      & -6.0      & -1.9      & -3.4      & -4.4      & -5.0      & -1.7      & -2.7      & -3.3      & -4.3       \\
	${\rm Si^{4+}}$               & -14.1     & -10.9     & -0.4      & -10.9     & -13.3     & -0.0      & -9.9      & -12.2     & -13.4     & -1.4      & -5.8      & -6.9      & -9.1      & -2.4      & -4.2      & -5.7      & -6.5      & -1.8      & -3.2      & -3.7      & -4.8       \\
	${\rm Si^{5+}}$               & -17.5     & -12.4     & -4.5      & -12.4     & -17.0     & -3.8      & -12.6     & -13.7     & -16.7     & -4.3      & -9.3      & -9.5      & -11.3     & -4.6      & -6.4      & -8.5      & -9.8      & -2.9      & -5.1      & -5.8      & -8.3       \\
	\hline
	${\rm S}$                     & -0.7      & -1.7      & -4.3      & -1.7      & -1.1      & -6.8      & -1.6      & -1.2      & -0.8      & -2.8      & -1.0      & -0.7      & -0.6      & -1.1      & -0.8      & -0.6      & -0.6      & -1.0      & -0.7      & -0.5      & -0.5       \\
	${\rm S^{+}}$                 & -0.1      & -0.1      & -1.2      & -0.1      & -0.0      & -3.3      & -0.0      & -0.0      & -0.1      & -0.7      & -0.1      & -0.1      & -0.1      & -0.2      & -0.1      & -0.1      & -0.1      & -0.2      & -0.1      & -0.2      & -0.2       \\
	${\rm S^{2+}}$                & -2.7      & -0.8      & -0.1      & -0.8      & -2.0      & -1.2      & -1.4      & -1.9      & -2.2      & -0.2      & -1.5      & -2.1      & -2.2      & -0.6      & -1.5      & -2.1      & -2.2      & -0.6      & -1.3      & -1.8      & -2.1       \\
	${\rm S^{3+}}$                & -8.3      & -6.0      & -1.0      & -6.0      & -7.5      & -0.3      & -5.8      & -7.2      & -7.5      & -0.9      & -4.1      & -5.3      & -6.2      & -1.9      & -3.4      & -4.4      & -5.0      & -1.6      & -2.7      & -3.3      & -4.3       \\
	${\rm S^{4+}}$                & -14.5     & -11.7     & -3.5      & -11.7     & -13.5     & -0.3      & -9.0      & -12.4     & -12.9     & -1.8      & -5.2      & -6.3      & -8.2      & -2.5      & -4.1      & -5.3      & -6.1      & -2.0      & -3.3      & -3.8      & -5.1       \\
	\hline
	${\rm Ca}$                    & -1.3      & -2.4      & -7.1      & -2.4      & -1.3      & -8.0      & -2.2      & -1.5      & -1.0      & -3.6      & -1.5      & -1.0      & -0.8      & -1.6      & -1.1      & -0.9      & -0.8      & -1.5      & -1.0      & -0.8      & -0.8       \\
	${\rm Ca^{+}}$                & -0.1      & -0.8      & -3.4      & -0.8      & -0.1      & -3.4      & -0.7      & -0.3      & -0.1      & -1.9      & -0.4      & -0.1      & -0.1      & -0.6      & -0.2      & -0.1      & -0.1      & -0.6      & -0.2      & -0.1      & -0.1       \\
	${\rm Ca^{2+}}$               & -1.0      & -0.1      & -0.1      & -0.1      & -0.6      & -0.6      & -0.1      & -0.3      & -1.1      & -0.1      & -0.3      & -0.8      & -1.3      & -0.2      & -0.5      & -0.9      & -1.3      & -0.2      & -0.5      & -0.9      & -1.2       \\
	${\rm Ca^{3+}}$               & -5.4      & -4.3      & -0.8      & -4.3      & -4.7      & -0.3      & -4.3      & -4.4      & -5.0      & -0.9      & -4.1      & -4.6      & -5.0      & -1.9      & -3.4      & -4.3      & -4.8      & -1.6      & -2.8      & -3.4      & -4.2       \\
	${\rm Ca^{4+}}$               & -11.1     & -9.5      & -2.7      & -9.5      & -10.0     & -0.8      & -9.5      & -9.5      & -9.7      & -2.1      & -6.1      & -7.1      & -9.1      & -2.9      & -4.6      & -6.1      & -7.1      & -2.2      & -3.7      & -4.2      & -5.7       \\
	\hline
	${\rm Fe}$                    & -4.0      & -3.7      & -6.7      & -3.7      & -4.0      & -9.2      & -3.8      & -3.9      & -4.0      & -4.6      & -3.9      & -4.0      & -4.0      & -4.0      & -3.9      & -4.0      & -4.0      & -4.0      & -3.9      & -4.0      & -4.0       \\
	${\rm Fe^{+}}$                & -0.0      & -0.2      & -3.2      & -0.2      & -0.0      & -4.9      & -0.1      & -0.1      & -0.0      & -1.2      & -0.1      & -0.0      & -0.0      & -0.2      & -0.1      & -0.0      & -0.0      & -0.3      & -0.1      & -0.0      & -0.0       \\
	${\rm Fe^{2+}}$               & -1.3      & -0.4      & -0.6      & -0.4      & -1.1      & -2.1      & -0.6      & -0.9      & -1.3      & -0.2      & -0.7      & -1.2      & -1.4      & -0.4      & -0.8      & -1.2      & -1.5      & -0.4      & -0.8      & -1.2      & -1.4       \\
	${\rm Fe^{3+}}$               & -5.9      & -4.7      & -0.2      & -4.7      & -5.7      & -0.6      & -4.9      & -5.4      & -5.9      & -0.5      & -4.0      & -5.0      & -5.5      & -1.5      & -3.1      & -4.1      & -4.6      & -1.3      & -2.4      & -3.0      & -3.8       \\
	${\rm Fe^{4+}}$               & -11.2     & -9.6      & -0.8      & -9.6      & -10.9     & -0.1      & -9.3      & -10.2     & -10.4     & -1.2      & -5.4      & -6.5      & -8.5      & -2.3      & -4.0      & -5.3      & -6.1      & -1.7      & -3.1      & -3.7      & -4.7       \\
\hline
\hline
\end{tabular}
\begin{tablenotes}
\item	{\bf Notes:} Abundances values are given as log$_{10}(F_i/F_t)$. 
\end{tablenotes}
\end{threeparttable}
}
\subsection{Table 5: U$=$10$^{-3}$}
\resizebox{0.95\textwidth}{!}{%
\begin{threeparttable}
	%\caption{Results for U=1.0$\times$10$^{-3}$.}
	\label{tab:uminus3}
	\begin{tabular}{|l||r|rrrr|rrrr|rrrr|rrrr|rrrr|}
	\hline
	\hline
	
	Col. (cm$^{-2}$)              & 1.0E17    & 2.9E17    & 1.0E17    & 2.9E17    & 1.0E18    & 1.4E17    & 7.1E17    & 1.4E18    & 7.1E18    & 1.4E18    & 4.3E18    & 1.4E19    & 4.3E19    & 5.7E18    & 1.4E19    & 5.7E19    & 1.4E20    & 1.0E19    & 2.9E19    & 1.0E20    & 2.9E20     \\
	$\sigma_{1D}$ (km s$^{-1}$)   & 3.5       & 11.5      & 11.5      & 11.5      & 11.5      & 20.2      & 20.2      & 20.2      & 20.2      & 34.6      & 34.6      & 34.6      & 34.6      & 46.2      & 46.2      & 46.2      & 46.2      & 57.7      & 57.7      & 57.7      & 57.7       \\
	T$_{DW}$ (10$^{4}$ K)         & 1.2       & 2.2       & 4.0       & 2.2       & 1.4       & 13.2      & 3.3       & 2.1       & 1.1       & 4.6       & 2.1       & 1.2       & 0.8       & 2.6       & 1.6       & 0.9       & 0.8       & 2.9       & 1.5       & 0.9       & 0.8        \\
	M$_{DW}$                      & 0.4       & 1.0       & 0.7       & 1.0       & 1.1       & 0.7       & 1.4       & 1.7       & 2.3       & 2.1       & 2.9       & 3.5       & 4.8       & 3.6       & 4.3       & 5.8       & 6.5       & 4.1       & 6.0       & 7.2       & 8.2        \\
	\hline
	\hline
	${\rm H}$                     & -2.1      & -2.3      & -3.2      & -2.3      & -2.1      & -4.9      & -2.7      & -2.1      & -1.8      & -2.6      & -1.8      & -1.5      & -1.5      & -1.7      & -1.6      & -1.5      & -1.5      & -1.7      & -1.6      & -1.5      & -1.5       \\
	${\rm H^+}$                   & -0.0      & -0.0      & -0.0      & -0.0      & -0.0      & -0.0      & -0.0      & -0.0      & -0.0      & -0.0      & -0.0      & -0.0      & -0.0      & -0.0      & -0.0      & -0.0      & -0.0      & -0.0      & -0.0      & -0.0      & -0.0       \\
\hline
	${\rm He}$                    & -2.0      & -2.3      & -2.7      & -2.3      & -2.1      & -5.1      & -2.4      & -2.0      & -1.6      & -2.4      & -1.7      & -1.3      & -1.2      & -1.5      & -1.3      & -1.2      & -1.1      & -1.4      & -1.3      & -1.1      & -1.1       \\
	${\rm He^{+}}$                & -0.1      & -0.2      & -0.4      & -0.2      & -0.2      & -1.7      & -0.2      & -0.2      & -0.2      & -0.3      & -0.2      & -0.2      & -0.2      & -0.2      & -0.2      & -0.2      & -0.2      & -0.2      & -0.2      & -0.2      & -0.2       \\
	${\rm He^{2+}}$               & -0.6      & -0.4      & -0.2      & -0.4      & -0.4      & -0.0      & -0.4      & -0.4      & -0.5      & -0.3      & -0.5      & -0.6      & -0.6      & -0.5      & -0.6      & -0.6      & -0.6      & -0.5      & -0.6      & -0.6      & -0.6       \\
\hline
	${\rm C}$                     & -3.9      & -3.2      & -2.9      & -3.2      & -3.6      & -6.0      & -2.9      & -3.0      & -3.1      & -3.0      & -2.7      & -2.6      & -2.5      & -2.6      & -2.6      & -2.5      & -2.5      & -2.5      & -2.5      & -2.4      & -2.4       \\
	${\rm C^{+}}$                 & -0.7      & -0.7      & -0.5      & -0.7      & -0.7      & -2.8      & -0.5      & -0.6      & -0.5      & -0.6      & -0.5      & -0.4      & -0.3      & -0.4      & -0.4      & -0.3      & -0.3      & -0.4      & -0.3      & -0.3      & -0.3       \\
	${\rm C^{2+}}$                & -0.1      & -0.1      & -0.2      & -0.1      & -0.1      & -0.9      & -0.2      & -0.2      & -0.2      & -0.2      & -0.2      & -0.3      & -0.3      & -0.2      & -0.3      & -0.3      & -0.3      & -0.3      & -0.3      & -0.3      & -0.3       \\
	${\rm C^{3+}}$                & -1.4      & -1.4      & -1.5      & -1.4      & -1.4      & -0.5      & -1.5      & -1.5      & -1.4      & -1.1      & -1.5      & -1.5      & -1.5      & -1.4      & -1.5      & -1.5      & -1.5      & -1.3      & -1.5      & -1.5      & -1.5       \\
	${\rm C^{4+}}$                & -3.0      & -2.8      & -2.3      & -2.8      & -2.9      & -0.3      & -2.6      & -2.7      & -2.5      & -1.6      & -2.4      & -2.4      & -2.2      & -1.9      & -2.2      & -2.2      & -2.1      & -1.4      & -2.1      & -2.1      & -2.0       \\
	${\rm C^{5+}}$                & -5.5      & -5.0      & -4.5      & -5.0      & -5.1      & -2.6      & -4.8      & -4.9      & -4.4      & -3.7      & -4.4      & -4.2      & -3.8      & -4.0      & -4.1      & -3.7      & -3.5      & -3.4      & -3.8      & -3.6      & -3.4       \\
\hline
	${\rm N}$                     & -3.2      & -3.2      & -3.1      & -3.2      & -3.1      & -5.5      & -3.0      & -2.7      & -2.3      & -2.9      & -2.2      & -1.8      & -1.7      & -2.1      & -1.9      & -1.7      & -1.6      & -2.0      & -1.8      & -1.6      & -1.6       \\
	${\rm N^{+}}$                 & -0.7      & -0.7      & -0.6      & -0.7      & -0.6      & -2.3      & -0.6      & -0.5      & -0.4      & -0.6      & -0.4      & -0.3      & -0.3      & -0.3      & -0.3      & -0.3      & -0.3      & -0.3      & -0.3      & -0.3      & -0.3       \\
	${\rm N^{2+}}$                & -0.1      & -0.1      & -0.2      & -0.1      & -0.1      & -0.6      & -0.2      & -0.2      & -0.3      & -0.2      & -0.3      & -0.4      & -0.4      & -0.3      & -0.4      & -0.4      & -0.4      & -0.4      & -0.4      & -0.4      & -0.4       \\
	${\rm N^{3+}}$                & -1.2      & -1.2      & -1.0      & -1.2      & -1.2      & -0.2      & -1.1      & -1.2      & -1.2      & -0.9      & -1.2      & -1.2      & -1.2      & -1.2      & -1.2      & -1.2      & -1.3      & -1.1      & -1.3      & -1.3      & -1.3       \\
	${\rm N^{4+}}$                & -2.9      & -2.8      & -2.7      & -2.8      & -2.8      & -1.1      & -2.7      & -2.8      & -2.5      & -2.3      & -2.6      & -2.4      & -2.2      & -2.3      & -2.4      & -2.2      & -2.2      & -2.0      & -2.3      & -2.2      & -2.2       \\
	${\rm N^{5+}}$                & -4.7      & -4.3      & -4.4      & -4.3      & -4.3      & -2.4      & -4.2      & -4.2      & -3.7      & -3.0      & -3.6      & -3.4      & -3.1      & -2.9      & -3.3      & -3.0      & -2.8      & -1.9      & -2.9      & -2.9      & -2.8       \\
	${\rm N^{6+}}$                & -7.6      & -7.0      & -7.2      & -7.0      & -7.0      & -5.1      & -6.8      & -6.7      & -5.9      & -5.4      & -5.9      & -5.5      & -4.9      & -5.2      & -5.4      & -4.8      & -4.4      & -4.2      & -4.9      & -4.6      & -4.3       \\
\hline
	${\rm O}$                     & -2.7      & -3.0      & -3.7      & -3.0      & -2.6      & -5.6      & -3.3      & -2.4      & -1.9      & -2.9      & -1.9      & -1.6      & -1.6      & -1.8      & -1.6      & -1.6      & -1.5      & -1.7      & -1.6      & -1.5      & -1.5       \\
	${\rm O^{+}}$                 & -0.6      & -0.7      & -0.8      & -0.7      & -0.5      & -2.2      & -0.7      & -0.5      & -0.3      & -0.7      & -0.3      & -0.3      & -0.3      & -0.3      & -0.3      & -0.2      & -0.2      & -0.3      & -0.2      & -0.2      & -0.2       \\
	${\rm O^{2+}}$                & -0.2      & -0.2      & -0.2      & -0.2      & -0.2      & -0.6      & -0.2      & -0.3      & -0.4      & -0.2      & -0.4      & -0.5      & -0.5      & -0.4      & -0.5      & -0.5      & -0.6      & -0.4      & -0.5      & -0.5      & -0.6       \\
	${\rm O^{3+}}$                & -1.1      & -0.9      & -0.7      & -0.9      & -1.1      & -0.2      & -0.8      & -1.0      & -1.1      & -0.8      & -1.1      & -1.1      & -1.1      & -1.0      & -1.1      & -1.1      & -1.1      & -1.0      & -1.1      & -1.1      & -1.1       \\
	${\rm O^{4+}}$                & -2.5      & -2.3      & -2.0      & -2.3      & -2.4      & -1.2      & -2.1      & -2.3      & -2.2      & -2.0      & -2.2      & -2.1      & -2.0      & -2.1      & -2.1      & -2.0      & -2.0      & -1.8      & -2.0      & -2.0      & -2.0       \\
	${\rm O^{5+}}$                & -4.3      & -4.0      & -4.0      & -4.0      & -4.0      & -3.1      & -4.0      & -4.0      & -3.5      & -3.7      & -3.8      & -3.3      & -3.0      & -3.4      & -3.4      & -3.0      & -2.8      & -2.6      & -3.1      & -2.9      & -2.8       \\
	${\rm O^{6+}}$                & -6.5      & -6.0      & -6.1      & -6.0      & -6.0      & -5.2      & -5.9      & -5.8      & -5.0      & -4.7      & -5.1      & -4.7      & -4.1      & -4.2      & -4.5      & -4.1      & -3.7      & -2.7      & -4.0      & -3.9      & -3.7       \\
	${\rm O^{7+}}$                & -9.8      & -9.0      & -9.3      & -9.0      & -9.0      & -8.4      & -8.8      & -8.6      & -7.5      & -7.4      & -7.6      & -7.0      & -6.0      & -6.7      & -6.9      & -6.0      & -5.4      & -5.3      & -6.1      & -5.8      & -5.3       \\
\hline
	${\rm Ne}$                    & -3.9      & -4.3      & -4.4      & -4.3      & -3.9      & -5.4      & -4.3      & -3.6      & -2.8      & -3.8      & -2.8      & -2.1      & -2.0      & -2.6      & -2.2      & -2.0      & -1.9      & -2.4      & -2.1      & -1.9      & -1.8       \\
	${\rm Ne^{+}}$                & -1.2      & -1.4      & -1.4      & -1.4      & -1.3      & -2.0      & -1.4      & -1.2      & -0.9      & -1.1      & -0.9      & -0.6      & -0.6      & -0.8      & -0.7      & -0.6      & -0.5      & -0.7      & -0.6      & -0.5      & -0.5       \\
	${\rm Ne^{2+}}$               & -0.1      & -0.1      & -0.2      & -0.1      & -0.1      & -0.5      & -0.2      & -0.1      & -0.1      & -0.2      & -0.1      & -0.2      & -0.2      & -0.2      & -0.2      & -0.2      & -0.2      & -0.2      & -0.2      & -0.2      & -0.2       \\
	${\rm Ne^{3+}}$               & -0.9      & -0.8      & -0.5      & -0.8      & -0.9      & -0.2      & -0.6      & -0.9      & -1.0      & -0.6      & -0.9      & -1.0      & -1.0      & -0.9      & -1.0      & -1.0      & -1.1      & -0.9      & -1.0      & -1.1      & -1.1       \\
	${\rm Ne^{4+}}$               & -2.4      & -2.1      & -1.5      & -2.1      & -2.3      & -1.0      & -1.7      & -2.0      & -1.9      & -1.6      & -1.8      & -1.8      & -1.8      & -1.7      & -1.7      & -1.7      & -1.7      & -1.6      & -1.7      & -1.7      & -1.7       \\
	${\rm Ne^{5+}}$               & -4.1      & -3.7      & -3.1      & -3.7      & -4.0      & -2.7      & -3.2      & -3.4      & -3.2      & -3.0      & -3.1      & -2.9      & -2.8      & -2.9      & -2.8      & -2.7      & -2.6      & -2.5      & -2.7      & -2.7      & -2.6       \\
	${\rm Ne^{6+}}$               & -6.8      & -6.1      & -5.4      & -6.1      & -6.4      & -5.0      & -5.4      & -5.6      & -5.1      & -5.1      & -5.0      & -4.5      & -4.2      & -4.6      & -4.4      & -4.1      & -3.9      & -3.5      & -4.1      & -4.0      & -3.9       \\
	${\rm Ne^{7+}}$               & -9.3      & -8.7      & -8.0      & -8.7      & -8.9      & -7.7      & -8.0      & -8.0      & -7.0      & -7.4      & -7.1      & -6.3      & -5.5      & -6.5      & -6.2      & -5.6      & -5.1      & -4.6      & -5.6      & -5.3      & -5.0       \\
	${\rm Ne^{8+}}$               & -12.3     & -11.4     & -10.9     & -11.4     & -11.5     & -10.6     & -10.6     & -10.4     & -8.9      & -9.6      & -9.0      & -8.2      & -6.9      & -7.9      & -8.1      & -7.0      & -6.2      & -5.6      & -7.1      & -6.6      & -6.0       \\
	${\rm Ne^{9+}}$               & -16.0     & -14.9     & -14.5     & -14.9     & -15.0     & -14.3     & -14.0     & -13.7     & -11.6     & -12.7     & -11.9     & -10.8     & -9.1      & -10.5     & -10.7     & -9.2      & -7.9      & -8.5      & -9.5      & -8.7      & -7.6       \\
\hline
	${\rm Na}$                    & -5.1      & -6.3      & -7.2      & -6.3      & -5.1      & -7.3      & -6.5      & -4.9      & -3.5      & -5.3      & -3.8      & -2.7      & -2.4      & -3.6      & -2.9      & -2.4      & -2.3      & -3.2      & -2.7      & -2.3      & -2.2       \\
	${\rm Na^{+}}$                & -1.2      & -1.4      & -1.4      & -1.4      & -1.2      & -1.8      & -1.4      & -1.2      & -0.9      & -1.2      & -0.9      & -0.6      & -0.6      & -0.8      & -0.7      & -0.6      & -0.5      & -0.8      & -0.6      & -0.5      & -0.5       \\
	${\rm Na^{2+}}$               & -0.0      & -0.0      & -0.0      & -0.0      & -0.0      & -0.0      & -0.0      & -0.0      & -0.1      & -0.0      & -0.1      & -0.1      & -0.1      & -0.1      & -0.1      & -0.1      & -0.2      & -0.1      & -0.1      & -0.2      & -0.2       \\
\hline
	${\rm Mg}$                    & -2.2      & -3.5      & -5.8      & -3.5      & -2.6      & -8.6      & -4.4      & -3.0      & -2.4      & -3.9      & -2.6      & -2.2      & -2.3      & -2.5      & -2.2      & -2.2      & -2.3      & -2.4      & -2.2      & -2.2      & -2.2       \\
	${\rm Mg^{+}}$                & -0.4      & -1.5      & -3.1      & -1.5      & -0.9      & -4.6      & -2.3      & -1.2      & -0.6      & -2.1      & -0.8      & -0.4      & -0.3      & -0.8      & -0.5      & -0.4      & -0.3      & -0.7      & -0.5      & -0.3      & -0.3       \\
	${\rm Mg^{2+}}$               & -0.3      & -0.1      & -0.2      & -0.1      & -0.1      & -0.6      & -0.2      & -0.1      & -0.2      & -0.2      & -0.2      & -0.3      & -0.4      & -0.2      & -0.3      & -0.4      & -0.4      & -0.2      & -0.3      & -0.4      & -0.4       \\
	${\rm Mg^{3+}}$               & -1.1      & -0.8      & -0.4      & -0.8      & -1.0      & -0.1      & -0.5      & -0.8      & -0.9      & -0.5      & -0.8      & -0.9      & -0.9      & -0.8      & -0.9      & -0.9      & -0.9      & -0.7      & -0.9      & -0.9      & -0.9       \\
\hline
	${\rm Si}$                    & -4.9      & -4.7      & -4.7      & -4.7      & -4.7      & -8.8      & -4.7      & -4.5      & -4.1      & -4.7      & -4.1      & -3.6      & -3.5      & -3.9      & -3.7      & -3.5      & -3.4      & -3.8      & -3.6      & -3.4      & -3.4       \\
	${\rm Si^{+}}$                & -0.6      & -1.0      & -1.4      & -1.0      & -0.7      & -5.1      & -1.2      & -0.8      & -0.3      & -1.2      & -0.5      & -0.3      & -0.2      & -0.5      & -0.4      & -0.2      & -0.2      & -0.5      & -0.3      & -0.2      & -0.2       \\
	${\rm Si^{2+}}$               & -0.1      & -0.1      & -0.0      & -0.1      & -0.1      & -2.7      & -0.1      & -0.1      & -0.3      & -0.2      & -0.2      & -0.4      & -0.5      & -0.2      & -0.3      & -0.5      & -0.5      & -0.3      & -0.3      & -0.5      & -0.5       \\
	${\rm Si^{3+}}$               & -1.2      & -1.4      & -1.4      & -1.4      & -1.3      & -1.7      & -1.3      & -1.4      & -1.4      & -0.8      & -1.3      & -1.5      & -1.5      & -1.2      & -1.4      & -1.5      & -1.5      & -1.2      & -1.4      & -1.5      & -1.4       \\
	${\rm Si^{4+}}$               & -3.3      & -3.3      & -2.7      & -3.3      & -3.3      & -0.1      & -2.6      & -3.0      & -2.9      & -0.9      & -2.1      & -2.5      & -2.5      & -1.5      & -2.1      & -2.3      & -2.3      & -1.2      & -2.0      & -2.3      & -2.3       \\
	${\rm Si^{5+}}$               & -4.5      & -4.1      & -3.3      & -4.1      & -4.1      & -1.0      & -3.3      & -3.8      & -3.4      & -1.7      & -2.6      & -2.9      & -2.7      & -2.2      & -2.6      & -2.5      & -2.4      & -1.8      & -2.4      & -2.5      & -2.3       \\
\hline
	${\rm S}$                     & -5.2      & -4.8      & -4.4      & -4.8      & -5.0      & -7.2      & -4.5      & -4.5      & -4.2      & -4.5      & -4.0      & -3.5      & -3.4      & -3.8      & -3.6      & -3.4      & -3.3      & -3.7      & -3.5      & -3.3      & -3.2       \\
	${\rm S^{+}}$                 & -1.2      & -1.0      & -1.0      & -1.0      & -1.1      & -3.6      & -1.0      & -0.9      & -0.8      & -1.1      & -0.8      & -0.6      & -0.6      & -0.7      & -0.6      & -0.5      & -0.5      & -0.7      & -0.6      & -0.5      & -0.5       \\
	${\rm S^{2+}}$                & -0.1      & -0.1      & -0.1      & -0.1      & -0.1      & -1.3      & -0.1      & -0.1      & -0.1      & -0.2      & -0.1      & -0.2      & -0.2      & -0.2      & -0.2      & -0.2      & -0.2      & -0.2      & -0.2      & -0.2      & -0.2       \\
	${\rm S^{3+}}$                & -1.0      & -0.9      & -1.0      & -0.9      & -1.0      & -0.4      & -0.9      & -1.0      & -1.0      & -0.6      & -1.1      & -1.1      & -1.1      & -1.0      & -1.1      & -1.1      & -1.1      & -1.0      & -1.1      & -1.1      & -1.1       \\
	${\rm S^{4+}}$                & -2.5      & -2.3      & -2.6      & -2.3      & -2.4      & -0.3      & -2.4      & -2.4      & -2.1      & -1.4      & -2.1      & -2.0      & -1.7      & -1.8      & -2.0      & -1.8      & -1.6      & -1.5      & -1.9      & -1.7      & -1.6       \\
\hline
	${\rm Ca}$                    & -4.6      & -5.5      & -7.2      & -5.5      & -4.8      & -8.2      & -6.1      & -4.7      & -3.9      & -5.4      & -4.0      & -3.4      & -3.5      & -3.7      & -3.5      & -3.4      & -3.5      & -3.6      & -3.4      & -3.4      & -3.5       \\
	${\rm Ca^{+}}$                & -1.6      & -2.5      & -3.6      & -2.5      & -1.8      & -3.6      & -3.0      & -1.9      & -1.3      & -2.6      & -1.4      & -0.9      & -0.8      & -1.3      & -1.0      & -0.8      & -0.7      & -1.2      & -0.9      & -0.8      & -0.7       \\
	${\rm Ca^{2+}}$               & -0.2      & -0.3      & -0.3      & -0.3      & -0.2      & -0.7      & -0.3      & -0.2      & -0.2      & -0.2      & -0.2      & -0.2      & -0.2      & -0.2      & -0.2      & -0.2      & -0.3      & -0.2      & -0.2      & -0.2      & -0.3       \\
	${\rm Ca^{3+}}$               & -0.4      & -0.4      & -0.4      & -0.4      & -0.4      & -0.3      & -0.4      & -0.5      & -0.5      & -0.4      & -0.6      & -0.6      & -0.6      & -0.6      & -0.6      & -0.7      & -0.7      & -0.6      & -0.7      & -0.7      & -0.7       \\
	${\rm Ca^{4+}}$               & -1.6      & -1.4      & -1.1      & -1.4      & -1.5      & -0.6      & -1.2      & -1.4      & -1.3      & -1.3      & -1.5      & -1.3      & -1.2      & -1.4      & -1.4      & -1.3      & -1.2      & -1.3      & -1.4      & -1.3      & -1.2       \\
\hline
	${\rm Fe}$                    & -5.5      & -5.7      & -7.1      & -5.7      & -5.3      & -9.5      & -6.2      & -5.2      & -4.9      & -5.8      & -4.8      & -4.7      & -4.7      & -4.8      & -4.7      & -4.6      & -4.6      & -4.7      & -4.7      & -4.6      & -4.6       \\
	${\rm Fe^{+}}$                & -1.8      & -2.4      & -3.7      & -2.4      & -1.8      & -5.2      & -2.9      & -1.7      & -0.9      & -2.2      & -1.1      & -0.6      & -0.5      & -1.0      & -0.8      & -0.5      & -0.4      & -0.9      & -0.7      & -0.5      & -0.4       \\
	${\rm Fe^{2+}}$               & -0.4      & -0.5      & -1.1      & -0.5      & -0.4      & -2.3      & -0.7      & -0.4      & -0.4      & -0.7      & -0.3      & -0.4      & -0.5      & -0.4      & -0.4      & -0.5      & -0.5      & -0.4      & -0.4      & -0.5      & -0.5       \\
	${\rm Fe^{3+}}$               & -0.3      & -0.3      & -0.3      & -0.3      & -0.3      & -0.7      & -0.3      & -0.4      & -0.5      & -0.3      & -0.5      & -0.6      & -0.6      & -0.5      & -0.6      & -0.7      & -0.7      & -0.5      & -0.6      & -0.7      & -0.7       \\
	${\rm Fe^{4+}}$               & -1.0      & -0.9      & -0.4      & -0.9      & -1.0      & -0.1      & -0.5      & -0.8      & -0.9      & -0.5      & -0.8      & -0.9      & -0.9      & -0.8      & -0.9      & -0.9      & -0.9      & -0.7      & -0.9      & -0.9      & -0.9       \\
\hline
\hline
\end{tabular}
\begin{tablenotes}
\item	{\bf Notes:} Abundances values are given as log$_{10}(F_i/F_t)$. 
\end{tablenotes}
\end{threeparttable}
}
\subsection{Table 6: U$=$10$^{-1}$}
\resizebox{0.95\textwidth}{!}{%
\begin{threeparttable}
	%\caption{Results for U=1.0$\times$10$^{-1}$.}
	\label{tab:uminus1}
	\begin{tabular}{|l||r|rrrr|rrrr|rrrr|rrrr|rrrr|}
	\hline
	Col. (cm$^{-2}$)              & 1.0E17    & 2.9E17    & 1.0E17    & 2.9E17    & 1.0E18    & 1.4E17    & 7.1E17    & 1.4E18    & 7.1E18    & 1.4E18    & 4.3E18    & 1.4E19    & 4.3E19    & 5.7E18    & 1.4E19    & 5.7E19    & 1.4E20    & 1.0E19    & 2.9E19    & 1.0E20    & 2.9E20     \\
	$\sigma_{1D}$ (km s$^{-1}$)   & 3.5       & 11.5      & 11.5      & 11.5      & 11.5      & 20.2      & 20.2      & 20.2      & 20.2      & 34.6      & 34.6      & 34.6      & 34.6      & 46.2      & 46.2      & 46.2      & 46.2      & 57.7      & 57.7      & 57.7      & 57.7       \\
	T$_{DW}$ (10$^{4}$ K)         & 2.4       & 4.5       & 7.0       & 4.5       & 3.3       & 26.5      & 7.5       & 5.5       & 2.9       & 79.6      & 6.1       & 3.1       & 2.4       & 9.4       & 4.5       & 2.5       & 2.2       & 12.0      & 4.0       & 2.5       & 2.1        \\
	M$_{DW}$                      & 0.3       & 0.7       & 0.5       & 0.7       & 0.8       & 0.5       & 0.9       & 1.1       & 1.5       & 0.5       & 1.7       & 2.3       & 2.7       & 1.8       & 2.8       & 3.3       & 3.8       & 2.0       & 3.6       & 4.2       & 4.5        \\
	\hline
	\hline
	${\rm H}$                     & -4.3      & -4.5      & -4.8      & -4.5      & -4.4      & -5.7      & -4.8      & -4.6      & -4.2      & -6.4      & -4.4      & -3.8      & -3.7      & -4.2      & -3.8      & -3.6      & -3.5      & -4.0      & -3.6      & -3.4      & -3.4       \\
	${\rm H^+}$                   & -0.0      & -0.0      & -0.0      & -0.0      & -0.0      & 0.0       & -0.0      & -0.0      & -0.0      & 0.0       & -0.0      & -0.0      & -0.0      & -0.0      & -0.0      & -0.0      & -0.0      & -0.0      & -0.0      & -0.0      & -0.0       \\
	\hline
	${\rm He}$                    & -6.1      & -6.5      & -6.6      & -6.5      & -6.3      & -7.2      & -6.5      & -6.4      & -5.8      & -8.9      & -5.8      & -4.6      & -4.5      & -5.2      & -4.6      & -4.2      & -4.1      & -4.5      & -4.3      & -4.0      & -3.9       \\
	${\rm He^{+}}$                & -2.0      & -2.2      & -2.4      & -2.2      & -2.1      & -3.3      & -2.4      & -2.2      & -2.0      & -4.3      & -2.0      & -1.6      & -1.5      & -1.9      & -1.6      & -1.4      & -1.4      & -1.8      & -1.4      & -1.3      & -1.3       \\
	${\rm He^{2+}}$               & -0.0      & -0.0      & -0.0      & -0.0      & -0.0      & -0.0      & -0.0      & -0.0      & -0.0      & -0.0      & -0.0      & -0.0      & -0.0      & -0.0      & -0.0      & -0.0      & -0.0      & -0.0      & -0.0      & -0.0      & -0.0       \\
	\hline
	${\rm C}$                     & -9.3      & -8.2      & -8.2      & -8.2      & -8.5      & -11.5     & -8.1      & -7.8      & -7.6      & -15.7     & -6.7      & -6.3      & -6.3      & -6.1      & -5.8      & -6.0      & -6.0      & -5.6      & -5.7      & -5.8      & -5.9       \\
	${\rm C^{+}}$                 & -4.8      & -4.3      & -4.4      & -4.3      & -4.4      & -7.3      & -4.4      & -4.1      & -3.8      & -10.8     & -3.3      & -2.7      & -2.7      & -2.9      & -2.6      & -2.5      & -2.4      & -2.6      & -2.4      & -2.3      & -2.2       \\
	${\rm C^{2+}}$                & -2.3      & -2.3      & -2.4      & -2.3      & -2.2      & -4.5      & -2.5      & -2.3      & -1.8      & -7.0      & -1.7      & -1.2      & -1.1      & -1.4      & -1.1      & -1.0      & -0.9      & -1.3      & -1.0      & -0.9      & -0.8       \\
	${\rm C^{3+}}$                & -1.4      & -1.6      & -1.7      & -1.6      & -1.5      & -2.7      & -1.8      & -1.7      & -1.3      & -4.0      & -1.4      & -1.0      & -0.9      & -1.3      & -1.0      & -0.9      & -0.8      & -1.2      & -0.9      & -0.8      & -0.8       \\
	${\rm C^{4+}}$                & -0.2      & -0.3      & -0.3      & -0.3      & -0.3      & -0.4      & -0.4      & -0.3      & -0.3      & -0.9      & -0.3      & -0.3      & -0.3      & -0.3      & -0.3      & -0.3      & -0.3      & -0.4      & -0.3      & -0.3      & -0.3       \\
	${\rm C^{5+}}$                & -0.4      & -0.3      & -0.3      & -0.3      & -0.4      & -0.2      & -0.3      & -0.3      & -0.4      & -0.1      & -0.3      & -0.5      & -0.5      & -0.3      & -0.5      & -0.6      & -0.6      & -0.3      & -0.5      & -0.6      & -0.6       \\
	\hline
	${\rm N}$                     & -10.1     & -8.7      & -8.2      & -8.7      & -9.1      & -11.0     & -8.1      & -8.0      & -8.1      & -17.0     & -6.9      & -6.3      & -6.2      & -6.3      & -6.0      & -5.7      & -5.5      & -5.3      & -5.7      & -5.3      & -5.2       \\
	${\rm N^{+}}$                 & -5.6      & -4.9      & -4.4      & -4.9      & -5.1      & -6.8      & -4.4      & -4.4      & -4.4      & -12.1     & -3.6      & -3.1      & -3.0      & -3.1      & -2.9      & -2.8      & -2.7      & -2.8      & -2.7      & -2.5      & -2.5       \\
	${\rm N^{2+}}$                & -2.7      & -2.5      & -2.3      & -2.5      & -2.6      & -4.2      & -2.4      & -2.3      & -2.1      & -8.5      & -1.8      & -1.3      & -1.2      & -1.5      & -1.2      & -1.1      & -1.0      & -1.3      & -1.1      & -0.9      & -0.9       \\
	${\rm N^{3+}}$                & -1.3      & -1.3      & -1.2      & -1.3      & -1.3      & -2.4      & -1.4      & -1.3      & -1.1      & -5.6      & -1.0      & -0.8      & -0.7      & -0.9      & -0.7      & -0.7      & -0.7      & -0.9      & -0.7      & -0.6      & -0.6       \\
	${\rm N^{4+}}$                & -0.8      & -1.0      & -1.0      & -1.0      & -0.9      & -1.5      & -1.1      & -1.0      & -0.8      & -3.2      & -0.9      & -0.7      & -0.7      & -1.0      & -0.8      & -0.7      & -0.7      & -1.0      & -0.8      & -0.7      & -0.7       \\
	${\rm N^{5+}}$                & -0.2      & -0.2      & -0.2      & -0.2      & -0.2      & -0.2      & -0.2      & -0.2      & -0.2      & -0.5      & -0.3      & -0.4      & -0.4      & -0.3      & -0.4      & -0.4      & -0.4      & -0.4      & -0.4      & -0.5      & -0.5       \\
	${\rm N^{6+}}$                & -0.8      & -0.7      & -0.7      & -0.7      & -0.7      & -0.5      & -0.5      & -0.6      & -0.8      & -0.2      & -0.6      & -0.8      & -0.9      & -0.6      & -0.8      & -0.9      & -0.9      & -0.6      & -0.8      & -0.9      & -0.9       \\
	\hline
	${\rm O}$                     & -10.9     & -9.5      & -8.7      & -9.5      & -10.1     & -10.3     & -8.4      & -8.5      & -8.8      & -17.9     & -7.4      & -6.2      & -5.9      & -6.7      & -6.0      & -5.4      & -5.1      & -4.9      & -5.5      & -4.9      & -5.0       \\
	${\rm O^{+}}$                 & -6.1      & -5.5      & -4.9      & -5.5      & -5.8      & -6.1      & -4.7      & -4.8      & -4.9      & -12.9     & -4.0      & -3.2      & -3.1      & -3.4      & -3.1      & -2.8      & -2.7      & -2.9      & -2.8      & -2.5      & -2.5       \\
	${\rm O^{2+}}$                & -3.1      & -3.0      & -2.5      & -3.0      & -3.1      & -3.4      & -2.5      & -2.6      & -2.5      & -9.4      & -2.0      & -1.6      & -1.5      & -1.7      & -1.5      & -1.3      & -1.3      & -1.5      & -1.3      & -1.2      & -1.1       \\
	${\rm O^{3+}}$                & -1.4      & -1.4      & -1.2      & -1.4      & -1.4      & -1.7      & -1.2      & -1.2      & -1.1      & -6.6      & -1.0      & -0.8      & -0.7      & -0.8      & -0.7      & -0.7      & -0.6      & -0.8      & -0.7      & -0.6      & -0.6       \\
	${\rm O^{4+}}$                & -0.7      & -0.7      & -0.5      & -0.7      & -0.7      & -0.8      & -0.7      & -0.7      & -0.6      & -4.4      & -0.6      & -0.6      & -0.6      & -0.6      & -0.6      & -0.6      & -0.6      & -0.7      & -0.6      & -0.6      & -0.6       \\
	${\rm O^{5+}}$                & -0.5      & -0.6      & -0.6      & -0.6      & -0.6      & -0.6      & -0.8      & -0.7      & -0.6      & -2.7      & -0.7      & -0.7      & -0.6      & -0.8      & -0.7      & -0.7      & -0.7      & -0.9      & -0.7      & -0.7      & -0.7       \\
	${\rm O^{6+}}$                & -0.4      & -0.4      & -0.4      & -0.4      & -0.4      & -0.3      & -0.3      & -0.4      & -0.4      & -0.4      & -0.4      & -0.6      & -0.6      & -0.5      & -0.6      & -0.6      & -0.6      & -0.5      & -0.6      & -0.7      & -0.7       \\
	${\rm O^{7+}}$                & -1.4      & -1.2      & -1.3      & -1.2      & -1.2      & -1.1      & -1.0      & -1.1      & -1.3      & -0.2      & -1.1      & -1.2      & -1.3      & -1.0      & -1.2      & -1.3      & -1.3      & -0.9      & -1.2      & -1.2      & -1.3       \\
	\hline
	${\rm Ne}$                    & -12.1     & -11.9     & -10.9     & -11.9     & -12.2     & -10.5     & -10.0     & -10.3     & -10.4     & -17.2     & -8.8      & -7.2      & -6.9      & -7.8      & -7.1      & -6.1      & -6.1      & -6.1      & -6.5      & -5.9      & -5.6       \\
	${\rm Ne^{+}}$                & -7.2      & -7.3      & -6.7      & -7.3      & -7.3      & -6.5      & -6.1      & -6.4      & -6.1      & -12.5     & -5.1      & -3.9      & -3.7      & -4.3      & -3.8      & -3.4      & -3.3      & -3.6      & -3.5      & -3.1      & -3.0       \\
	${\rm Ne^{2+}}$               & -3.8      & -4.1      & -3.9      & -4.1      & -4.0      & -3.8      & -3.5      & -3.7      & -3.2      & -8.9      & -2.8      & -1.8      & -1.7      & -2.2      & -1.8      & -1.5      & -1.4      & -1.9      & -1.6      & -1.3      & -1.3       \\
	${\rm Ne^{3+}}$               & -1.9      & -2.1      & -2.1      & -2.1      & -2.0      & -1.9      & -1.9      & -2.0      & -1.6      & -6.0      & -1.5      & -1.1      & -1.0      & -1.2      & -1.0      & -0.9      & -0.9      & -1.1      & -0.9      & -0.8      & -0.8       \\
	${\rm Ne^{4+}}$               & -0.7      & -0.8      & -0.8      & -0.8      & -0.8      & -0.8      & -0.8      & -0.8      & -0.6      & -3.9      & -0.6      & -0.5      & -0.5      & -0.6      & -0.5      & -0.5      & -0.5      & -0.6      & -0.5      & -0.5      & -0.5       \\
	${\rm Ne^{5+}}$               & -0.3      & -0.3      & -0.3      & -0.3      & -0.3      & -0.4      & -0.3      & -0.3      & -0.3      & -2.4      & -0.4      & -0.4      & -0.4      & -0.4      & -0.4      & -0.5      & -0.5      & -0.5      & -0.5      & -0.5      & -0.6       \\
	${\rm Ne^{6+}}$               & -0.7      & -0.6      & -0.5      & -0.6      & -0.6      & -0.5      & -0.6      & -0.6      & -0.7      & -1.5      & -0.7      & -0.9      & -0.9      & -0.7      & -0.9      & -0.9      & -1.0      & -0.8      & -0.9      & -1.0      & -1.0       \\
	${\rm Ne^{7+}}$               & -1.3      & -1.2      & -1.2      & -1.2      & -1.2      & -1.1      & -1.2      & -1.2      & -1.2      & -1.0      & -1.2      & -1.3      & -1.3      & -1.2      & -1.3      & -1.3      & -1.3      & -1.2      & -1.3      & -1.3      & -1.3       \\
	${\rm Ne^{8+}}$               & -2.0      & -1.7      & -1.8      & -1.7      & -1.8      & -1.7      & -1.6      & -1.6      & -1.7      & -0.2      & -1.5      & -1.5      & -1.6      & -1.4      & -1.5      & -1.5      & -1.5      & -1.2      & -1.4      & -1.5      & -1.5       \\
	${\rm Ne^{9+}}$               & -3.5      & -3.1      & -3.3      & -3.1      & -3.2      & -3.0      & -2.8      & -2.9      & -3.0      & -0.7      & -2.6      & -2.4      & -2.4      & -2.4      & -2.2      & -2.3      & -2.2      & -2.0      & -2.1      & -2.1      & -2.1       \\
	\hline
	${\rm Na}$                    & -8.9      & -9.6      & -9.1      & -9.6      & -9.3      & -8.5      & -8.9      & -9.2      & -8.8      & -9.6      & -8.7      & -7.6      & -7.4      & -8.2      & -7.6      & -6.4      & -6.5      & -7.1      & -7.0      & -6.3      & -5.9       \\
	${\rm Na^{+}}$                & -3.4      & -3.6      & -3.3      & -3.6      & -3.5      & -2.9      & -3.2      & -3.4      & -3.3      & -3.6      & -3.1      & -2.9      & -2.8      & -2.9      & -2.8      & -2.7      & -2.6      & -2.7      & -2.7      & -2.5      & -2.5       \\
	${\rm Na^{2+}}$               & -0.0      & -0.0      & -0.0      & -0.0      & -0.0      & -0.0      & -0.0      & -0.0      & -0.0      & -0.0      & -0.0      & -0.0      & -0.0      & -0.0      & -0.0      & -0.0      & -0.0      & -0.0      & -0.0      & -0.0      & -0.0       \\
	\hline
	${\rm Mg}$                    & -8.6      & -9.4      & -10.3     & -9.4      & -8.8      & -10.7     & -10.2     & -9.6      & -7.9      & -12.5     & -7.6      & -5.5      & -5.3      & -6.4      & -5.5      & -5.0      & -4.9      & -5.1      & -5.1      & -4.7      & -4.8       \\
	${\rm Mg^{+}}$                & -5.4      & -6.1      & -6.6      & -6.1      & -5.7      & -6.2      & -6.6      & -6.2      & -5.1      & -7.2      & -5.1      & -3.4      & -3.2      & -4.1      & -3.4      & -2.9      & -2.8      & -3.3      & -3.1      & -2.6      & -2.6       \\
	${\rm Mg^{2+}}$               & -2.5      & -2.7      & -2.7      & -2.7      & -2.6      & -2.2      & -2.6      & -2.6      & -2.3      & -2.9      & -2.3      & -1.7      & -1.6      & -2.0      & -1.7      & -1.4      & -1.4      & -1.8      & -1.5      & -1.3      & -1.3       \\
	${\rm Mg^{3+}}$               & -0.0      & -0.0      & -0.0      & -0.0      & -0.0      & -0.0      & -0.0      & -0.0      & -0.0      & -0.0      & -0.0      & -0.0      & -0.0      & -0.0      & -0.0      & -0.0      & -0.0      & -0.0      & -0.0      & -0.0      & -0.0       \\
	\hline
	${\rm Si}$                    & -10.4     & -9.6      & -9.5      & -9.6      & -9.6      & -14.2     & -10.2     & -9.5      & -8.6      & -17.3     & -8.0      & -7.2      & -7.2      & -7.4      & -7.0      & -6.8      & -6.8      & -6.9      & -6.7      & -6.6      & -6.5       \\
	${\rm Si^{+}}$                & -4.8      & -4.6      & -4.6      & -4.6      & -4.5      & -9.0      & -5.3      & -4.7      & -3.8      & -11.5     & -3.5      & -2.5      & -2.5      & -2.9      & -2.5      & -2.2      & -2.2      & -2.6      & -2.3      & -2.0      & -2.0       \\
	${\rm Si^{2+}}$               & -3.0      & -2.9      & -2.7      & -2.9      & -2.9      & -6.0      & -3.4      & -2.9      & -2.2      & -7.6      & -2.0      & -1.2      & -1.2      & -1.5      & -1.2      & -1.0      & -1.0      & -1.4      & -1.0      & -0.9      & -0.9       \\
	${\rm Si^{3+}}$               & -2.4      & -2.8      & -2.4      & -2.8      & -2.6      & -3.8      & -3.0      & -2.8      & -2.1      & -4.5      & -2.2      & -1.4      & -1.3      & -1.8      & -1.4      & -1.2      & -1.1      & -1.6      & -1.2      & -1.1      & -1.0       \\
	${\rm Si^{4+}}$               & -1.3      & -1.5      & -1.6      & -1.5      & -1.4      & -1.3      & -1.6      & -1.5      & -1.2      & -1.5      & -1.3      & -1.1      & -1.1      & -1.2      & -1.1      & -1.0      & -1.0      & -1.2      & -1.1      & -1.0      & -1.0       \\
	${\rm Si^{5+}}$               & -0.0      & -0.0      & -0.0      & -0.0      & -0.0      & -0.0      & -0.0      & -0.0      & -0.0      & -0.0      & -0.0      & -0.1      & -0.1      & -0.0      & -0.1      & -0.1      & -0.1      & -0.1      & -0.1      & -0.2      & -0.2       \\
	\hline
	${\rm S}$                     & -11.4     & -8.4      & -7.7      & -8.4      & -9.3      & -11.0     & -7.7      & -7.7      & -8.4      & -14.6     & -7.3      & -7.7      & -7.9      & -7.0      & -7.1      & -7.5      & -7.4      & -6.8      & -7.0      & -7.2      & -7.2       \\
	${\rm S^{+}}$                 & -5.6      & -3.5      & -3.1      & -3.5      & -4.2      & -6.1      & -3.1      & -3.1      & -3.7      & -9.2      & -2.9      & -3.2      & -3.3      & -2.8      & -2.9      & -3.1      & -3.1      & -2.7      & -2.8      & -2.9      & -2.9       \\
	${\rm S^{2+}}$                & -2.7      & -1.4      & -1.0      & -1.4      & -1.9      & -3.2      & -1.1      & -1.1      & -1.7      & -5.4      & -1.1      & -1.3      & -1.4      & -1.1      & -1.1      & -1.3      & -1.3      & -1.1      & -1.1      & -1.2      & -1.2       \\
	${\rm S^{3+}}$                & -1.2      & -0.5      & -0.4      & -0.5      & -0.7      & -1.3      & -0.4      & -0.5      & -0.8      & -2.5      & -0.5      & -0.7      & -0.8      & -0.5      & -0.6      & -0.7      & -0.8      & -0.6      & -0.6      & -0.7      & -0.8       \\
	${\rm S^{4+}}$                & -0.0      & -0.2      & -0.3      & -0.2      & -0.1      & -0.0      & -0.3      & -0.2      & -0.1      & -0.0      & -0.2      & -0.1      & -0.1      & -0.2      & -0.2      & -0.1      & -0.1      & -0.2      & -0.2      & -0.1      & -0.1       \\
	\hline
	${\rm Ca}$                    & -12.7     & -12.5     & -10.9     & -12.5     & -12.9     & -11.1     & -10.3     & -10.8     & -11.4     & -14.4     & -9.9      & -8.3      & -8.0      & -9.1      & -8.2      & -7.6      & -7.4      & -7.0      & -7.7      & -7.2      & -7.3       \\
	${\rm Ca^{+}}$                & -7.9      & -7.6      & -5.8      & -7.6      & -8.1      & -5.3      & -5.2      & -5.8      & -7.0      & -7.9      & -5.1      & -4.8      & -4.6      & -4.7      & -4.7      & -4.2      & -4.0      & -4.1      & -4.3      & -3.8      & -3.8       \\
	${\rm Ca^{2+}}$               & -4.2      & -4.0      & -2.6      & -4.0      & -4.3      & -2.2      & -2.1      & -2.7      & -3.6      & -4.3      & -2.1      & -2.4      & -2.4      & -1.7      & -2.1      & -2.1      & -2.1      & -1.7      & -2.0      & -1.9      & -1.9       \\
	${\rm Ca^{3+}}$               & -1.8      & -1.7      & -1.1      & -1.7      & -1.9      & -0.9      & -0.9      & -1.3      & -1.6      & -2.0      & -1.0      & -1.2      & -1.2      & -0.8      & -1.0      & -1.1      & -1.0      & -0.8      & -1.0      & -1.0      & -1.0       \\
	${\rm Ca^{4+}}$               & -0.0      & -0.0      & -0.0      & -0.0      & -0.0      & -0.1      & -0.1      & -0.0      & -0.0      & -0.0      & -0.0      & -0.0      & -0.0      & -0.1      & -0.0      & -0.0      & -0.0      & -0.1      & -0.1      & -0.1      & -0.1       \\
	\hline
	${\rm Fe}$                    & -14.3     & -13.4     & -12.5     & -13.4     & -13.8     & -12.8     & -12.0     & -12.3     & -12.7     & -15.8     & -11.3     & -9.3      & -8.9      & -10.2     & -9.1      & -8.3      & -8.2      & -7.8      & -8.5      & -8.0      & -7.9       \\
	${\rm Fe^{+}}$                & -9.7      & -9.0      & -8.0      & -9.0      & -9.4      & -7.5      & -7.4      & -7.9      & -8.6      & -9.9      & -7.1      & -5.5      & -5.1      & -6.4      & -5.4      & -4.7      & -4.3      & -4.0      & -4.9      & -4.1      & -4.2       \\
	${\rm Fe^{2+}}$               & -5.7      & -5.7      & -4.9      & -5.7      & -5.7      & -4.1      & -4.4      & -4.9      & -5.2      & -5.8      & -4.2      & -3.4      & -3.2      & -3.7      & -3.3      & -2.9      & -2.7      & -2.9      & -2.9      & -2.5      & -2.6       \\
	${\rm Fe^{3+}}$               & -2.4      & -2.5      & -2.2      & -2.5      & -2.4      & -1.7      & -2.0      & -2.3      & -2.2      & -2.6      & -1.9      & -1.7      & -1.6      & -1.6      & -1.6      & -1.5      & -1.4      & -1.6      & -1.5      & -1.3      & -1.3       \\
	${\rm Fe^{4+}}$               & -0.0      & -0.0      & -0.0      & -0.0      & -0.0      & -0.0      & -0.0      & -0.0      & -0.0      & -0.0      & -0.0      & -0.0      & -0.0      & -0.0      & -0.0      & -0.0      & -0.0      & -0.0      & -0.0      & -0.0      & -0.0       \\
	\hline
\hline
\end{tabular}
\begin{tablenotes}
\item	{\bf Notes:} Abundances values are given as log$_{10}(F_i/F_t)$. 
\end{tablenotes}
\end{threeparttable}
}

\end{document}